\documentstyle[12pt,epsf,epsfig]{article} \hoffset -0.5in \textwidth 6.00in 
\textheight
8.5in  \parskip 7pt \openup1\jot \parindent=0.5in
\newskip\humongous \humongous=0pt plus 1000pt minus 1000pt
\def\caja{\mathsurround=0pt}
\def\eqalign#1{\,\vcenter{\openup1\jot \caja
        \ialign{\strut \hfil$\displaystyle{##}$&$
        \displaystyle{{}##}$\hfil\crcr#1\crcr}}\,}
\newif\ifdtup
\def\panorama{\global\dtuptrue \openup1\jot \caja
        \everycr{\noalign{\ifdtup \global\dtupfalse
        \vskip-\lineskiplimit \vskip\normallineskiplimit
        \else \penalty\interdisplaylinepenalty \fi}}}
\def\eqalignno#1{\panorama \tabskip=\humongous 
        \halign to\displaywidth{\hfil$\displaystyle{##}$
        \tabskip=0pt&$\displaystyle{{}##}$\hfil 
        \tabskip=\humongous&\llap{$##$}\tabskip=0pt
        \crcr#1\crcr}}

\def\eqright #1\cr{\noalign{\hfill$\displaystyle{{}#1}$}}
\def\eqleft #1\cr{\noalign{\noindent$\displaystyle{{}#1}$\hfill}}

\def\oldreffmt#1{\rlap{[#1]} \hbox to 2\parindent{}}

\def\figfmt#1{\rlap{Figure {#1}} \hbox to 1in{}}

\def\VEV#1{\left\langle #1\right\rangle}

\def\sectioneq{\def\theequation{\thesection.\arabic{equation}}{\let
\holdsection=\section\def\section{\setcounter{equation}{0}\holdsection}}}%

\newcounter{holdequation}

\def\num{(\refstepcounter{equation}\theequation)}
\def\auto{\eqno(\refstepcounter{equation}\theequation)}
\def\begineq #1\endeq{$$ \refstepcounter{equation}\eqalign{#1}\eqno
	(\theequation) $$}
\def\contlimit{\,{\hbox{$\longrightarrow$}\kern-1.8em\lower1ex
\hbox{${\scriptstyle (a\rightarrow0)}$}}\,}
\def\centeron#1#2{{\setbox0=\hbox{#1}\setbox1=\hbox{#2}\ifdim
\wd1>\wd0\kern.5\wd1\kern-.5\wd0\fi
\copy0\kern-.5\wd0\kern-.5\wd1\copy1\ifdim\wd0>\wd1
\kern.5\wd0\kern-.5\wd1\fi}}
\def\centerover#1#2{\centeron{#1}{\setbox0=\hbox{#1}\setbox
1=\hbox{#2}\raise\ht0\hbox{\raise\dp1\hbox{\copy1}}}}
\def\centerunder#1#2{\centeron{#1}{\setbox0=\hbox{#1}\setbox
1=\hbox{#2}\lower\dp0\hbox{\lower\ht1\hbox{\copy1}}}}
\def\lsim{\;\centeron{\raise.35ex\hbox{$<$}}{\lower.65ex\hbox
{$\sim$}}\;}
\def\gsim{\;\centeron{\raise.35ex\hbox{$>$}}{\lower.65ex\hbox
{$\sim$}}\;}
\def\st#1{\centeron{$#1$}{$/$}}

\def\super#1{\ifmmode \hbox{\textsuper{#1}}\else\textsuper{#1}\fi}
\def\textsuper#1{\newcount\holdspacefactor\holdspacefactor=\spacefactor
$^{#1}$\spacefactor=\holdspacefactor}

\def\getcite#1,{\advance\citenumber by1
\def\getcitearg{#1}\def\lastarg{@}
\ifnum\citenumber=1
\ref{#1}\let\next=\getcite\else\ifx\getcitearg\lastarg\let\next=\relax
\else ,\ref{#1}\let\next=\getcite\fi\fi\next}

\def\pom{{\rm P\kern -0.53em\llap I\,}}
\def\spom{{\rm P\kern -0.36em\llap \small I\,}}
\def\sspom{{\rm P\kern -0.33em\llap \footnotesize I\,}}

\relax
\def\num{(\refstepcounter{equation}\theequation)}
\def\auto{\eqno(\refstepcounter{equation}\theequation)}
\def\begineq #1\endeq{$$ \refstepcounter{equation}\eqalign{#1}\eqno
	(\theequation) $$}
\def\contlimit{\,{\hbox{$\longrightarrow$}\kern-1.8em\lower1ex
\hbox{${\scriptstyle (a\rightarrow0)}$}}\,}
\def\centeron#1#2{{\setbox0=\hbox{#1}\setbox1=\hbox{#2}\ifdim
\wd1>\wd0\kern.5\wd1\kern-.5\wd0\fi
\copy0\kern-.5\wd0\kern-.5\wd1\copy1\ifdim\wd0>\wd1
\kern.5\wd0\kern-.5\wd1\fi}}
\def\centerover#1#2{\centeron{#1}{\setbox0=\hbox{#1}\setbox
1=\hbox{#2}\raise\ht0\hbox{\raise\dp1\hbox{\copy1}}}}
\def\centerunder#1#2{\centeron{#1}{\setbox0=\hbox{#1}\setbox
1=\hbox{#2}\lower\dp0\hbox{\lower\ht1\hbox{\copy1}}}}
\def\lsim{\;\centeron{\raise.35ex\hbox{$<$}}{\lower.65ex\hbox
{$\sim$}}\;}
\def\gsim{\;\centeron{\raise.35ex\hbox{$>$}}{\lower.65ex\hbox
{$\sim$}}\;}
\def\st#1{\centeron{$#1$}{$/$}}

\def\super#1{\ifmmode \hbox{\textsuper{#1}}\else\textsuper{#1}\fi}
\def\textsuper#1{\newcount\holdspacefactor\holdspacefactor=\spacefactor
$^{#1}$\spacefactor=\holdspacefactor}

\def\getcite#1,{\advance\citenumber by1
\ifnum\citenumber=1
\ref{#1}\let\next=\getcite\else\ifx#1@\let\next=\relax
\else ,\ref{#1}\let\next=\getcite\fi\fi\next}

\def\upon #1/#2 {{\textstyle{#1\over #2}}}
\relax
\renewcommand{\thefootnote}{\fnsymbol{footnote}} 

\def\mainhead#1{\setcounter{equation}{0}\addtocounter{section}{1}
  \vbox{\begin{center}\large\bf #1\end{center}}\nobreak\par}
\sectioneq
\def\subhead#1{\bigskip\vbox{\noindent\bf #1}\nobreak\par}

\def\til#1{\centeron{\hbox{$#1$}}{\lower 2ex\hbox{$\char'176$}}}
\def\tild#1{\centeron{\hbox{$\,#1$}}{\lower 2.5ex\hbox{$\char'176$}}}
\def\sumtil{\centeron{\hbox{$\displaystyle\sum$}}{\lower
-1.5ex\hbox{$\widetilde{\phantom{xx}}$}}}
\def\sumtilt{\sum^{\raisebox{-.15mm}{\hspace{-1.75mm}$\widetilde{}$}}\ }

\def\kbar{\underline{k}}

\def\pom{{\rm P\kern -0.53em\llap I\,}}
\def\spom{{\rm P\kern -0.36em\llap \small I\,}}
\def\sspom{{\rm P\kern -0.33em\llap \footnotesize I\,}}

\newcommand{\bit}{\begin{itemize}}
\newcommand{\eit}{\end{itemize}}

\newcommand{\beq}{\begin{equation}}
\newcommand{\eeq}{\end{equation}}
\newcommand{\beqa}{\begin{eqnarray}}
\newcommand{\eeqa}{\end{eqnarray}}

\begin{document} 
\begin{titlepage} 

\rightline{\vbox{\halign{&#\hfil\cr
&ANL-HEP-PR-97-95\cr
&\today\cr}}} 
\vspace{0.25in} 

\begin{center} 
 
{\large\bf 
CONFINEMENT AND THE SUPERCRITICAL POMERON IN QCD }\footnote{Work 
supported by the U.S.
Department of Energy, Division of High Energy Physics, \newline Contracts
W-31-109-ENG-38 and DEFG05-86-ER-40272} 
\medskip

Alan. R. White\footnote{arw@hep.anl.gov }

\vskip 0.6cm

\centerline{High Energy Physics Division}
\centerline{Argonne National Laboratory}
\centerline{9700 South Cass, Il 60439, USA.}
\vspace{0.5cm}

\end{center}

\begin{abstract} 

Deep-inelastic diffractive scaling violations have provided fundamental
insight into the QCD pomeron, suggesting a single gluon inner structure rather
than that of a perturbative two-gluon bound state. This paper derives 
a high-energy, transverse 
momentum cut-off, confining solution of QCD. The pomeron, in first 
approximation, is a single reggeized gluon plus a ``wee parton'' component
that compensates for the color and particle properties of the gluon. 
This solution corresponds to a supercritical phase of Reggeon Field Theory.

Beginning with the multi-regge behavior of massive quark and gluon
amplitudes, reggeon unitarity is used to derive a reggeon diagram 
description of a wide class of multi-regge amplitudes, including those 
describing the formation and scattering of bound-state Regge poles. When quark
and gluon masses are taken to zero, a logarithmic divergence is produced by 
helicity-flip reggeon interactions containing the infra-red quark triangle 
anomaly. With the gauge symmetry partially broken, this divergence selects the
bound states and amplitudes of a confining theory. Both the pomeron and
hadrons have an anomalous color parity wee-parton component. For the pomeron
the wee parton component determines that it carries negative color charge
parity and that the leading singularity is an isolated Regge pole. 

\end{abstract}

\renewcommand{\thefootnote}{\arabic{footnote}} \end{titlepage}

\mainhead{1. INTRODUCTION}

This is the first of two articles that will report our recent progress in 
``understanding the pomeron in QCD''. A complete understanding of the 
pomeron requires no more or less than solving the theory at high-energy.
While high-energy can be expected to keep the theory as close as possible to
perturbation theory, nevertheless the non-perturbative properties of
confinement and chiral symmetry breaking must emerge. Therefore this paper
(and that following) necessarily also reports progress in ``understanding
confinement and chiral symmetry breaking''. 

Our formalism is entirely based within the high-energy S-Matrix. We start
with the multi-regge behavior of massive quarks and gluons and arrive at the
S-Matrix for hadrons via an extended analysis of infra-red divergences
within multi-regge amplitudes. Rather than appearing as consequences of a
non-perturbative vacuum, both confinement and chiral symmetry breaking are
properties of the bound-state (Regge pole) spectrum. It is a crucial
strength of the multi-regge formalism that we can
simultaneously study the formation of bound states and their scattering
amplitudes. Hadrons, and the pomeron by which they scatter, emerge together
as Regge pole states at spacelike momentum transfer. Indeed, there is a
close link between confinement, chiral symmetry breaking, and the Regge pole
property of the pomeron. 

The main purpose of this first paper is to establish the relationship, that we
initially suggested over seventeen years ago\cite{arw}, between a 
supercritical pomeron phase of Reggeon Field Theory\cite{gr} (RFT) and a
confining solution of QCD with the gauge symmetry broken to SU(2)
(``partially-broken QCD''). In this phase the pomeron is, approximately, an
SU(2) singlet reggeized gluon plus 
a ``wee parton'' component that compensates for the 
particle properties of the gluon. The restoration of SU(3) gauge 
symmetry is directly related to the critical behavior\cite{cri} of the pomeron.
However, in the RFT formalism, the 
transverse momentum cut-off is a relevant parameter at the critical phase 
transition. This implies that the supercritical phase 
can appear with the full gauge symmetry if a physical cut-off is present.
Alternatively, the large $Q^2$ of deep-inelastic scattering can be 
viewed as introducing a (local) lower tranverse momentum cut-off 
which effectively removes the critical behavior altogether and (locally) 
keeps the theory in the 
supercritical phase as the full gauge symmetry is restored. 

We will postpone, until the second paper, almost all discussion of the many
issues of principle and interpretation involved in connecting our results to
other, more conventional, field theory formalisms. However, 
if our results can be interpreted within a field-theoretic framework, 
it is likely to be that of light-cone quantization. In this formalism it is
hoped\cite{kw} that the zero mode (zero longitudinal momentum) 
component of physical states can reproduce the non-trivial vacuum properties
of confinement and chiral symmetry breaking. At infinite momentum the ``zero
modes'' are simply the ``wee partons'' - carrying finite momentum. 
Correspondingly, in our solution of partially-broken QCD, both the pomeron
and hadrons have a zero momentum component
which we refer to as a ``wee-parton component''.  This component, which in
the past we have called a ``reggeon condensate'', is closely related to the
fermion anomaly and carries ``anomalous'' color parity (i.e. it contains
vector-like multi-gluon combinations carrying positive color parity, c.f.
the three gluon component of the winding-number current.) 

The anomalous color parity of the wee parton component determines that the
pomeron carries negative color charge parity overall and also that it's 
leading singularity is a Regge pole with a trajectory that is exchange
degenerate with that of a massive, reggeized, gluon. There is confinement in
that the states carry color-zero and have a completeness property and also 
there are no massless multigluon states. Note that the BFKL pomeron\cite{bfkl}
appears in the positive color parity sector. Our analysis implies that it
does not couple to the physical states. As we will discuss in detail in the
second paper, 
the color parity property of the wee parton component also determines the
chiral symmetry breaking nature of the hadron spectrum. In fact, without
chiral symmetry breaking it would be inconsistent for a negative color
parity pomeron to describe total cross-sections and the BFKL pomeron would 
not decouple. While it may eventually be
possible to formulate our solution in terms of a light-cone quantization
procedure which leads directly to the correct properties of physical states,
we would like to emphasize that we have been able to understand the physics
of the wee-parton component only by determining the role of the
fermion anomaly in the construction of the fully unitary, high-energy, 
multiparticle S-Matrix. This is a very complicated and intricate problem 
which it is hard to imagine studying outside of the multi-regge
framework we use.
 
The discovery of deep-inelastic scaling provided the impetus for the original 
development of the parton model and underlaid the formulation of QCD as the 
theory of the strong interaction. Deep-inelastic scaling violations now 
provide much of the information on short-distance partonic structure that is
the basis for the application of perturbative QCD to a wide range of
hadronic physics. We believe that the observation\cite{hera} of diffractive
deep-inelastic scattering at HERA will turn out to be almost as significant in
developing an understanding of how QCD describes strong interaction physics.
This is because it tells us how the parton model operates beyond the
simplest short-distance processes and, in doing so, provides vital
information on the wee-parton component of physical states. The pomeron
which, as we have already implied, is deeply tied to the 
long distance dynamics of confinement and chiral symmetry breaking, is 
studied experimentally at short distances. By analysing diffractive scaling
violations H1 have shown\cite{h1} that, in deep-inelastic scattering, the
pomeron behaves like a single gluon (rather than the perturbative two-gluon
bound state BFKL pomeron\cite{bfkl}). Within perturbative QCD, gauge
invariance makes this is a very difficult property to realize. 
From our perspective, the H1 analysis \cite{h1} implies\cite{arw3} that at
intermediate $Q^2$ values the pomeron is effectively in the supercritical
phase. The phenomenon can also be understood directly within QCD, once the
physics of the wee-parton component is incorporated\cite{arw3}. 

We first suggested that the pomeron could appear as a single
(reggeized) gluon in \cite{arw} . The idea that the pomeron should carry 
negative color parity and that this is closely tied to chiral symmetry 
breaking was also present. Although this long paper was accepted for
publication, the journal insisted it be split in two. After eventually 
conceding this point, we then decided that further development was needed
before ``final'' publication. We first attempted to do this in \cite{arwlc}
by (partially) recasting the S-Matrix language of \cite{arw} in the more
field-theoretic language of light-cone quantization. However, essentially
because of problems with our treatment of the fermion anomaly, the results 
were still unsatisfactory. We then returned to the S-Matrix formalism of
multi-regge theory and, in two lengthy articles\cite{arw1,arw2}, laid out
what we hoped could be developed into a complete dynamical understanding of
the pomeron in QCD. As in our original paper\cite{arw} (and the present 
paper), our aim was to use general multi-regge theory to carry out a
combined infra-red and multi-regge limit analysis. The essential idea being
always the association of supercritical RFT with partially-broken QCD and
the identification of the critical pomeron phase-transition\cite{cri} with
the restoration of the full gauge symmetry. 

Unfortunately the arguments presented in \cite{arw2} were still very
incomplete. Even so, they gave a fundamentally different picture of the
pomeron to what might be called the conventional, perturbative, BFKL
picture\cite{bfkl}. 
In addition to the incompleteness of the arguments, the techniques we were
using were (and still are) unfamiliar to most theorists studying QCD. The
analysis also depended on our version of the supercritical pomeron which
was the subject of heated controversy in the pre-QCD years of
RFT\cite{scrft}. As a result, we anticipated that the validity of our
arguments would take many years of theoretical study to resolve. We
certainly did not anticipate that experiment could play a role in what we
regarded as fundamentally a (deep) theoretical issue. 

Remarkably, as we discussed above, it now appears that experiment is
providing significant support for our picture. The experimental results have
encouraged us to return to our earlier work and make another major effort to
put it on firmer ground and to make it accessible. The outcome is the
present article (and it's successor). This time around we believe we really
have solved the problem. A major reason for the incompleteness of our 
earlier work was ignorance as to how to construct the complicated reggeon 
diagrams that are necessary to discuss the simultaneous formation and
scattering of bound-states. The solution of this problem via reggeon 
unitarity and the realization of the special role played by ``helicity-flip
vertices'' is, we believe, a significant achievement of the early Sections
of this paper. Helicity-flip vertices only appear as interactions 
coupling dynamically different reggeon channels. They do not appear as 
interactions within the normal reggeon diagrams that, for example, generate 
pomeron RFT. The other central difficulty in our previous work was that, 
although we understood qualitatively that the fermion anomaly should have a
crucial infra-red dynamical role, we were unable to pin down specifically
how this is the case. The inter-relation with ultra-violet regularization
seemed inevitably to lead off into unresolvable field-theoretic
complications. In fact the solution of the formal reggeon diagram problem
has led us to the realization that the anomaly enters just in the
helicity-flip vertices. In our new development the anomaly plays a
straightforward infra-red role (although ultra-violet 
regularization is still involved). As a result, it is clear that 
the infra-red divergence phenomenon we have been searching for is 
(when the gauge symmetry is broken to SU(2)) a very simple overall ``volume''
divergence directly related to confinement. Although the global
picture we presented in our previous papers re-emerges, the details are
different in very important ways. 

RFT is not a conventional field theory. It is really just a diagrammatic 
technique set in field-theoretic language\cite{gr}. Since it has a non-hermitian
interaction, it is not apparent that there is any kind of ``vacuum state''
in the theory. As a result the physical meaning of a ``vacuum 
expectation value'' for the pomeron field, together with the consequent 
``vacuum production of pomerons'', has always been particularly
elusive. This was, at least partially, responsible for the disagreement
about the nature of the supercritical phase\cite{scrft}. The pomeron field
effectively describes the ``wee particle'' distribution in a scattering
hadron. Therefore it is natural that a vacuum expectation value for this field
could be associated with a zero mode contribution in the light-cone language
and so represent non-trivial vacuum properties of the underlying theory. If
this is the case, then the physical context for our supercritical 
solution, which does involve a pomeron vacuum expectation value, is an
underlying theory with a non-trivial vacuum. In particular, an understanding
of the QCD pomeron may be essential. (Technically, it is the presence of
helicity-flip vertices containing the anomaly which provides a meaning 
for reggeon vacuum production.) Since this was certainly not 
available at the time of the controversy concerning the nature of the
supercritical phase it is, perhaps, not surprising that the issue remained
unresolved. Conversely, as we will see is indeed the case, the
supercritical pomeron may be a valuable high-energy formalism for describing
the role of the vacuum properties of QCD. 

In this first paper we will concentrate on the development and application
of the S-Matrix technical machinery that is the basis for our arguments. As
we noted above, we want to reserve all 
field-theoretic discussion of the interpretation and significance of our 
results for the second paper. For the purposes of this paper we could even 
define QCD as the massless limit of a theory of massive, reggeized, vector
particles (gluons) with SU(3) quantum numbers, whose interactions satisfy
(reggeon) Ward identities as a condition of gauge invariance, and which
couple to quarks with the usual vector interaction. In practise, though, we 
will use Feynman diagrams as a direct tool to construct the reggeon 
interactions we discuss. The infra-red problems we consider involve taking a
subset, or all of, the gluon masses to zero and also taking the quark mass
to zero. Since the solution of reggeon unitarity  by reggeon diagrams is an
infra-red approximation, a (gauge-invariant) transverse momentum cut-off is
always implicitly present in our analysis. Consequently we 
could\cite{arw,arw2}
specifically formulate our discussion in terms of the Higgs mechanism for
spontaneous symmetry breaking and appeal to complimentarity\cite{fs} to
justify using the massless limit to define QCD. However, for this first
paper, we will minimize references to specific field-theoretic assumptions
that could be made since, in our experience, this often serves only to
confuse the reader as to the issues involved. 

For related reasons, we will reserve discussion of a number of topics for
the second paper. These include chiral symmetry-breaking and the quark
bound-state spectrum, deep-inelastic diffractive scaling violations - the
implications for perturbative QCD and the parton model, the RFT formulation
of both the supercritical pomeron and the critical pomeron, the restoration
of full SU(3) symmetry and the dependence on the
quark flavor spectrum. Our aim in this paper
is to simply expose the infra-red massless quark problem related to the
anomaly and to show, in a self-contained manner, that this leads to a
confining solution of partially-broken QCD. We will identify all the
elements of supercritical pomeron behaviour but, as we just implied, we
will not discuss the RFT formulation in any detail. 

\newpage

\mainhead{2. OUTLINE OF THE ARGUMENTS}

If a theory is ``reggeized'', that is all the particles lie on Regge
trajectories, it is not unreasonable to expect that the full S-Matrix is
then determined by the corresponding tree diagrams. If all the multiparticle
amplitudes containing the poles due to the stable particles of the theory
can be found, reggeization should imply there is no subtraction ambiguity in
constructing the full 
amplitudes of the theory dispersively via unitarity. (In practise there is no
formulation of such a program. Although recent ``unitarity based''
calculations\cite{bdk} of loop amplitudes in QCD and supersymmetric gauge
theories partially illustrate the principle. The loop expansion for
string theories is perhaps an illustration of the essence of the argument.) 
Reggeization also implies that 
the tree amplitudes can be found by studying the behaviour of all
multiparticle amplitudes in multi-regge limits. The leading Regge pole
trajectories in each quantum number $t$-channel are directly associated with
a corresponding particle (or resonance) and, at the particle poles, Regge 
pole amplitudes give the corresponding particle amplitudes. Since QCD is
believed to be a bound state theory in which all the particle states lie on
Regge trajectories, studying multi-regge limits should be a direct way to
study the particle spectrum. 

In the vacuum quantum number $t$-channel, however, the leading Regge pole
is the pomeron. The pomeron is even signature and probably (in our view) has 
no particles on it's trajectory. The pomeron determines, in particular, the 
high-energy elastic scattering amplitudes of the particles in the theory. In
this, and the following paper, we will see that we can extract both the
particle spectrum and the high-energy amplitudes that correspond to the
pomeron, by studying multi-regge limits.

During the period that quantum field theory was out of vogue, very extensive
analyticity methods were developed\cite{arw1,sw} to study multi-regge
behavior and its inter-relation with unitarity. The analyticity domains
for multiparticle amplitudes derived within the formalisms of ``Axiomatic
Field Theory'' and ``Axiomatic S-Matrix Theory' were the basis for this
abstract analysis. All the assumptions made within these formalisms are
expected to be valid in a completely massive spontaneously-broken gauge
theory and, as we discussed in the Introduction, the S-Matrix of such a
theory can be thought of as the starting point for our analysis of QCD. 

The abstract formalism remains little known and so in Section 3 we both
summarize and develop the contents of our previous 
papers\cite{arw1}. We emphasize those results required for the rest of the
paper. The most important point, which we do not elaborate explicitly in this
paper, is that there are relatively simple, many-variable, domains of
analyticity in the multi-regge asymptotic regime and 
corresponding multiparticle dispersion relations are valid. Consequently,
generalized ``Sommerfeld-Watson representations'' exist which imply that all
multiparticle asymptotic behavior is strongly constrained by ``cross-channel"
multiparticle unitarity continued in complex angular momenta
and helicity variables. These constraints are embodied in the general
``reggeon unitarity equations'', which hold in every complex angular 
momentum and helicity plane and control multi-regge exchanges in all 
amplitudes. These equations were first proposed in \cite{gpt}. At the time 
they were a 
remarkable ``all-orders'' generalization of results found in lowest-order field 
theory models of Regge cut behaviour. However, the full dispersion theory 
basis for multi-regge theory had to be developed before the validity and 
generality of the reggeon unitarity  equations could be established\cite{arw1}.
Given the reggeization of gluons and quarks, the (essentially) factorizing
nature of the reggeon unitarity equations implies the very powerful
consequence that the multi-regge behavior of all QCD multiparticle
amplitudes is built up from elementary components, many of which are already
known from existing calculations of elastic scattering production processes.

In Section 4 we apply the general formalism of Section 3 to the 
special case of triple-regge kinematics. For our purposes, it is important 
that the conventional ``triple-regge'' limit of the 
one-particle inclusive cross-section is only the simplest kinematical 
situation in which triple-regge behavior appears. We show that in the full
triple-regge limit, and also in what we term a helicity-flip 
helicity-pole limit, new ``helicity-flip vertices'' appear. These vertices 
are generated by amplitudes with distinctive combinations of invariant cuts. 
We also formulate the additional limits in terms of large light-cone 
momenta. This is important in Section 5 for building up 
the very complicated multi-reggeon diagrams that we use in later Sections.

The initial discussion in Section 5 is concerned with the similarity between RFT
pomeron diagrams and the reggeon diagrams that describe Regge limit 
calculations in QCD. Both sets of diagrams can 
be regarded as explicit solutions of the reggeon unitarity equations.
The remainder of the Section is devoted to the task of constructing the
reggeon diagrams that in QCD will contain the bound state hadron and pomeron
behaviour that we are looking for. The essential point is that in a general 
class of limits, that we call ``maximal helicity-pole limits'', only a single 
analytically-continued multiparticle partial-wave amplitude  appears,
related to leading-helicity particle amplitude. Such partial-wave 
amplitudes straightforwardly satisfy reggeon unitarity equations in each
$t$-channel and, as a result, have a reggeon diagram description 
in terms of two-dimensional transverse momentum integrals. We show,
however, that when a helicity-flip vertex is involved the reduction to 
transverse momentum integrals is more subtle. In this case, if a light-cone
description of the limits is formulated, a correlated light-like vector is
necessarily part of the ``physical transverse plane''. This 
longitudinal component vanishes with the corresponding transverse momentum.

We begin our QCD analysis in Section 6. We show first how elementary 
quark-reggeon couplings are obtained by calculating successive on-shell 
scatterings of fast quarks. We then discuss the derivation of reggeon Ward 
identities from gauge invariance gluon Ward identities. We show that 
quark scattering reggeon diagrams have infra-red divergences and trace 
the related failure of reggeon Ward identities to the restricted Regge limit
kinematics of on-shell elastic scattering. After discussing how the reggeon
Ward identities are satisfied in high-order reggeon interactions we note
that there is an ultra-violet divergence problem in the quark-loops
contributing to triple-Regge vertices. To obtain the reggeon Ward identities
for massive quark loops, it is necessary to introduce Pauli-Villars 
regulator fermions. These provide a unitarity-violating ultra-violet cut-off 
in the quark sector which we ultimately remove only after the massless quark
limit is taken. 

In Section 7 we show how the triangle quark loop diagram appears in 
triple-Regge helicity-flip vertices coupling multi-reggeon states. We show 
that the presence of the triangle singularity leads to a non-uniformity in 
the massless and zero transverse momentum limits for such vertices. We 
identify the momentum and color structure of this ``anomaly''. As 
we discuss, it is essentially the infra-red appearance of the U(1) axial 
anomaly. It's appearance in reggeon diagrams is a subtle effect, related
to the presence of non-local infra-red axial-like couplings for
multi-reggeon states. We show that anomalous color parity reggeon states 
(with distinct color parity and signature) must be involved. 

The infra-red divergence phenomenon producing confinement is described in 
Section 8. We show that in the limit of zero quark mass the triangle
anomaly, combined with the Pauli-Villars regularization procedure, leads to
the violation of reggeon Ward identities in a complicated set of reggeon
diagrams. In such diagrams helicity flip interactions of anomalous reggeon
states accompany the non-flip interactions of normal reggeon states. We
argue that the resulting logarithmic divergence cancels in the sum of such
diagrams when the gauge symmetry is unbroken. However, when the gauge
symmetry of QCD is partially-broken to SU(2), the divergence does not
cancel but rather selects the ``physical amplitudes''. The physical states we
identify contain massive SU(2) singlet reggeons with a zero momentum
anomalous odderon component that acts like a background wee parton 
component, or ``reggeon  condensate''. We show that we have 
a ``confinement phenomenon'' in that two initial physical 
reggeon states states scatter only into arbitrary numbers of the same 
physical states. We also have confinement
in the sense that, in the gluon sector, we have only massive 
reggeon states composed of elementary Regge pole constituents. We postpone 
discussion of the quark states and chiral symmetry breaking until the 
next paper. 

It is presumably important that because the zero transverse momenta in the
reggeon condensate are implicitly accompanied by longitudinal zero momenta,
the condensate can potentially be understood as a zero-mode effect in 
light-cone quantization and as a wee parton component at infinite
momentum. We summarize our confining solution of partially-broken QCD as
containing exchange
degenerate even and odd signature reggeons together with 
vacuum production of
multi-reggeon states. These are the defining characteristics of
supercritical pomeron RFT. 

\newpage

\mainhead{3. MULTI-REGGE LIMITS AND REGGEON UNITARITY}

In this Section we describe the general multi-regge theory that will underly 
the analysis and arguments of this paper. In many cases a more 
extensive discussion of the subjects we cover can be found in \cite{arw1}
and a very useful background review is provided by \cite{bdw}. However, as we
noted in the previous Section, we will also need additional elements that
were not adequately described in \cite{arw1}. 
We first describe the general kinematics and partial-wave analyses which are
the basis of multi-regge theory. 

\subhead{3.1 Toller Diagrams and Little Group Variables}

To describe the most general Regge behavior of a multiparticle amplitude we
first introduce a set of angular variables. For a given amplitude,
there are many possible sets, each associated with a distinct Toller
diagram. A Toller diagram is simply a tree diagram with only
three-point vertices. 

Denoting the external momenta for an $N$-point amplitude by $P_i$,
$i=1,\ldots, N$, we begin by drawing a Toller diagram and introducing internal 
momenta $Q_j$, $j=1,\ldots,N-3$ for each
internal line of the diagram as illustrated in Fig.~3.1. 

\begin{center}
\leavevmode
\epsfxsize=3.5in
\epsffile{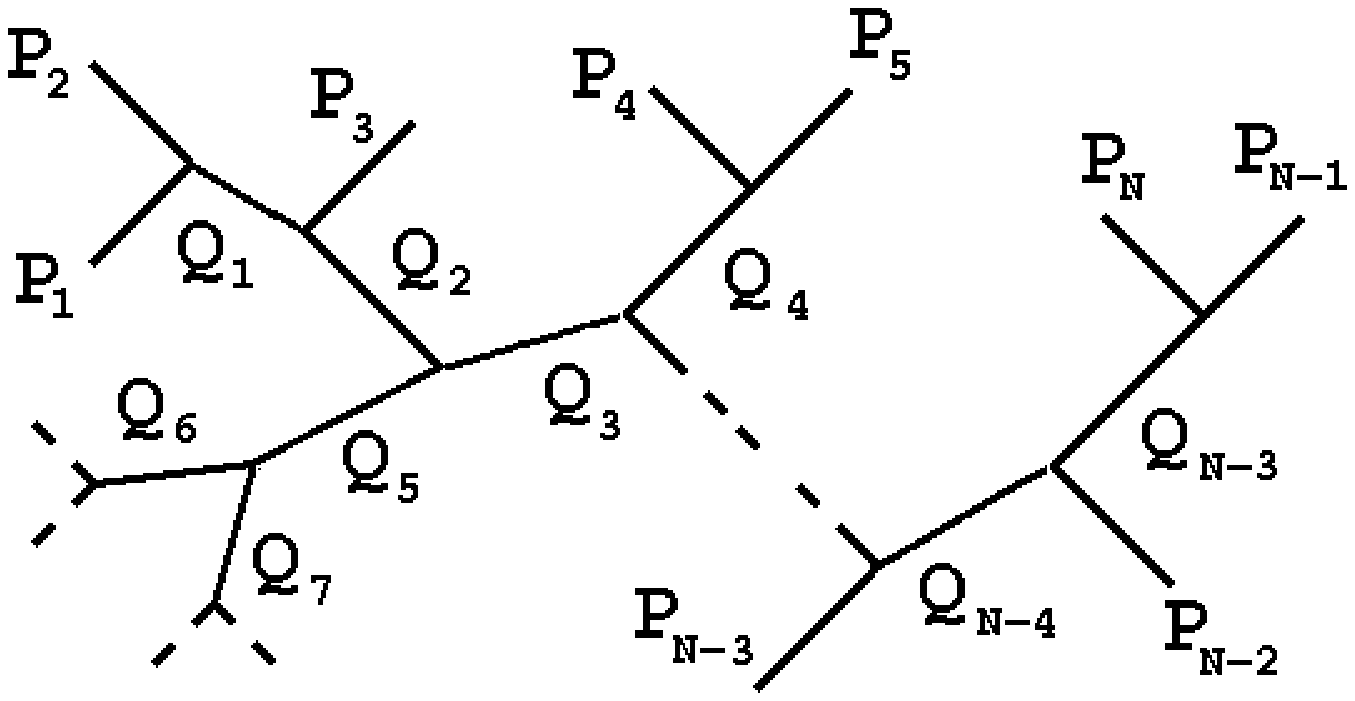}

Fig.~3.1 A Toller Diagram for the N-point Amplitude
\end{center}
The $Q_j$ are
defined by imposing momentum conservation at each vertex.
Next we introduce three standard Lorentz frames at each vertex,
in each of
which one of the three momenta entering the vertex has a standard form
- chosen according to some convention. We then 
denote as $g_j$ the Lorentz transformation---associated with the internal
line $j$---which transforms between the two standard frames, in which $Q_j$
has the standard form, defined respectively at the two vertices to which the
line $j$ is attached. Since $Q_j$ has the same form, say $Q^0_j$, in both
standard frames, $g_j$ necessarily belongs to the little group of $Q^0_j$
implying that 
$$
\eqalign{
g_j&\in {}\hbox{SO(2,1) if $Q_j$ is spacelike}\cr
g_j&\in {}\hbox{SO(3) if $Q_j$ is timelike}.}\auto\label{st}
$$

We also introduce the Lorentz transformations $\zeta_{jk}$ transforming
between the standard frames defined for $Q_j$ and $Q_k$ respectively at
the same vertex. Note that $\zeta_{jk}$ is a function of $t_j=Q^2_j$,
$t_k=Q^2_k$ and $t_\ell=(Q_j+Q_k)^2$ only. We can clearly combine the $g_j$ and
$\zeta_{jk}$ (together with $\zeta_{ij}$ transformations defined analogously
to the $\zeta_{jk}$, but at external vertices) to determine any of the
external momenta in any of the standard frames associated with the Toller
diagram. For an $N$-point amplitude $M_N$ we can therefore write 
$$
M_N(P_1,\ldots,P_N)\equiv
M_N\left(t_1,\ldots,t_{N-3},g_1,\ldots,g_{N-3}\right).\auto\label{amp}
$$

If we initially consider all the $Q_j$ to be timelike then we can use the
SO(3) parametrization 
$$
\eqalign{g=u_z(\mu)u_x(\theta)u_z(\nu)}\qquad
~~~~~~\eqalign{0&\leq \theta<\pi\cr 0&\leq \nu,\ \mu\leq 2\pi},\auto\label{par}
$$
where $u_z$ and $u_x$ are respectively rotations about the $z$ and $x$ axes.
We can also take all the
$\zeta_{jk}$ and $\zeta_{ij}$ to be boosts $a_z(\zeta)$ in the $z$-$t$ plane.
In this case the $u_z$ rotations clearly commute with the $a_z$ and as a
result the external invariant variables depend only on combinations
$w_{jk}=\mu_j-\mu_k$ of azimuthal angles. The net effect is that the
angular variables for each Toller diagram reduce always to the $(3N-10)$
independent variables needed to describe an $N$-point amplitude. There are
always 
$$
\left.
\eqalign{
&(N-3)\hbox{~~~$t_i$ variables }(\equiv Q^2_i)\cr
&(N-3)\hbox{~~~$z_j$ variables }(\equiv \cos\theta_j)\cr
&(N-4)\hbox{~~~$u_{jk}$ variables}(\equiv e^{i\omega_{jk}})}\right\}
\eqalign{
(3N-10)\hbox{~~variables}.}\auto\label{3n10}
$$
For each Toller diagram the $t_j$, $z_j$, and $u_{jk}$ variables are an
unconstrained Lorentz invariant set of variables for an $N$-point
amplitude. 

We will also make use of two parameterizations of ${\rm SO}(2,1)$. The 
first corresponds directly to the SO(3) parametrization (\ref{par}) (with 
cos$\theta \to$ cosh$\beta$) i.e.
$$
\eqalign{
g=u_z(\mu)a_x(\beta)u_z(\nu)}\qquad
~~~~~~\eqalign{
&-\infty<\beta<\infty\cr
&0\leq \mu,\nu\leq 2\pi},\auto\label{3.24}
$$
where $a_x$ is now a boost in the $x-t$ plane. An alternative 
parameterization is
$$                              
\eqalign{
g=u_z(\mu) a_x(\beta)a_y(\gamma)}\qquad
~~~~~~~\eqalign{
&-\infty<\beta,\,\gamma<\infty\cr
         &0\leq \mu\leq 2\pi.}\auto\label{3.37}
$$

\subhead{3.2 Invariants and Angular Variables}

A general multi-regge limit is defined, via a particular Toller diagram, as 
$$
z_1,~z_2,~~...~z_{N-3}~~~\to ~ \infty~, ~~\forall ~t_i, u_{jk} ~fixed
\auto\label{rl}
$$
A variety of ``helicity-pole limits'' in which some combination of the $z_j$
and $u_{jk}$ variables are taken large can also be discussed. The reason for 
the helicity-pole name will be clear after we introduce Sommerfeld Watson 
representations. ``Maximal helicity-pole limits'' in which (in a sense we
will discuss later) the maximum number of $u_{jk}$ variables are taken large
will play an important role in our discussion. The significance of maximal
helicity-pole limits is that they can be used to isolate a single,
analytically continued, ``helicity amplitude''. A multi-regge limit, in
general, has contributions from many different helicity amplitudes. 

It is straightforward to calculate the behavior of channel invariants in
terms of the angular variables. An explicit example, the
six-point amplitude and the angular variables corresponding 
to the Toller diagram of Fig.~3.2, can be
found in the Appendix of \cite{sw}. 
The parametrization (\ref{3.24}) is used and the specific standard frames 
are essentially those we have described. 
\begin{center}
\leavevmode
\epsfxsize=3in
\epsffile{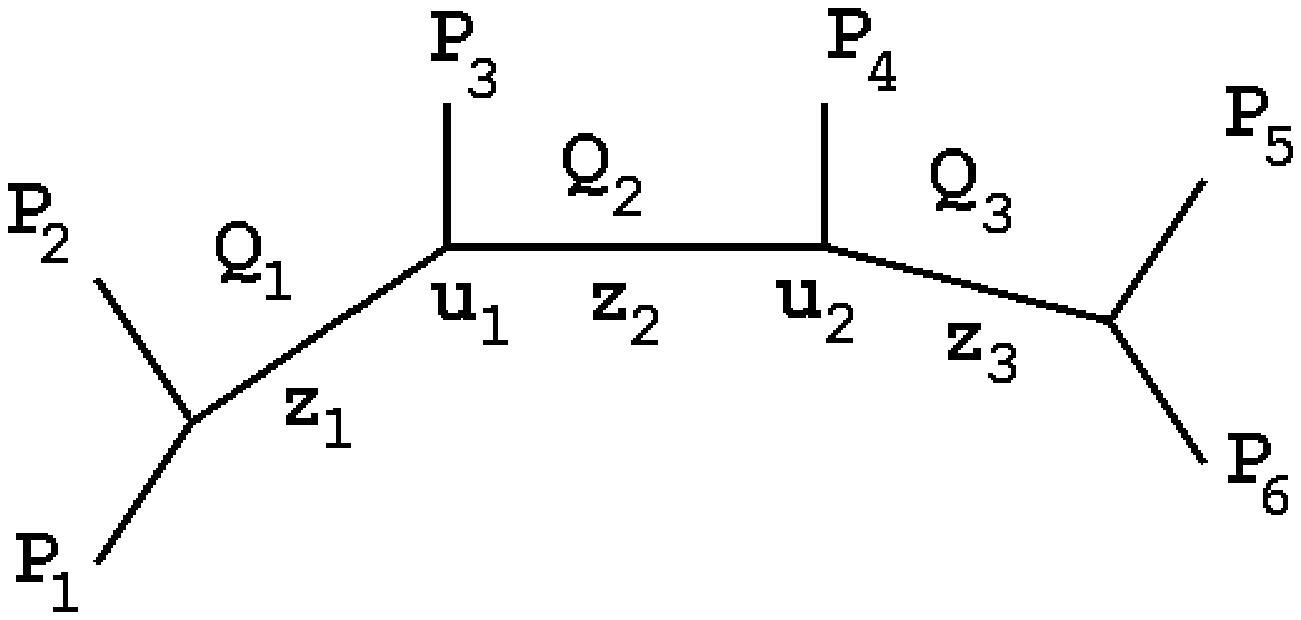}

Fig.~3.2 A Toller Diagram for the 6-point Amplitude
\end{center}
We can also list 
a few of the most important features that appear in general.

\begin{itemize}
\item[3A)] If we write $z_j={1\over 2}\left(v_j+v_j^{-1}\right)$ 
(i.e. $v_j=e^{i\theta j}$) and define $u_{jk}$ as above then all factors of
$i$ in expressions for invariants (coming from $\sin\theta_j$ and
$\sin\omega_{jk}$) cancel. The relation between all invariants and the $u$'s
and $v$'s is real and analytic.
\item[3B)] When all the $z_j$'s are large (or all the $v_j$'s ) we obtain
for $s_{mn}=\left(p_m + p_n\right)^2$
$$
\eqalign{
s_{mn}~
\sim \sinh \zeta_{mj_1}v_{j_1}&\left(\cosh\zeta_{j_1j_2}+
\cos\omega_{j_1j_2}\right)v_{j_2}\cdots v_{j_{r-1}}\cr
\times&\left(\cosh\zeta_{j_{r-1}j_r}+\cos\omega_{j_{r-1}j_r}\right)
v_{j_r}\sinh\zeta_{j_rn},}\auto\label{as1}
$$
where $j_1,\,j_2,\ldots,j_s$ is the set of internal lines of the tree diagram 
linking the two external momenta. As a result, for any invariant $s_{mn\cdots
r}=\left(p_m+p_n+\ldots+p_r\right)^2$ we obtain
$$
s_{mn \ldots r} \;\centerunder{$\large\sim$}{\raisebox{-4mm}{
$z_j\to
\infty\ \forall_j$}}\;~        
f(\til{t},\til{\omega})z_{j_1}z_{j_2}
\cdots z_{j_s},\auto\label{as2}
$$
where now $j_1,\,j_2,\ldots,j_s$ denotes the longest path through
the tree diagram linking any two of the external momenta contained in
$s_{mn\cdots r}$.
\item[3C)] When all the $u_{jk}$'s are large we similarly obtain 
$$
\eqalign{
s_{mn}~\sim &\sinh\zeta_{mj_1}\sin\theta_{j_1}u_{j_1,j_2}(\cos\theta
_{j_2}+1)u_{j_2,j_3}\cdots u_{j_{r-2},j_{r-1}}\cr
&\times\left(\cos\theta_{j_{r-1}}+1\right) u_{j_{r-1},j_r}\sin\theta_{j_r}\sinh
\zeta_{j_rn}.}\auto\label{as3}
$$
Again the leading behavior of any $s_{mn\cdots r}$ is obtained from the 
two-particles linked by the longest path through the tree diagram.
 
\end{itemize}
 
It is important, although we will make little reference to it, that the
singularities of amplitudes as functions of the invariant variables 
have a similar asymptotic structure in terms of either the $z_j$ variables
or the $u_{jk}$ variables. 

\subhead{3.3 Hexagraphs, Direct-Channels and Cross-Channels}

While the Toller diagram is sufficient to introduce angular variables, there 
are many analytic and kinematic properties of amplitudes for which it is 
very useful to introduce a further set of related ``tree diagrams''
called ``hexagraphs''. There are many hexagraphs for each Toller diagram.

A hexagraph is necessarily drawn in a plane. It
has the same number of vertices as the parent Toller diagram but each
internal line of the Toller diagram is replaced by a line containing both
horizontal and sloping elements. The complete set of hexagraphs
corresponding to a Toller Diagram is constructed as follows. 

We begin by substituting for each of the vertices
of the Toller diagram, the sets of vertices shown in Fig.~3.2, in each of 
which one of the $Q_i$ is attached to a horizontal line

\begin{center}
\leavevmode
\epsfxsize=3.5in
\epsffile{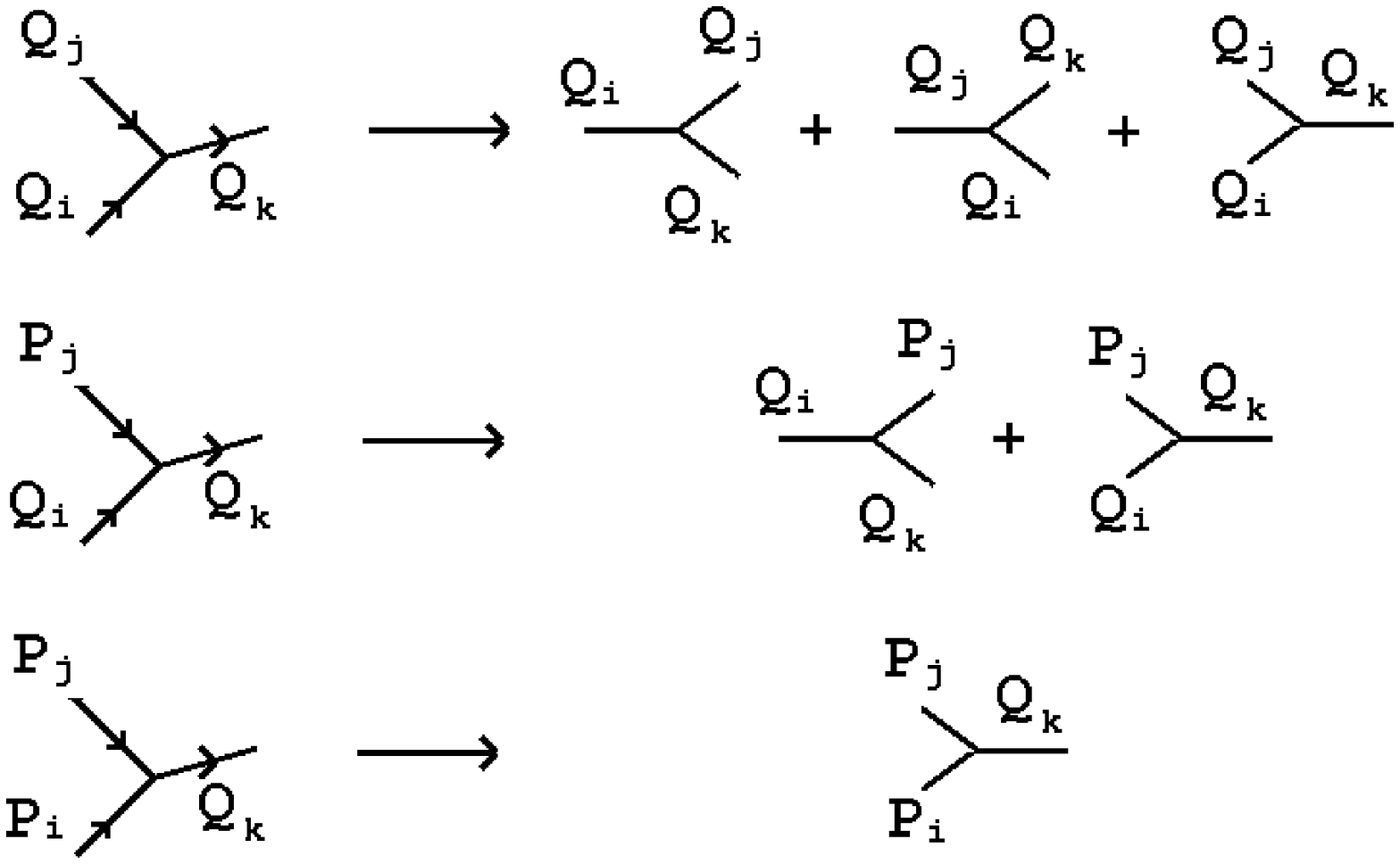}

Fig.~3.3 Hexagraph Vertices from Toller Diagram Vertices. 
\end{center}
(As illustrated the number of vertices substituted depends on the number of
external lines entering the vertex.) We next join the available vertices
with horizontal lines in all possible manners, forming projections on the
plane. Fom the set of graphs obtained we generate further graphs by 
``twisting'' each graph about each internal horizontal line. Twisting
rotates all of that part of the graph attached to one end of the horizontal 
line by $180^o$ relative to the remainder of the graph - turning it
upside-down in the plane. We continue ``twisting'' until no new graphs are
obtained. 

Examples of hexagraphs obtained from the Toller diagram of Fig.~3.1 are
shown in Fig.~3.4. 

\begin{center}
\leavevmode
\epsfxsize=5in
\epsffile{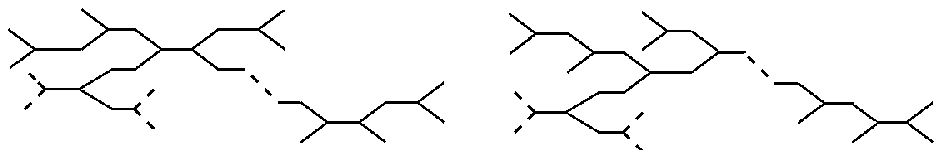}

Fig.~3.4 Examples of Hexagraphs obtained from the Toller Diagram of Fig.~3.1.
\end{center}
One use of a hexagraph is to generalise the elastic scattering concepts of 
the ``$s$-channel'',  or  ``direct-channel'', physical region and 
the ``$t$-channel'', or ``cross-channel'', physical region. Each hexagraph
simultaneously describes an ``$s$-channel'' physical region in which all the
$Q_i$ of the Toller diagram are spacelike and a ``$t$-channel'' physical
region in which all the $Q_i$ are timelike. (Of course, there are also
additional channels in which some $Q_i$ are timelike and some are spacelike
but we will not discuss them specifically.) The 
direct-channel is obtained by interpreting the diagram as describing 
scattering particles entering from the bottom of the diagram and exiting
at the top. The cross-channel is obtained by interpreting the diagram as 
describing scattering particles entering from the left of the diagram and 
exiting to the right. (Since we do not consider 
scattering processes as distinct that differ by an overall TCP 
transformation, we do not consider hexagraphs as distinct that differ only 
by the complete vertical, or horizontal, reflection corresponding to a TCP
transformation of the corresponding direct-channel or cross-channel. As a
result it is irrelevant whether the 
scattering particles enter from the bottom or the top in the direct channel
or whether they enter from the left or the right in the 
cross-channel.) Note that the same cross-channel is described by a class of
distinct direct-channel hexagraphs related by ``twisting''. As we describe 
further below, the process of twisting a hexagraph about a horizontal line
defines the multiparticle generalization of signature. 

The angular variables can be straightforwardly introduced in any physical
region by the procedure described in sub-section 3.1. In a cross-channel 
$$
t_j\geq 4m^2,\quad -1\leq z_j\leq 1,\quad -1\leq \cos\omega_{jk}\leq 1
\auto\label{cro}
$$
For a direct channel the situation is more complicated. Even if all the
$Q_j$ meeting at a vertex are spacelike, the vertex may lie in either a
spacelike or a 
timelike plane (i.e. $\lambda(t_i,t_j,t_k) 
{\raisebox{1mm}{\centerunder{$\scriptscriptstyle 
<$}{$\scriptscriptstyle > $}}} 0$ where $\lambda(t_i,t_j,t_k) = t_i^2 + t_j^2
+ t_k^2 - 2t_it_j - 2t_jt_k - 2t_kt_i$). In that part of a direct-channel in
which all the $Q_j$ are spacelike and all the internal vertices
are timelike 
$$
t_j<0,\quad z_j\geq 1\hbox{ or }\leq 1\quad -1\leq \cos\omega_{jk}\leq 1
\auto\label{dir}
$$
In this kinematic configuration, the multi-regge limit is a physical limit 
but a helicity-pole limit is unphysical. For those parts of a direct channel
where a vertex is spacelike, the physical region is parametrized by the
$\omega_{jk}$ angles becoming boosts as in (\ref{3.37}). In this case both
Regge and helicity-pole limits are physical region limits. 

We will use hexagraphs to describe more and more information as we proceed.
In particular we can associate each $\theta_j$ and each $t_j$ 
with the corresponding horizontal line of the hexagraph, while the 
independent $\omega_{jk}$ can always be associated (in an obvious manner)
with the internal sloping lines. (This association can also be made for the
conjugate $J_j,n_k$ and $n_k'$ variables that we introduce below. It will be
illustrated in Fig.~3.5.) We can then associate a ``twist'' about a
horizontal line of a hexagraph with a change of sign of the corresponding
$z_j$ and also, for a sloping line attached directly to this line 
(not via a vertex), with a change of sign of the corresponding $u_{jk}$.
This is how twisting is used in defining signature. 

\subhead{3.4 Partial-Wave Expansions}

In a cross-channel all the little groups are SO(3). For a general function
$f(g)$ on SO(3) we can write $$ 
f(g)=\sum^\infty_{J=0}\,\sum_{|n|,|n'|<J}D^J_{nn'}(g)a_{J nn'},
\auto\label{pw}
$$
where the $D^J_{nn'}(g)$ are representation functions. For the parametrization 
(\ref{par}) 
$$
D^J_{nn'}(g)=e^{in\mu}d^J_{nn'}(\theta)e^{in'\nu}.\auto\label{dfn}
$$
where the $d^J_{nn'}(\theta)$ are well-known special functions. 
From (\ref{amp}) we can write 
$$
\eqalign{
M_N(\til{t},g_1,\ldots,g_{N-3})={}&\sum^\infty_{J_1=0}\,
\sum_{|n_1|,|n'_1|<J_1}\ldots\sum^\infty_{J_{N-3}=0}~~\,
\sum^\infty_{|n_{N-3}|,|n'_{N-3}|<J_{N-3}}\cr
&D^{J_1}_{n_1n'_1}(g_1)\ldots D^{J_{N-3}}_{n_{N-3},n'_{N-3}}(g_{N-3})
a_{J_1,n_1,n'_1,\ldots,J_{N-3},n_{N-3},n'_{N-3}}(\til{t}).}
\auto\label{pw1}
$$
Since, as we have discussed, each $M_N$ depends only on combinations of the 
azimuthal angles $\mu_j$ and $\nu_j$ there is a related constraint on the
sums over $n_j$ and $n'_j$ in (\ref{pw1}). With the particular 
convention that, at the vertex where lines $j,k,\ell$ meet, the Lorentz
transformations $g_j,g_k,g_\ell$ are defined to transform from this
particular vertex to adjacent vertices this constraint takes the
form 
$$
n_j+n_k+n_\ell=0.\auto\label{hc}
$$
After this constraint is imposed there are $(N-4)$ independent
$n$ and $n'$ indices in (\ref{pw1}) (considering spinless external particles)
which are ``conjugate'' to the $(N-4)$ independent azimuthal angles 
$\omega_{jk}$ introduced above. The $j,n$ and $n'$ variables can be 
associated with the lines of a hexagraph as illustrated in Fig.~3.5

\begin{center}
\leavevmode
\epsfxsize=5in
\epsffile{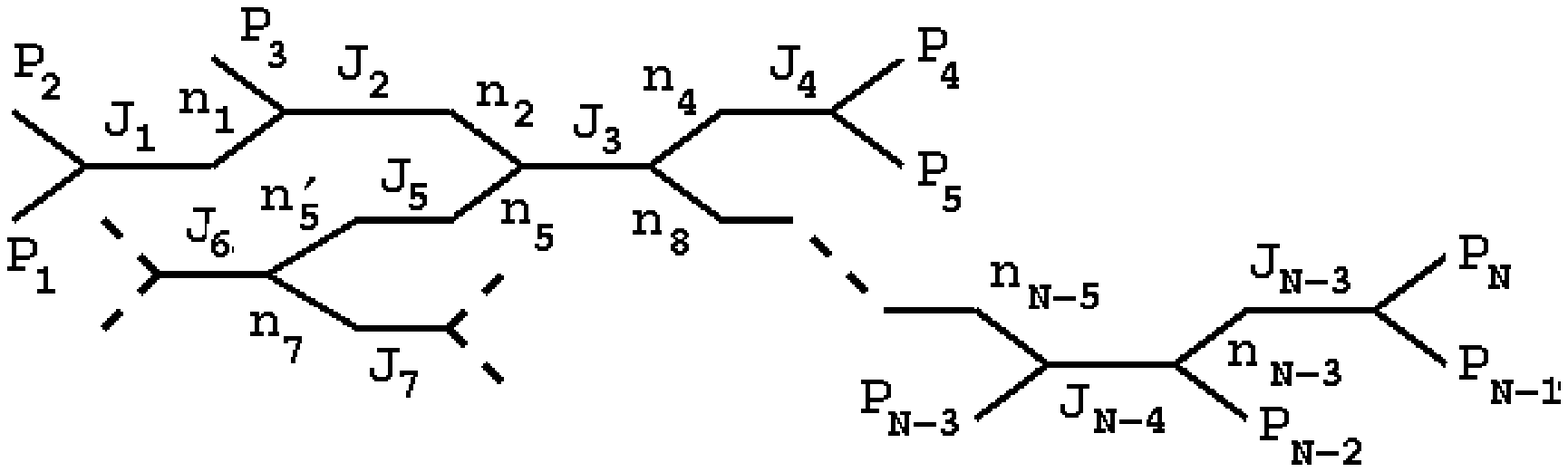}

Fig.~3.5 Association of $j,n $ and $n'$ Indices with the Lines of a 
Hexagraph.

\end{center}

To use the partial-wave expansion to discuss Regge behavior in Regge and 
helicity-pole limits in direct channels we first define continuations of the
partial-wave amplitudes $a_{\til{\scriptstyle J},\til{\scriptstyle n},
\til{\scriptstyle n'}}(\til{t})$ to complex 
values of the angular momenta $J_i$ and the helicities $n_i, n'_i$. This 
will enable us to transform (some of) the summations in (\ref{pw1}) into
integrals via the Sommerfeld-Watson transformation. For this purpose it is
neccessary to break the full amplitude $M_N$ down into spectral components
containing distinct multiple discontinuities in the invariant variables that
are large in the limit discussed. This is achieved by writing an
(asymptotic) dispersion relation in the $z_j$ variables. As we noted in 
SEction 2, the existence of such dispersion relations is actually the
fundamental core of our development of multi-regge theory. However, since 
an understanding of their derivation is not necessary for the purposes of this 
paper we simply go straight to the result. A extended description of the
general derivation can be found in \cite{arw1} and the particular
example corresponding to the Toller diagram of Fig.~3.2 is discussed in
detail in \cite{sw}. 

\subhead{3.5 Asymptotic Dispersion Relations}

A primary purpose of the hexagraph notation is to describe the spectral
contributions to the asymptotic dispersion relation, for an amplitude $M_N$,
obtained by simultaneously dispersing in all the $z_j$ variables of the
parent Toller diagram. 
By introducing the concept of a ``cut'' through a hexagraph we can use such
cuts to describe invariant channels in which there is a discontinuity or
``cut''. For each hexagraph we define an ``allowable'' direct-channel
discontinuity to be in any sub-channel, defined by a sub-set of the external
particles, such that the minimal ``cut'' drawn through the graph connecting
all the particles involved enters and exits only between a pair of
sloping lines. Some allowable cuts of the upper hexagraph in Fig.~3.4 are
shown in Fig.~3.6 

\begin{center}
\leavevmode
\epsfxsize=4.5in
\epsffile{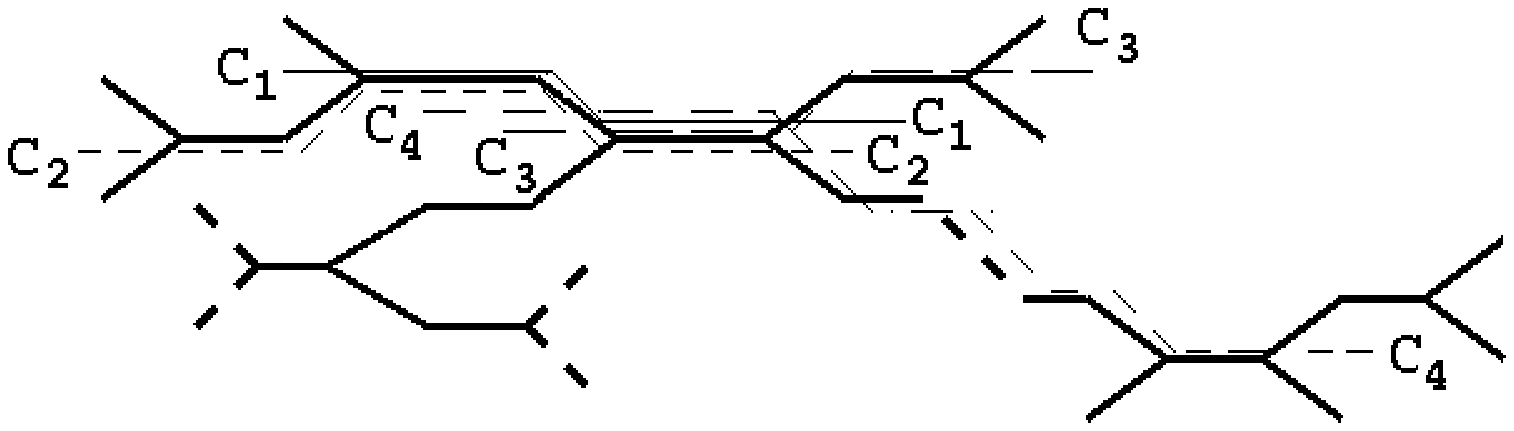}

Fig.~3.6 Allowable Cuts through the Hexagraph of Fig.~3.3.
\end{center}

The asymptotic dispersion relation takes the form
$$
M(p_1,\ldots p_N)=\sum_{H\in T}M^H(p_1,\ldots,p_N)+M^0~,\auto\label{dis}
$$
where the sum is over all hexagraphs $H$ generated by the Toller diagram $T$
and $M^0$ contains only non-leading multi-regge behavior. Each
``hexagraphical component'' $M^H$ is further written as
$$
M^H=\sum_{C\in H}M^C(p_1,\ldots,p_N)~
,\auto\label{dis1}
$$
where now the sum is over all sets $C$ of ($N-3$) non-overlapping cuts which
are (all) allowable cuts of the hexagraph. (In the simplest graphs there 
will be only one set $C$). The ($N-3$) cuts must be ``asymptotically 
distinct'' when all the $z_j$ variables are large. If we denote the
invariant cuts of a particular set $C$ as $(s_1,\ldots,s_{N-3})$ then 
$$  
M^C(p_1,\ldots,p_N)={1\over (2\pi i)^{N-3}}
\int {ds'_1\ldots ds'_{N-3}\Delta^C(\til{t},
\til{w},s'_1,s'_2,\ldots,s'_{n-3})\over
(s'_1-s_1)(s'_2-s_2)\ldots (s'_{N-3}-s_{N-3})},\auto\label{dis2}
$$
where 
$$                                   
\Delta^C (\til{t},
\til{w},s_1\ldots,s_{n-3})=
\sum_{\epsilon}(-1)^{\epsilon}M(\til{t},
\til{w},s_1 \pm i0,s_2\pm i0,\ldots,s_{N-3}\pm i0),
\auto\label{dis4}
$$
The sum over $\epsilon$ is over all combinations of $+$ and $-$ signs in 
(\ref{dis4}) and $(-1)^{\epsilon}$ is positive when the number of $+$ signs 
is even. In writing (\ref{dis2}) 
the asymptotic relation (\ref{as2}) has been used to change variables from
$z_1,\ldots,z_{N-3}$ to $s_1,\ldots,s_{N-3}$. We note again that an explicit
example of an asymptotic dispersion relation is described in full detail in
\cite{sw}. 

\subhead{3.6 Froissart-Gribov Continuations and Signature }

Each hexagraph spectral component $M^H$ has simultaneous cuts in only N-3 
large invariants. As we will see, the invariant cuts are reflected directly 
in the form that multi-regge behavior takes. Each cut is associated with
a particular power behavior. Correspondingly, the multi-regge behavior 
of a spectral component is obtained by SW transforming only N-3 of the 
angular-momentum and helicity sums in (\ref{pw1}). Indeed, unique
Froisart-Gribov (F-G) continuations in the complex plane can only be made
for the relevant indices. An 
important property of the hexagraph notation is that it classifies together
all those sets of cuts for which continuations in the same helicity and
angular momentum variables can be made. The construction of F-G 
continuations is described in detail in \cite{arw1}. Here we will simply
give the rules for determining the continuations that exist for
a particular hexagraph amplitude.

We first need to define T, D and V subgraphs of a hexagraph as in Fig.~3.7 

\begin{center}
\leavevmode
\epsfxsize=4in
\epsffile{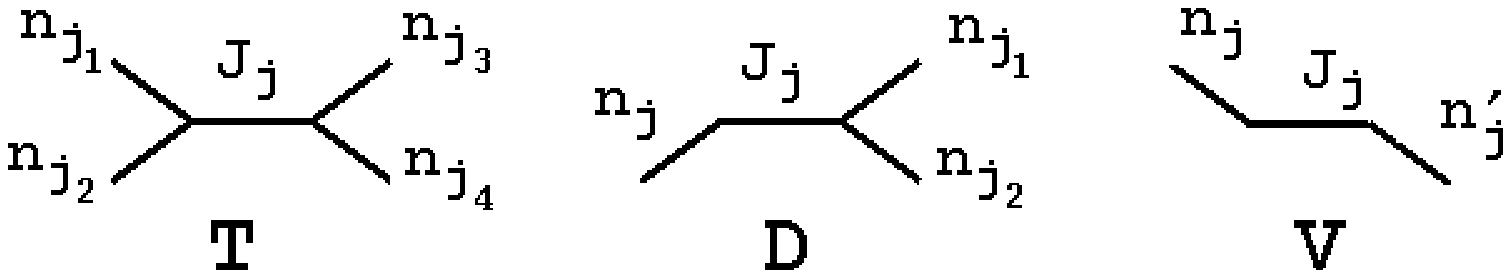}

Fig.~3.7 T, D and V subgraphs of a Hexagraph.
\end{center}
It is obvious how hexagraphs, such as those of Fig.~3.4, break up into
subgraphs of this form. The continuation rules are that in each $V_j$ we
take $n_j$ complex with ($J_j-n_j$) and ($n_j-n'_j$) held fixed at integer
values. In each $D_j$ we take $n_j$ complex with ($J_j-n_j$) held
fixed at an integer value.  In each $T_j$ we take $J_j$ complex, 
independently of all the $n_{j_i}$. These rules imply that the helicity
labels, which are attached to sloping lines of the hexagraph, are always
coupled to (that is differ only by an integer from) the angular momentum
associated with the corresponding horizontal line of the hexagraph. 

An important point for all continuations is that they are made 
separately for positive and negative helicities and also for positive and
negative helicity differences. That is, for 
$n_j~ {\raisebox{1mm}{\centerunder{$\scriptscriptstyle >$}{$\scriptscriptstyle 
<$}}}~ n'_j$ 
for each $V_j$, for $n_j ~{\raisebox{1mm}{\centerunder{$\scriptscriptstyle 
>$}{$\scriptscriptstyle <$}}}~ (n_{j_1} \pm n_{j_2})$ for each $D_j$ and for 
$ (n_{j_1} \pm n_{j_2})~ {\raisebox{1mm}{\centerunder{$\scriptscriptstyle 
>$}{$\scriptscriptstyle <$}}} ~(n_{j_3} \pm n_{j_4})$ for each $T_j$. 
We will use a convention in which if $n_{j_1}$ and $n_{j_2}$ have the same
sign, this implies they have opposite sign helicities in the $t_j$-channel
center of mass. (In a direct channel this would correspond to helicity sign
conservation). Continuations from values of $n_{j_1}$ and $n_{j_2}$ with the
opposite sign will be referred to as ``helicity-flip'' continuations and
will be crucial in what follows. 

As in elementary Regge theory it is necessary to introduce signature to
obtain well-defined F-G continuations. In the analytic procedure we are
following, signatured amplitudes are obtained by adding or subtracting
the dispersion relation spectral components corresponding to those
hexagraphs differing simply by a twist, about the corresponding horizontal line 
for a continuation in a $J_j$,  and about the horizontal line to which the 
corresponding sloping line is attached for a continuation in $n_j$. 
This definition also separates ``even'' and ``odd'' terms in the relevant 
series appearing in the 
partial-wave expansion. As we described above a single twist 
changes the sign of the angular variable (associated with the line about
which the twist is made) whose conjugate variable ($J_j$ or $n_{j}$) is
taken complex. 

We shall also utilise the following, equivalent, ``group-theoretic''
definition of signature, since in general it is easier to implement.
Beginning with an N-point amplitude in a particular direct channel, we
form the positive (or negative) signatured amplitude, with respect to a
particular internal line of a Toller diagram, by adding (or subtracting) the
amplitude obtained by making a complete TCP transformation on all external
particles connected (through the diagram) to one end of the internal line.
The fully signatured amplitude is formed by carrying out this procedure for
all internal lines of the Toller diagram. In this way signature is
introduced at the amplitude level without introducing spectral components.
It is an operation defined directly on the external states. Although the 
equivalence of the two definitions has only been proven in the simplest 
cases, we have no reason to doubt that the equivalence is true in general, 
and we will assume this to be the case. Of course, to
understand the implications of signature for phases etc. it is necessary to
utilise the analytic formulation. 

It is interesting to note 
that, in the case when no $V_j$'s are present in the hexagraph, the total
cross-channel angular momentum is continued to complex values, together with
all the helicities of (cross-channel) subchannels. In no case is the angular
momentum of a subchannel continued separately from the helicity. When $V_j$'s
are present the total angular momentum of the cross-channel is not used
as a variable.  Instead, the scattering can be regarded as made up of
subprocesses for which the total angular momenta and subchannel helicities
are analytically continued. 

\subhead{3.7 Sommerfeld-Watson Representations, Multi-Regge and 
Helicity-pole Limit Amplitudes}

The process of first defining a S-W transformation on the partial-wave
expansion for a hexagraph amplitude and then 
studying asymptotic limits is sufficiently complicated that it is difficult 
to give a general description. We give a general idea of the procedure by
considering simple examples. We will study further examples in the
following Section. As we remarked earlier, we will be particularly
interested in ``maximal helicity-pole limits''. For hexagraphs with no V
subgraphs, a maximal helicity-pole limit is simply defined by taking all the
azimuthal $u_{ij}$ variables large. When a V subgraph is involved , only one 
combination of the two azimuthal angles associated with the central line of 
the graph is taken large. The maximal number of helicity poles is still 
involved and a single partial-wave amplitude is isolated.

We consider specifically the Toller diagram for the six-point function 
shown in Fig.~3.2. 
This is the Toller diagram for which the asymptotic dispersion relation is
derived in \cite{sw}. There are 4 basic hexagraphs which after twisting 
gives a total of 32 hexagraphs. 
A full discussion of the S-W representation for all the 
hexagraphs and their use in all asymptotic limits is given in
\cite{arw1}. Here, for illustration, we concentrate on two 
of the basic graphs. Consider first the hexagraph shown in Fig.~3.8. 
\begin{center}
\leavevmode
\epsfxsize=4.5in
\epsffile{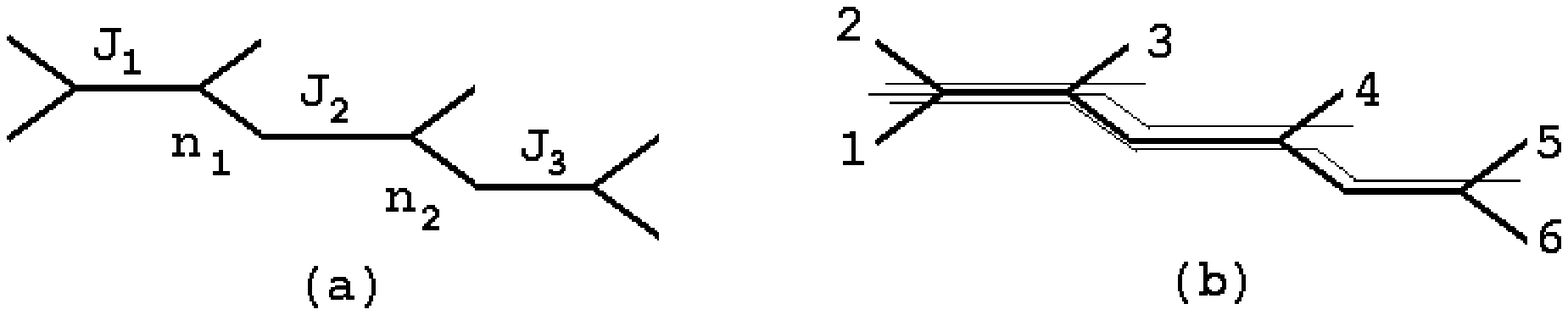}

Fig.~3.8 A Hexagraph from the Toller Diagram of Fig.~3.2 (a) $J$ and $n$ 
Variables (b) Cuts

\end{center}
With the $J$ and $n$ variables as illustrated, the partial-wave expansion has 
the form
$$
A_H(z_1,z_2,z_3,u_1,u_2,t_1,t_2,t_3)=
\sum_{\til{\scriptstyle J}\til{\scriptstyle n}} 
d^{J_1}_{0,n_1}(z_1)u_1^{n_1}d^{J_2}_{n_1,n_2}
(z_2)u_2^{n_2}d^{J_3}_{n_2,0}(z_3)a_{\til{\scriptstyle J}\til{\scriptstyle
n}}(\til{t}).
\auto\label{5.39}
$$
The hexagraph contains one $T$-graph and two $D$-graphs and the above rules
determine that from $n_1, n_2 > 0$  (signatured) F-G continuations can be 
made to complex $J_1$, $n_1$ and $n_2$ in the three complex half-planes 
$$
{\rm Re}(J_1-n_1)\geq 0,\quad {\rm Re}(n_1-n_2)\geq 0,\quad {\rm
Re}\,n_2\geq0,\auto\label{5.44}
$$
while $J_2 - n_1$ and $J_3 -n_2$ are held fixed at integer values. For the
present we omit the complications of signature in order to more simply
illustrate other features. The S-W transform of that part of
(\ref{5.39}) satisfying (\ref{5.44}) is then 
$$                
\eqalign{                                  
A_H={}&{1\over 8}\int_{C_{n_2}}
{dn_2\,u_2^{n_2}\over \sin\pi
n_2}~\int_{C_{n_1}} {dn_1\,u_1^{n_1}\over
\sin\pi(n_1-n_2)}~\int_{C_{J_1}} {dJ_1~d^{J_1}_{0,n_1}(z_1)\over
\sin\pi(J_1-n_1)}\cr
&~~~\times~   \sum^\infty_{{\scriptstyle J_2-n_1=N_1=0\atop
\scriptstyle J_3-n_2=N_2=0}}
d^{J_2}_{n_1,n_2}(z_2)d^{J_3}_{n_2,0}(z_3)a_{N_2N_3}(J_1,n_1,n_2,
\til{t})\cr
& ~+~~\sumtil_{\til{\scriptstyle J}\til{\scriptstyle
n}}d^{J_1}_{0,n_1}(z_1)u_1^{n_1}d^{J_2}_{n_1,n_2}
(z_2)u_2^{n_2}d^{J_3}_{n_2,0}(z_3)a_{\til{\scriptstyle J}\til{\scriptstyle
n}}(\til{t}
,}\auto\label{5.45}
$$
where $ C_{n_2}, C_{n_1}$  and $C_{J_1}$ are parallel to the imaginary axis.
The sum $\sumtilt$ is over that part
of (\ref{5.39}) not satisfying (\ref{5.44}). 

We will show first that the representation (\ref{5.45}) is sufficient to study 
the ``maximal helicity-pole limit''
$$
z_1,~ u_1,~ u_2\to\infty
\auto\label{hpl}
$$
with $Z_2$ and $z_3$ (and $t_1, t_2,t_3$) kept fixed. The cut structure of
$A_H$ is staightforwardly represented asymptotically by the S-W integrals as
follows. Asymptotically, the invariant cuts of Fig.~3.8(b) appear in the
angular variables via 
$$
\eqalign{
~~~~~~~~s_{23}&=(p_2+p_3)^2\sim ~~~z_1\cr
s_{234}&=(p_2+p_3+p_4)^2~\sim ~~y_{12}~~\equiv \left[\left(z^2_1-1\right)^{1/2}
\left(z^2_2-1\right)^{1/2}\right]u_1\cr
s_{16}&=(p_1+p_6)^2~~\sim ~~y_{123}~~\equiv \left[\left(z^2_1-1\right)^{1/2}
(z_2 + 1)\left(z^2_3-1\right)^{1/2}\right]u_1u_2}
\auto\label{5.42}
$$
We can rewrite (\ref{5.45}) in the form
$$                                         
\eqalign{           
~~~~~~~~~~A_H{}=&\int{dn_2dn_1dJ_1\over \sin\pi
n_2\sin\pi(n_1-n_2)\sin\pi(J_1-n_1)}
y^{n_2}_{123}\,y^{n_1-n_2}_{12}P^{J_1-n_1}(z_1)\cr
&\times
\sum^\infty_{N_1,N_2=0}P^{N_1}(z_2)P^{N_2}(z_3)a_{N_1N_2}(J_1,n_1,n_2,
\til{t}) ~~+~~~~\sumtil 
}
\auto\label{5.47}
$$
where
$$                                   
p^{j-n}(z) ~=~~ {1 \over 2}(1+z)^{-n-n'\over 2}(1-z)^{n'-n\over
2}d^j_{nn'}(z)\quad n>n'
\auto\label{5.38b}
$$
is a polynomial for integer $j-n=N$. In the form (\ref{5.47}), it is 
clear that each of the asymptotic cuts of $A_H$ is directly represented by one 
of the S-W integrals. Since $A_H$ has no singularities in the 
remaining variables, the sums over $N_1$ and $N_2$ (of 
polynomials) will be convergent in the asymptotic region. 

An asymptotic expansion for the limit (\ref{hpl}) can be obtained by pulling
the $J_1$, $n_1$ and $n_2$ contours to the left in (\ref{5.47}) - provided 
positive power singluarities are encountered. The $\tilde{\sum}$ 
contribution gives only inverse powers of either $u_1$ or $u_1 u_2$.
(We will not describe the subtleties  of introducing second-type
representation functions etc. that are necessary to obtain a true asymptotic
expansion). It can be shown\cite{arw1} from the analytically continued
unitarity equations that the Regge singularities of
$a_{N_1N_2}(J_1,n_1,n_2)$ occur at values of $J_1,J_2=n_1+N_1$ and
$J_3=n_2+N_2$. In particular, if there are Regge poles at $J_1=\alpha_1$,
$J_2=\alpha_2$, $J_3=\alpha_3$, the leading behavior in the limit
(\ref{hpl}) arises from $N_1=N_2=0$. A Regge pole at $J_1 = \alpha_1$,
together with ``helicity-poles'' at  $n_1 = J_2= \alpha_2$ and 
$n_2 = J_3= \alpha_3$ give 
$$                                                                
A_H ~
\centerunder{$\large \sim$}{\raisebox{-5mm}{{$\scriptstyle z_1\to\infty
\atop {\scriptstyle u_2\to\infty\atop \scriptstyle u_3\to\infty}$}}}
~~{z_1^{\alpha_1-\alpha_2}y_{12}^{\alpha_2-\alpha_3}y^{\alpha_3}_{123}
~~\beta_{00}^{\alpha_1\alpha_2\alpha_3}\over \sin\pi\alpha_3\sin\pi(\alpha_2-
\alpha_3)\sin\pi(\alpha_1-\alpha_2) }
\auto\label{5.55}
$$
Note that this result holds whether or not $z_2$ and/or $z_3$ are large. The 
partial-wave amplitude with $N_1=N_2=0$ is selected provided only that the 
limit $u_1, u_2 \to \infty$ is taken. The limit is called a 
``helicity-pole limit'' because it is controlled (in part) by poles (or more 
generally singularities) in helicity planes. 

The denominator factors in (\ref{5.55}) give singularities in the $t_i$
variables that are determined by the consistency of the asymptotic cut
structure of $A_H$ with the Steinmann relations. To see this we use
(\ref{5.42}) to rewrite (\ref{5.55}) in the form 
$$
A_H~ {\large\sim}~~
s_{23}^{\alpha_1-\alpha_2}s_{236}^{\alpha_2-\alpha_3}s_{15}^{\alpha_3}
~{\beta_{00}^{\alpha_1\alpha_2\alpha_3}\over \sin\pi\alpha_3\sin\pi(\alpha_2-
\alpha_3)\sin\pi(\alpha_1-\alpha_2)}
\auto\label{5.56}
$$  
which implies that asymptotically
$$
\eqalignno{
\centerunder{\raisebox{2pt}{disc }}{$\scriptstyle S_{23}$} A_H&\sim
~~\sin\pi(\alpha_1-\alpha_2)A_H&\num\label{5.52}\cr
\centerunder{\raisebox{2pt}{disc }}{$\scriptstyle S_{236}$} A_H&\sim
~~\sin\pi(\alpha_2-\alpha_3)A_H&\num\label{5.53}\cr
\centerunder{\raisebox{2pt}{disc }}{$\scriptstyle S_{15}$} A_H&\sim
~~\sin\pi\alpha_3A_H.&\num\label{5.54}\cr}
$$
Consequently each discontinuity cancels one of the poles in the
$\alpha_j$-variables and as a result the triple discontinuity of $A_H$ has no
poles in the $t_j$-variables. The Steinmann relations imply this must be the
case. The Steinmann relations, which should be valid asymptotically, forbid
singularities in overlapping channels. 

To obtain a complete asymptotic expansion in the multi-regge limit 
$$
z_1,~z_2,~z_3~~ \to ~~\infty
\auto\label{mrl}
$$
(with $u_1$ and $u_2$ kept fixed) we must also 
S-W transform the sums with $n_2<0$ and/or $n_1 - n_2 <0$ in $\sumtil$.
If we again pull back the $J_1$, $n_2$ and $n_1$ contours appropriately we 
obtain
$$  
\eqalign{  &A_H  ~  \centerunder{\centerunder{ 
\centerunder{$ \sim$}{\raisebox{-2mm}{$\scriptstyle
z_1\to\infty $}}}{\raisebox{-2mm}{$\scriptstyle z_2\to\infty$}}}{
\raisebox{-2mm}{$\scriptstyle z_3\to\infty $}}~
\sum^{\infty}_{N_1,N_2=0} { P^{N_1}(z_2)P^{N_2}(z_3)\over
\sin\pi\alpha_3\sin\pi(\alpha_2-\alpha_3)\sin\pi(\alpha_1-\alpha_2)} 
~~~\times \cr
&\biggl[\beta_{N_1,N_2}^{\alpha_1\alpha_2\alpha_3}
z_1^{\alpha_1}\left(z_2u_1\right)^{\alpha_2-N_1}
\left(z_3u_2\right)^{\alpha_3-N_2} +
\beta_{N_1,N_2}^{\alpha_1-\alpha_2\alpha_3}
z_1^{\alpha_1}\left(z_2u_1^{-1}\right)^{\alpha_2-N_1}
\left(z_3u_2\right)^{\alpha_3-N_2} \cr
& + \beta_{N_1,N_2}^{\alpha_1\alpha_2 -\alpha_3}
z_1^{\alpha_1}\left(z_2u_1\right)^{\alpha_2-N_1}
\left(z_3u_2^{-1}\right)^{\alpha_3-N_2} +
\beta_{N_1,N_2}^{\alpha_1 -\alpha_2 -\alpha_3}
z_1^{\alpha_1}\left(z_2u_1^{-1}\right)^{\alpha_2-N_1}
\left(z_3u_2^{-1}\right)^{\alpha_3-N_2}\biggr] \cr
&~ \cr
&\hspace{-3mm}\centerunder{$\large \sim$}{$\phantom{z_3\to\infty}$}
{z_1^{\alpha_1}z_2^{\alpha_2}z_3^{\alpha_3}\over 
\sin\pi\alpha_3\sin\pi(\alpha_2-\alpha_3)\sin\pi(\alpha_1-\alpha_2)}
\sum^\infty_{N_1=N_2=0}\biggl[\beta_{N_1N_2}^{\alpha_1\alpha_2\alpha_3}
u_1^{\alpha_2-N_1}u_2^{\alpha_3-N_2} \cr
&+ \beta_{N_1N_2}^{\alpha_1 -\alpha_2\alpha_3}
u_1^{- \alpha_2-N_1}u_2^{\alpha_3-N_2} + 
\beta_{N_1N_2}^{\alpha_1\alpha_2-\alpha_3}
u_1^{\alpha_2-N_1}u_2^{-\alpha_3-N_2} + 
\beta_{N_1N_2}^{\alpha_1-\alpha_2-\alpha_3}
u_1^{-\alpha_2-N_1}u_2^{-\alpha_3-N_2}\biggr]. }
\auto\label{mrll}
$$
In terms of invariants we have the same result as (\ref{5.55}) but now
the vertex function contains infinite series of (analytically-continued) 
partial-wave helicity amplitudes. This illustrates the close
relationship between the $u_j$-dependence and $z_j$-dependence of amplitudes
in the asymptotic region which we referred to eariler. It is, as in this 
example, simply a consequence of the presence
of only $(N-3)$ cuts for $(2N-7)$ variables. 

By comparing (\ref{5.55}) and (\ref{mrll}) we see how a (maximal)
helicity-pole limit selects a single F-G partial-wave amplitude from the 
infinite series that appears in the multi-regge limit. This is 
important because the unitarity properties of a single F-G
partial-wave amplitude can be straightforwardly studied. Note that 
the helicity-pole limit (\ref{hpl}) is not a physical region limit although
for the more complicated hexagraphs studied in later Sections, analogous
limits will be physical.

Before we discuss the particle-pole properties of (\ref{5.55}) and 
(\ref{mrll}), we briefly discuss the S-W representation of a second hexagraph
associated with Fig.~3.2. We consider the hexagraph shown in Fig.~3.9.
\begin{center}
\leavevmode
\epsfxsize=2.5in
\epsffile{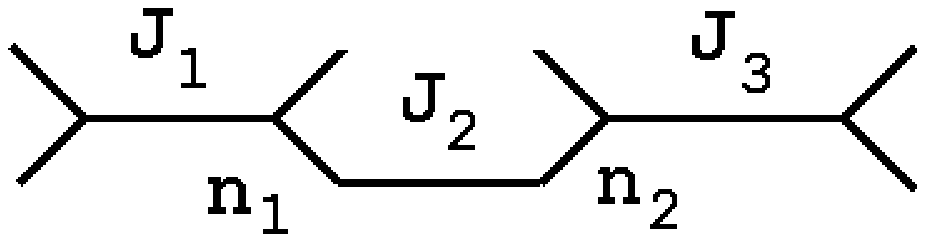}

Fig.~3.9 Another Hexagraph from the Toller Diagram of Fig.~3.2 
\end{center}
The partial-wave expansion of (\ref{5.39}) is again appropriate.
The hexagraph now contains one $V$-graph and two $T$-graphs and the above rules
determine that from $n_1, n_2 > 0$  (signatured) F-G continuations can be 
made to complex $J_1$, $J_3$ and $n_1$ in the three complex half-planes 
$$
{\rm Re}(J_1-n_1)\geq 0,\quad {\rm Re}(J_3-n_2)\geq 0,\quad {\rm
Re}\,n_1\geq0,\auto\label{h391}
$$
with $J_2 - n_1$ and $n_1 - n_2$ held fixed at integer values.
The S-W transform of 
that part of (\ref{5.39}) satisfying (\ref{h391}) is then 
$$                
\eqalign{                                  
A_H={}&{1\over 8}\int_{C_{n_1}}
{dn_1 (u_1u_2)^{n_1}\over \sin\pi
n_1}~\int_{C_{J_1}} {dJ_1 d^{J_1}_{0,n_1}(z_1)\over
\sin\pi(J_1-n_1)}~\int_{C_{J_3}} {dJ_3~d^{J_3}_{n_2,0}(z_3)\over
\sin\pi(J_3-n_2)}\cr
&~~~\times~   \sum^\infty_{{\scriptstyle J_2-n_1=N_1=0\atop
\scriptstyle n_1-n_2=N_2=0}}
d^{J_2}_{n_1,n_2}(z_2) u_2^{n_2 -n_1}a_{N_2N_3}(J_1,J_3,n_1,
\til{t})\cr
& ~+~~~~~~~~\sumtil_{\til{\scriptstyle J}\til{\scriptstyle
n}}d^{J_1}_{0,n_1}(z_1)u_1^{n_1}d^{J_2}_{n_1,n_2}
(z_2)u_2^{n_2}d^{J_3}_{n_2,0}(z_3)a_{\til{\scriptstyle J}\til{\scriptstyle
n}}(\til{t}) }
\auto\label{h392}
$$

We now consider the ``maximal helicity-pole limit''
$$
z_1, ~z_3, ~u_1u_2 ~~~ \to ~ \infty
\auto\label{h393}
$$
with $z_2$ and $u_1/u_2$ fixed. 
Regge poles at $J_1 = \alpha_1$ and $J_3 = \alpha_3$  contribute 
straightforwardly. If we take $N_1=N_2 = 0$, the Regge pole at  
$J_2 = \alpha_2$ appears as a helicity-pole at $n_1 = \alpha_2$ and we 
obtain, in analogy with (\ref{5.55})
$$                                                                
A_H ~
\centerunder{$\large \sim$}{\raisebox{-5mm}{{$\scriptstyle z_1\to\infty
\atop {\scriptstyle u_2\to\infty\atop \scriptstyle u_3\to\infty}$}}}
~~{z_1^{\alpha_1-\alpha_2}z_{3}^{\alpha_3-\alpha_2}y^{\alpha_2}_{123}
~~\beta_{00}^{\alpha_1\alpha_2\alpha_3}\over \sin\pi\alpha_2\sin\pi(\alpha_1-
\alpha_2)\sin\pi(\alpha_3-\alpha_2) }
\auto\label{h394}
$$
Again a single F-G partial-wave amplitude is isolated. Note that 
(\ref{h394}) continues to hold if $z_2$ is taken 
large.

We can use (\ref{5.55}), (\ref{mrll}) and (\ref{h394}) to illustrate some 
general properties of hexagraph multi-regge amplitudes.
Suppose, for simplicity, that the $\alpha_i$ are even-signature Regge 
trajectories giving a particle pole at $\alpha_i = 0$. We note first that 
(\ref{5.55}) contains a pole only at $\alpha_3 =0$. A pole 
at $\alpha_2 = 0$ appears if we first set $\alpha_3 =0$. In contrast 
(\ref{h394}) contains directly a pole at $\alpha_2 =0$. As we discussed, the 
pole structure in the $t_i$ variables relates directly to the analytic 
structure in the large invariant variables. Together
(\ref{5.55}) and (\ref{h394}) represent a general situation in very 
complicated hexagraphs. Particle poles occur in association with a V 
subgraph or with a D subgraph at the end of a ``cascade'' of D subgraphs.
Regge pole factorization gives that, in (\ref{h394})
$$ 
\beta_{00}^{\alpha_1 \alpha_2 \alpha_3} ~=~ \beta_{0}^{\alpha_1 \alpha_2}
\beta_{0}^{\alpha_2 \alpha_3} 
\auto\label{h395}
$$
and so, as the hexagraph of Fig.~3.9 suggests pictorially, at $\alpha_2 = 0$
the amplitude factorizes into a product of four-point amplitudes. The
factorization property (\ref{h395}) holds provided only that we pick out a
Regge pole in the $t_2$ channel. In general, we obtain full four-point
scattering amplitudes rather than just the Regge exchange amplitudes given
by (\ref{h395}). If we continue $\alpha_2$ to a non-zero even integer value
then the factorization of (\ref{h395}) gives the leading helicity four-point
amplitudes. Analagously, if we continue $\alpha_3$ to an even integer value in
(\ref{5.55}) we obtain the leading helicity amplitude at the particle pole. As
illustrated by (\ref{mrll}), a multi-regge limit amplitude in general gives a
sum over helicity amplitudes at a particle pole. 

Finally we note that we can also obtain leading helicity 
amplitudes with opposite signs for the $n_j$ involved by taking 
corresponding helicity-pole limits. For example, taking the limit $u_1 / u_2
\to \infty $ with $u_1 u_2 $ fixed in (\ref{h392}) and by taking $u_2 \to 0$
instead of $u_2 \to \infty$ in (\ref{5.45}). 

\subhead{3.8 Reggeon Unitarity}

The most important property of the F-G amplitudes is that they can 
effectively be used to analytically continue, in the complex $J_j$ and
$n_j$-planes, the cross-channel multiparticle unitarity equations in any
$t_i$ channel of any Toller diagram. This leads to a
set of ``Reggeon Unitarity'' equations for the discontinuities 
across multi-reggeon branch-cuts which appear in each of the complex angular 
momentum planes. These equations are crucial in enabling us to build the 
multi-regge behavior of QCD amplitudes on the basis of known results 
for elastic and production processes. We will only give a brief outline of 
the derivation of the reggeon unitarity equations here. (Note that in the
abstract analysis of this Section and the next Section we use reggeon to
refer to any Regge pole. In Section 5 we will use this term specifically for
an odd signature Regge pole with intercept near one, referring to an
even signature pole with intercept near one as a pomeron. From Section 6
onwards a reggeon will specifically be a reggeized gluon.) 

The discontinuity across the M-reggeon cut (i.e. the branch cut
due to the exchange of M Regge poles) in any J-plane is derived most simply
from the 2M particle discontinuity formula in the corresponding $t$-channel. 
The $t$-channel discontinuity is first expressed as a conventional unitarity
phase-space integral. By using a Toller diagram including the internal
particles, this phase-space integral $I_{2M}(t)$ can be written in the
form 
$$
I_{2M}(t)=i\int d\rho(t,t_1,\ldots t_j,\ldots)\int dg_J\prod_j
dg_j\auto\label{un1}
$$
where the $g_j$ are associated with lines of the Toller diagram and 
(apart from numerical factors)
$$
\int d\rho (t,t_1,.. t_j,..) =\int \prod_j dt_j
{\lambda^{1/2}(t,t_1,t_2)\over
t}{\lambda^{1/2}(t_1,t_3,t_4)\over t_1} ...
{\lambda^{1/2}(t_j,t_{j+1},t_{j+2})\over t_j} ..
\auto\label{un2}
$$
There is a $\lambda$-function for each internal
vertex, including those involving the internal particles (for which the
corresponding ``$t_j$'' is the mass$^2$). The integration region is defined
by
$$                                                                 
\lambda(t_j,t_{j+1},t_{j+2})\geq 0\qquad \forall~ j
\auto\label{un3}
$$

It can be shown\cite{arw1} that the unitarity integral generates Regge cut
behavior only when particular multiple discontinuities are present in the
amplitudes appearing in the integral. The necessary discontinuities are
present when (and only when) the amplitudes correspond to hexagraphs having
a ``cascade'' structure of D subgraphs with respect to the internal
phase-space, as illustrated in Fig.~3.10. (The subleties in isolating
hexagrah product contributions are discussed in \cite{arw1}, 
we will not discuss them here). 

\begin{center}
\leavevmode
\epsfxsize=3in
\epsffile{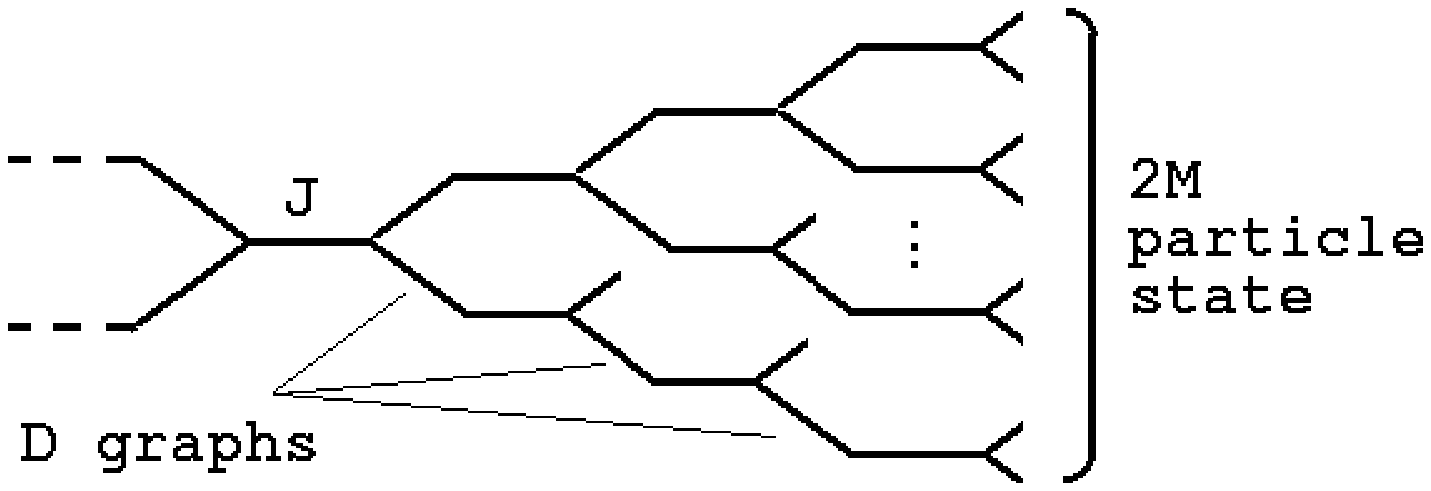}

Fig.~3.10 A ``Cascade'' of D Graphs for 2M Particle Phase-space.
\end{center}
As a result, for the purposes of studying Regge cuts, we obtain a form
of hexagraph diagonalisation of the $t$-channel 2M-particle unitarity
integral 
$$
disc~ A^H~ =i\int d\rho\int dg\prod_j dg_j\,A^{H_L}(g,\ldots,g_j,\ldots)
A^{H_R}\left(g^{-1}g_J,\ldots,g_j,\ldots\right),
\auto\label{un4}
$$
where $H_L$ and $H_R$ have the necessary cascade structure. For example, if
H is the hexagraph shown in Fig.~3.11

\begin{center}
\leavevmode
\epsfxsize=3in
\epsffile{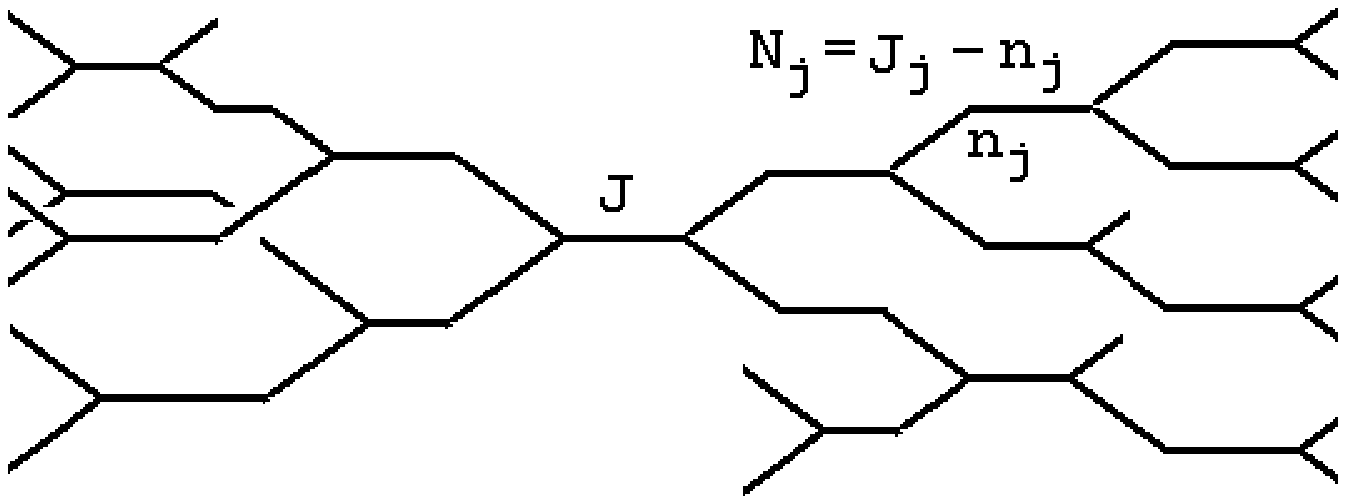}

Fig.~3.11 A Hexagraph H. 
\end{center}
the hexagraphs $H_L$ and $H_R$ have
the form illustrated in Fig.~3.12, i.e. $H_L$ and $H_R$ are formed from $H$
by splitting $H$ in two at the J line and substituting a product of D
cascades that connect to the intermediate particle state. 

\begin{center}
\leavevmode
\epsfxsize=4.5in
\epsffile{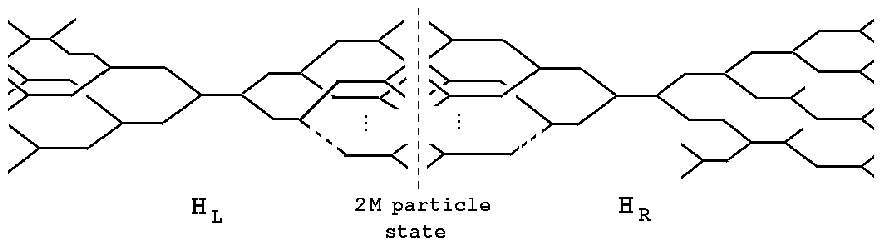}

Fig.~3.12 The Product of Hexagraphs in the Discontinuity Formula.
\end{center}
(\ref{un4}) can then be diagonalised by
partial-wave projection, i.e. (suppressing all the external hexagraph 
angular momenta and helicity labels) 
$$
disc~ a^H_J~=~i\int d\rho \sum_{\til{N},\til{n}} ~
a^{H_L}_{J\til{N}\til{n}}~a^{H_R}_{J\til{N}\til{n}}
\auto\label{un5}
$$
The summation shown is over all internal
helicity labels $\til{n}$ and angular momenta $\til{N}=\til{J}-\til{n}$
of all the $D$-graphs in the phase-space part of $H_L$ and $H_R$.

The partial-wave equations (\ref{un5}) can be analytically-continued to complex 
values of the external angular momenta and helicities by converting the 
internal sums to integrals having the S-W form. The M-reggeon cut is generated 
in the analytically continued equations by a combination of M Regge poles, the 
phase-space boundaries (\ref{un3}) and ``nonsense poles'' for each of the 
D-graph vertices. In the notation of Fig.~3.5, the nonsense poles are at 
$$
J_j=|n_j|~=~n_{j_1} + n_{j_2} - 1 
\auto\label{non1}
$$
when $n_{j_1}$ and $ n_{j_2}$ are positive or at 
$$
J_j=|n_j|~=~-n_{j_1} - n_{j_2} - 1 
\auto\label{non2}
$$
when both $n_{j_1}$ and $
 n_{j_2}$ are negative. If the Regge poles are identical then the relevant 
boundary of the phase space is at 
$$
\sqrt{t_i} = \sqrt{t_j} + \sqrt{t_k}  ~~~~\forall i, j, k
\auto\label{sqrt}
$$
This, combined with all the nonsense conditions, gives a trajectory 
$$
J ~=~ \alpha_M(t) ~=~ M \alpha(t/M^2) - M +1
\auto\label{mtrj}
$$   
As we stated earlier, in our notation 
$-n_{j_2}$ is the helicity in the $t$-channel 
center-of-mass frame. It is very important in what 
follows that there is no nonsense pole contribution from $n_{j_1}$ positive 
and $ n_{j_2}$ negative or from $n_{j_1}$ negative and $ n_{j_2}$ positive.
(These are not ``nonsense'' states''). Therefore ``helicity-flip'' partial-wave
continuations, from opposite-sign $n_{j_i}$ at an internal vertex, do not
contribute to the generation of Regge cuts. 
(We stated earlier that we will refer to amplitudes which have 
$n_{j_1} = -n_{j_2}$ as ``helicity-flip'' amplitudes. Such amplitudes are
``non-flip'' in the $t$-channel center of mass. However, for massless
particles, helicity is reversed in going from the $s$ to the $t$ channel and
so $t$-channel non-flip amplitudes correspond to $s$-channel helicity-flip
amplitudes. Ultimately it is $s$-channel helicity properties that will
interest us.) 

Consider now the hexagraph H and consider specifically the M-reggeon cut in
the J channel associated with the central T subgraph of Fig.~3.11. We denote
by $a \centerunder{\raisebox{3.5mm}{${\scriptstyle H,\til{
\scriptstyle \tau},\til{\scriptstyle >}}$}} 
{${\scriptstyle J \til{\scriptstyle N} \til{\scriptstyle n}}$}$ the 
 signatured F-G amplitude associated with H. All the helicities that are 
continued to complex values are now denoted by $\til{n}$, while $\til{N}$ 
denotes all the $N_j=J_j - n_j$ that are kept fixed at integer values,  
$\til{\tau} = (\tau_J,...,\tau_{n_j},...)$ are the signature labels, with
$\tau_J$ given by the product of the signatures of the contributing $M$ 
reggeons, and 
$\til{\scriptstyle >}$ denotes all the
${\raisebox{1mm}{\centerunder{$\scriptscriptstyle >$}{$\scriptscriptstyle
<$}}}$ labels describing the signs of helicities and helicity differences
from which the continuation is made. The discontinuity formulae involves the
product of nonsense/Regge pole amplitudes extracted from the F-G amplitudes
for the hexagraphs $H_L$ and $H_R$ of Fig.~3.12. The
discontinuity formula is then
$$
\eqalign{ disc
\centerunder{\raisebox{3.5mm}{$~$}}
{$J=\alpha_M(t)$}~~ 
a \centerunder{\raisebox{3.5mm}{${\scriptstyle H,\til{
\scriptstyle \tau},\til{\scriptstyle >}}$}} 
{${\scriptstyle J \til{\scriptstyle N} \til{\scriptstyle n}}$}
~=~ {\xi}_{M} \int d\hat\rho~&
A {\centerunder{\raisebox{3.5mm}{$\scriptstyle H_L,\til{\scriptstyle \tau},
\til{\scriptstyle >}$}}
{${\til{\alpha}~~~}$}}(J^+)
A {\centerunder{\raisebox{3.5mm}{$\scriptstyle H_R,\til{\scriptstyle \tau},
\til{\scriptstyle >}$}}
{${\til{\alpha}~~~}$}}(J^-)\cr
&{\delta\left(J-1-\sum^M_{k=1}
(\alpha_k-1)\right)\over \sin{\pi\over 2}(\alpha_1-\tau'_1)\ldots\sin{\pi\over
2}(\alpha_m-\tau'_M)}.}
\auto\label{run}
$$
where $\int d \hat\rho $ has the same form as (\ref{un2}) except that only 
Regge pole energies are integrated over (the integration over the masses of 
the pairs of particles has been eliminated by using elastic unitarity). 
${\xi}_{M}$ is a (relatively
complicated) signature factor that we will give simple appoximations for
in Section 5.  $\tau'=(\tau + 1)/2$, and $ A
{\centerunder{\raisebox{2.5mm}{$\scriptstyle H_L,\til{\scriptstyle \tau},
\til{\scriptstyle >}$}}{$ \raisebox{2mm}{\til{\alpha}~~~}$}}(J^+)$ is a
``nonsense'' reggeon scattering amplitude extracted from {\small$ 
a \centerunder{\raisebox{2.5mm}{${\scriptstyle H_L,\til{
\scriptstyle \tau},\til{\scriptstyle >}}$}} 
{${\scriptstyle J \til{\scriptstyle N} \til{\scriptstyle n}}$}$ }
and evaluated above the Regge cut at $J = \alpha_M (t)$. {\small $A
{\centerunder{\raisebox{2.5mm}{$\scriptstyle H_R,\til{\scriptstyle \tau},
\til{\scriptstyle >}$}} {${\til{\alpha}~~~}$}}(J^-)$} is the same amplitude
evaluated below the cut. 

For the introduction of pomeron and reggeon diagrams in Section 5 
it is important that the phase-space integration 
$\int d \hat\rho $ in (\ref{run}) can be modified by extracting the
``threshold behavior'' of the the nonsense amplitudes at the phase-space 
boundaries (\ref{sqrt}) i.e.
at the nonsense point $(n_j-n_{j+1} -n_{j+2})=-1$
$$
\eqalign{
A{\centerunder{\raisebox{3.5mm}{$\scriptstyle H_R,\til{\scriptstyle \tau},
\til{\scriptstyle >}$}}
{${\til{\alpha}~~~}$}}(J, t_1, \cdots, t_j,t_{j+1},t_{j+2}, \cdots)
~~\centerunder{$\sim$}{\raisebox{-5mm} 
{$\scriptstyle \lambda(t_j,t_{j+1},t_{j+2})~\to~0$}}~
~~&\biggl({\lambda(t_j,t_{j+1},t_{j+2}) \over t_j} \biggr)^{(n_j-n_{j+1} 
-n_{j+2})/2}\cr
 \sim ~~~~~~~~~~~&\biggl({\lambda(t_j,t_{j+1},t_{j+2}) \over t_j} \biggr)^{-1/2}
}
\auto\label{thr}
$$
We can then write 
$$
\int d \hat\rho (t,t_1,.. t_j,..) \to \int \prod_j dt_j
\lambda^{-1/2}(t,t_1,t_2)\lambda^{-1/2}(t_1,t_3,t_4) ...
\lambda^{-1/2}(t_j,t_{j+1},t_{j+2}) .. 
\auto\label{thr1}
$$

A discontinuity formula, essentially the same as (\ref{run}), also holds 
in any $n_j$-plane for $
a \centerunder{\raisebox{3.5mm}{${\scriptstyle H_L,\til{
\scriptstyle \tau},\til{\scriptstyle >}}$}} 
{${\scriptstyle J \til{\scriptstyle N} \til{\scriptstyle n}}$}$ 
except that the
hexagraphs $H_L$ and $H_R$ that are involved are obtained by inserting into
the j-line of H the same cascade structure that appears in Fig.~3.12.
This is illustrated in Fig.~3.13. 

\begin{center}
\leavevmode
\epsfxsize=4.5in
\epsffile{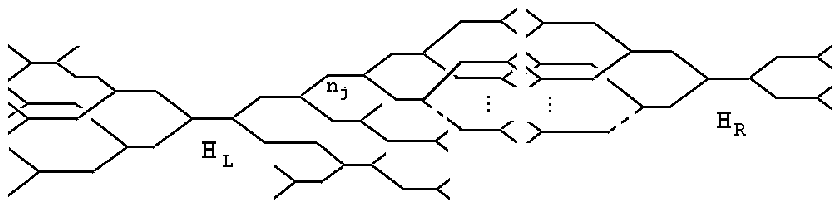}

Fig.~3.13  Another product of Hexagraphs. 
\end{center}

Similarly an analagous discontinuity formulae to (\ref{run}) holds
 in any complex angular momentum or 
helicity plane for any hexagraph F-G amplitude. The hexagraphs involved in 
the discontinuity formulae are simply found by introducing the relevant 
cascade structures as in Figs.~3.12 and 3.13. 
Before the advent of QCD it was understood that reggeon unitarity provides a 
general, model-independent, basis for a Reggeon Field Theory description of 
the pomeron. This will be elaborated in Section 5. However, only
a limited part of the full set of reggeon unitarity equations was exploited
historically. For the purpose of this paper, the full set of equations
(\ref{dis}) has another very important role.  Extensive results on the
reggeon diagram structure of elastic scattering have been derived by direct
calculation within QCD (at leading log, next-to-leading log, 
etc.)\cite{bfkl,bs,fl}. As we
will discuss, the power of the reggeon unitarity formulae is that they can
be used to directly extend these results to the multi-regge behavior of
arbitrarily complicated multiparticle scattering amplitudes. 

\newpage

\mainhead{4. TRIPLE-REGGE VERTICES AND LIMITS }

In this Section we specialize much of the discussion of the last Section to the 
various ``triple-regge'' limits of the six-particle amplitude. It is 
important that triple-regge kinematics are more general than the well-known
case of the large mass limit of the diffractive inclusive cross-section. There 
are ``triple-regge vertices'' which play a crucial role in our study of QCD but 
only appear in the more general triple-regge and helicity-pole limit
kinematics that we discuss below.

\subhead{4.1 Hexagraph Cuts and Limits}

We consider the Toller diagram shown in Fig.~4.1~. 

\begin{center}
\leavevmode
\epsfxsize=3.5in
\epsffile{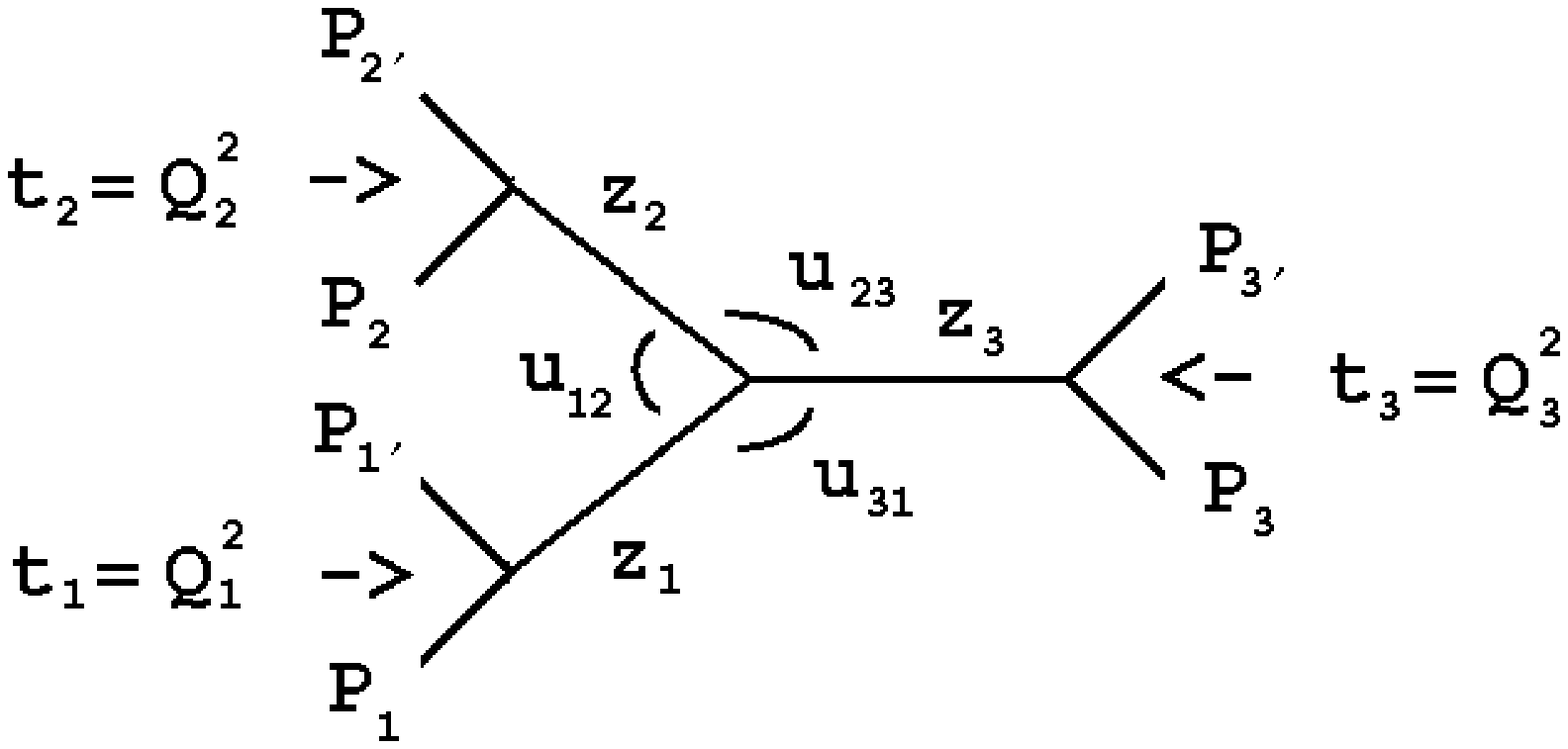}

Fig.~4.1 A Toller Diagram for $M_6$
\end{center}
As in (\ref{amp}) we write
$$
M_6(P_1,\ldots,P_6)\equiv
M_6\left(t_1,t_2,t_3,g_1,g_2,,g_3\right)
\auto\label{amp6}
$$
We initially take all the $t_i$ positive so that the $g_i$ are elements of 
SO(3). We also define each of the $g_i$ to transform from the central vertex
to the external vertex. If, for the moment, we take the external particles
to be spinless, the amplitude will be independent of the $\nu_i, ~i=1,2,3$
and will depend only on differences of the $\mu_i$. Therefore, if we define
$$ 
u_{12}~=~e^{i(\mu_1 - \mu_2)},~~~~~u_{23}~=~e^{i(\mu_2 - \mu_3)},~~~~~
u_{31}~=~e^{i(\mu_3 - \mu_1)},
\auto\label{azv}
$$
then 
$$
u_{12}u_{23}u_{31}~=~1
\auto\label{azv1}
$$
and we can take any two as independent variables. Combined with 
$t_1,t_2,t_3,$ and  $z_1,z_2,z_3$ this gives the appropriate eight 
independent variables.

The Toller diagram of Fig.~4.1 generates the set of hexagraphs shown in 
Fig.~4.2

\begin{center}
\leavevmode
\epsfxsize=4in
\epsffile{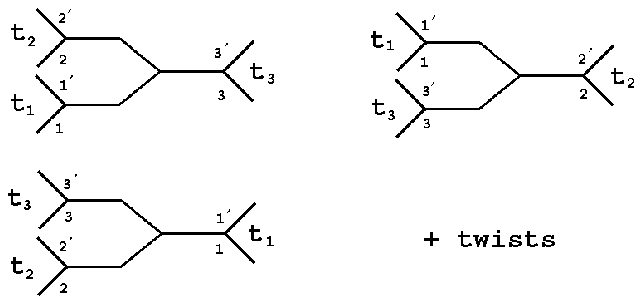}

Fig.~4.2 Hexagraphs Associated with the Toller Diagram of Fig.~4.1
\end{center}
Each hexagraph shown is one of $2\times2\times2=8$ related by twisting, 
where the twists are made about the three horizontal lines in the graphs.  
There are 24 hexagraphs in total. As we have described in the last
Section, each hexagraph corresponds to particular sets of allowable triple
discontinuities (in direct channel physical regions where the $t_i$ are 
negative). 

For the first hexagraph of Fig.~4.2, the allowable sets of cuts are as shown
in Fig.~4.3.

\begin{center}
\leavevmode
\epsfxsize=5in
\epsffile{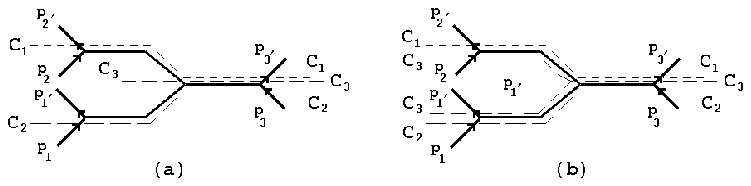}

Fig.~4.3 Hexagraph Cuts
\end{center}
The cuts of Fig.~4.3(a) are in the invariants
$$ 
\eqalign{ C_1~&\equiv s_{2'3'}~~[=~(P_{2'} +P_{3'})^2] \cr 
C_2 ~&\equiv s_{13}~~[=~(P_{1} +P_{3})^2] \cr
C_3 ~&\equiv s_{11'3}~~ [=~(P_1 - P_{1'} + P_3)^2]}
\auto\label{cuta}
$$
This set of cuts is well-known to be related to the one-particle
inclusive cross-section. The cuts of Fig.~4.3(b) are in the invariants
$$ 
C_1~\equiv s_{2'3'}~~~~C_2 ~\equiv s_{13}~~~~
C_3 ~\equiv s_{123}
\auto\label{cutb}
$$
This second set of cuts is less familiar but will play an important role in 
the following. For large $s_{12}$ and fixed $s_{23}, ~ s_{31}$
$$
s_{123}~= ~s_{1'2'3'}~\sim ~ s_{12} 
\auto\label{ap5}
$$
and so, asymptotically, the $s_{123}$ cut can be identified as an $s_{12}$ 
cut. The Steinmann relations forbid simultaneous cuts in $s_{2'3'}$, $ s_{31}$ 
and $ s_{12}$. However, we also have $s_{1'2} \sim s_{12'} \sim -s_{12}$ 
and simultaneous cuts in $s_{2'3'}$, $ s_{31}$ and $ s_{1'2}$ are allowed. In
the triple-regge direct-channel physical regions that we are interested in,
we can not have all three of $s_{12'}$, $s_{2'3'}$ and $ s_{31}$ positive.
Nevertheless, amplitudes with an $s_{1'2}$ cut, in addition to $s_{2'3'}$ and
$ s_{31}$ cuts, can be regarded as having a left-hand cut in $s_{123}$, even 
though it is unphysical, and
therefore as having the set of cuts (\ref{cutb}). This is important for the
quark loop amplitudes we discuss in Section 7. 

The full triple-regge limit associated with Fig.~4.1 is the multi-regge
limit of the form (\ref{rl}) i.e. 
$$
z_1,z_2,z_3 \rightarrow \infty~, ~~~~t_1,t_2,t_3,u_{31},u_{23}~~fixed
\auto\label{trl}
$$
We can also discuss triple-regge ``maximal helicity-pole limits'' involving the
$u_{ij}$. Since each hexagraph
naturally chooses particular pairs of the $u_{ij}$ as independent variables,
it is convenient (and dynamically significant) to associate the helicity-pole
limits with particular hexagraphs. For each hexagraph there are two distinct
helicity-pole limits.

To discuss the limits associated with the first hexagraph of Fig.~4.2 we first
simplify notation by writing $u_1\equiv u_{31}, u_2\equiv u_{23}$. 
We can then identify variables with the lines of the hexagraph as
illustrated in Fig.~4.4. 

\begin{center}
\leavevmode
\epsfxsize=3in
\epsffile{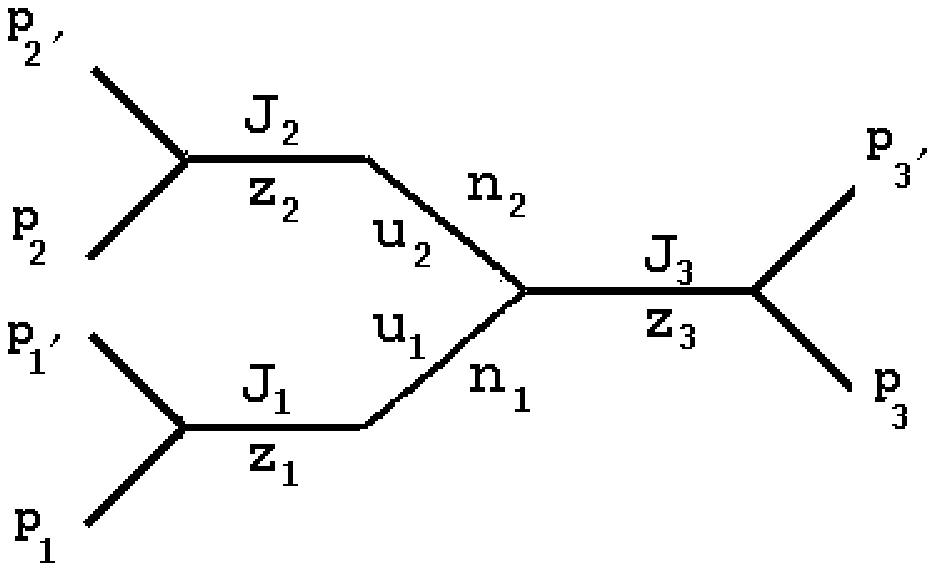}

Fig.~4.4 Hexagraph Notation
\end{center}
The first helicity-pole limit is 
$$ 
z_3,u_{1}, u_{2} \rightarrow \infty \ ~~~
({\rm or}\ u_{1}, u_{2} \rightarrow 0)
\auto\label{hp1} 
$$
This is the familiar ``triple-regge'' limit of the one-particle inclusive 
cross-section. The second helicity-pole limit is 
$$
z_3,u_{1},u_{2}^{-1} \rightarrow \infty \ ~~~
({\rm or}\ u_{1}, u_{2}^{-1} \rightarrow 0)
\auto\label{hp2}
$$
For reasons that will soon become apparent, we refer to the first limit as 
the ``non-flip limit'' and the second as the ``helicity-flip limit''.

From 3B) and 3C) we can see that the following approximations are
(essentially) uniformly valid in both helicity-pole limits, as well as the
triple-regge limit (\ref{trl}). 
$$
s_{13}~\sim~ s_{1'3'}~\sim -s_{13'}~\sim~ -s_{1'3}~~\sim 
~z_1z_3 (u_{1} + 1/u_{1})
\auto\label{ap1} 
$$
$$
s_{23} ~\sim~ s_{2'3'}~\sim -s_{23'}~\sim~ -s_{2'3}~~\sim 
~z_2z_3 (u_{2} + 1/u_{2})
\auto\label{ap2}
$$
$$
s_{11'3} ~\sim~ s_{2'23'}~\sim ~- s_{11'3'} ~\sim~- s_{2'23}~~ \sim ~z_3
\auto\label{ap3} 
$$
$$
s_{22'1} ~\sim~ s_{3'31'}~\sim ~- s_{22'1'} ~\sim~ - s_{3'31}~~ \sim ~z_1
\auto\label{ap31} 
$$
$$
s_{33'2} ~\sim~ s_{1'12'}~\sim ~- s_{33'2'} ~\sim~ - s_{1'12}~~  \sim ~z_2
\auto\label{ap32} 
$$
$$
s_{12}~ \sim s_{1'2'} \sim -s_{12'} \sim -s_{1'2} 
~\sim~ z_1z_2 (u_{1}/u_{2} + u_{2}/u_{1})
\auto\label{ap4} 
$$
Note that all 
invariants are unchanged when $u_1 \to 1/u_1, u_2 \to 1/u_2$. This is why 
the limits (\ref{hp1}) and (\ref{hp2}) have two equivalent definitions.

\subhead{4.2 Special Light-Cone Limits}

In later Sections it will be useful to have particular realizations of the 
limits defined in the previous sub-Section in terms of specific light-cone 
limits for the momenta involved.

We consider first the triple-regge limit. Since all three of 
$s_{12},~s_{23}$ and $s_{31}$ are large in this limit, $P_1,~P_2$ and $P_3$ 
should lie along distinct light-cones. In the notation of Fig.~4.1, we can 
define a particular version of the the triple-regge limit, which we
call a ``maximally non-planar'' limit, in which 
all three momenta are taken large and lightlike in orthogonal space
directions. We define 

\noindent {\large $L_1)$} \newline
\noindent \parbox{3in}{ 
$$
\eqalign{ P_1~\to&~ P_1^+~= ~(p_1,p_1,0,0)~,~~p_1 \to \infty \cr
P_2~\to&~ P_2^+~= ~(p_2,0,p_2,0)~,~~p_2 \to \infty \cr
P_3~\to&~ P_3^+~= ~(p_3,0,0,p_3)~,~~p_3 \to \infty  }
$$}
\parbox{3in}{
$$ \eqalign {
~~~Q_1~\to&~q_2-q_3 ~=~(0,0,q_2, -q_3)\cr
~~~Q_2~\to&~q_3-q_1 ~=~(0,-q_1,0,q_3)\cr
~~~Q_3~\to&~q_1-q_2 ~=~(0,q_1,-q_2,0)}
\auto\label{npl}
$$
}
(We omit the light-cone components of both the $P_i$ and the $Q_i$ that go
to zero asymptotically, but are necessary to put both initial and final
particles on mass-shell.) In terms of invariants, this limit gives 
$$
\eqalign{s_{12}~&=~(P_1 +P_2)^2~\to ~2p_1p_2~,~~~~
~~~~s_{23}~=~(P_2 +P_3)^2~\to ~2p_2p_3~,~~\cr
s_{31}~&=~(P_3 +P_1)^2~\to ~2p_3p_1~,~~
~~~~s_{122'}~=~(P_1+Q_2)^2~\to ~2p_1q_1~,~~\cr
s_{233'}~&=~(P_2+Q_3)^2~\to ~2p_2q_2~,~~
~~~~s_{311'}~=~(P_3+Q_1)^2~\to ~2p_3q_3~,~~}
\auto\label{npl1}
$$
and so can be identified with a triple-regge limit of the form 
(\ref{trl}) in which 
$$
p_1~\sim~z_1~,~~p_2~\sim~z_2~,~~p_3~\sim~z_3~.
\auto\label{npl2}
$$
This particular version of the triple-regge limit illustrates how the limit 
makes maximal use of four-dimensional minkowski space. To obtain exactly the
above momentum configuration we clearly have to choose particular values of
the $u_{i}$ and also go to a particular Lorentz frame. 

Next we give some different realizations of the ``helicity-flip''
helicity-pole limit (\ref{hp2}). The essential feature of this limit, 
compared to the triple limit is that, because $z_1$ and $z_2$ remain finite,
invariants such as $s_{33'1}$ and $s_{33'2}$ remain finite. We first define a
limit $L_2$, in which the finiteness of $s_{33'1}$ and $s_{33'2}$ is very 
simply achieved. In this limit $P_1$ and $P_2$ lie in the same plane, but have
opposite space momenta, and 
this plane is orthogonal to the transverse plane in which $Q_3, Q_2$ and
$Q_1$ lie. We define

\noindent {\large $L_2)$} \newline
\noindent \parbox{3in}{ 
$$
\eqalign{ P_1~\to&~ P_1^+~= ~(p_1,p_1,0,0)~,~p_1 \to \infty \cr
P_2~\to&~ P_2^-~= ~(p_2,-p_2,0,0)~,~p_2 \to \infty  \cr
P_3~\to&~ P_3^+~= ~(p_3,0,0,p_3)~,~~p_3 \to \infty  }
$$}
\parbox{3in}{
$$ \eqalign{
~~~Q_1~\to&~q_2-q_3 ~=~ (0,0,q_2,-q_3)\cr
~~~Q_2~\to&~q_3 -q_2' ~=~(0,0,-q_2',q_3)\cr
~~~Q_3~\to&~q_2'-q_2 ~=~(0,0,q_2' -q_2,0)}
\auto\label{np3}
$$}
In terms of invariants, this limit gives 
$$
\eqalign{&s_{12}~\to ~4p_1p_2~,~~~~
s_{23}~\to~ 2p_2p_3~,~~~~
s_{31}~\to ~2p_3p_1~,~~\cr
&s_{122'}~\st{\to} \infty~,~~~~
s_{233'}~\st{\to} \infty~,~~~~
s_{311'}~\to ~2p_3q_3~. }
\auto\label{npl4}
$$
Comparing with (\ref{hp2}) and (\ref{ap1})-(\ref{ap4}), we see that this
limit can be identified with the ``helicity-flip'' helicity-pole limit
(\ref{hp2}), with
$$
p_1~\sim ~u_1~, ~~p_2~\sim~u_2^{-1}~, ~~ p_3~\sim~z_3~,~
\auto\label{npl5}
$$
Again special values of the non-asymptotic angular variables (in this case 
$z_1$ and $z_2$) are implicitly involved. However, we will see in the next 
subsection that, in the leading asymptotic behavior, the dependence on these 
variables is determined by the S-W representation, as it was for the 
helicity-pole limit (\ref{5.55}). (For the triple-regge limit the dependence 
on the finite angular variables is expanded in infinite partial-wave series 
and therefore is unknown).

The following alternative 
realization of the helicity-flip limit will also be useful. In this case the 
finiteness of $s_{22'1}$ and $s_{33'2}$ is more subtle. We define 

\noindent {\large $L_2')$} \newline
\noindent \parbox{3in}{ 
$$
\eqalign{ P_1~\to&~ P_1^+~= ~(p_1,p_1,0,0)~,~p_1 \to \infty \cr
P_2~\to&~ P_2^+~= ~(p_2,0,0,p_2,)~,~p_2 \to \infty  \cr
P_3~\to&~ P_3^-~= ~(p_3,-p_3,0,0)~,~~p_3 \to \infty  }
$$}
\parbox{3in}{
$$ \eqalign{
~~~Q_1~\to&~q_2 - q_3 ~=~ (-q_3,-q_3,q_2,-q_3)\cr
~~~Q_2~\to&~q_3 - q_2' ~=~(q_3,q_3,- q_2',q_3
)\cr
~~~Q_3~\to&~q_2'-q_2 ~=~(0,0,q_2' -q_2,0)}
\auto\label{np31}
$$}
The behavior of invariants is essentially identical to (\ref{npl4}).
At first sight, the roles of $P_2$ and $P_3$ are simply interchanged in 
going from $L_2$ to $L_2'$. However, the crucial difference is that in
(\ref{np31}) the ``transverse momenta'' $Q_1$ and $Q_2$ have ``finite
light-like components'' out of the ``transverse plane'', i.e. the 2-3 plane.
Most importantly, if the transverse components of $Q_1$ and $Q_2$ vanish, 
then the light-like component must vanish also.
It will become more significant in the next Section that we always identify
the transverse plane as the 2-3 plane. 
(Note that the limits for each of $P_1$ and $P_2$ can be taken to be any linear
combination of $P_1^+$ and $P_2^+$ and, provided they are not parallel, the 
result will be the same. Consequently the roles of $P_1$ and $P_2$ can be
smoothly interchanged.) 

Finally we give two corresponding realizations of the ``non-flip'' limit. 
In this case, if $u_1 \sim u_2$, then $ s_{12}$ and $s_{12'}$  are also 
finite. This allows $P_1$ and $P_2$ to have parallel limiting values.
We first define a limit, $L_3$, in which $P_3$ lies along a different
light-cone. We define 

\noindent {\large $L_3)$} \newline
\noindent \parbox{3in}{ 
$$
\eqalign{ P_1~\to&~ P_1^+~= ~(p_1,p_1,0,0)~,~ p_1 \to \infty\cr
P_2~\to&~ P_2^+~= ~(p_2,p_2,0,,0)~,~p_2 \to \infty   \cr
P_3~\to&~ P_3^+~= ~(p_3,0,0,p_3)~,~ p_3 \to \infty  }
$$}
\parbox{3in}{
$$ \eqalign{
~~~Q_1~\to&~q_2-q_3 ~= ~(0,0,q_2,-q_3)\cr
~~~Q_2~\to&~~q_3 - q_2' ~=~(0,0,-q_2',q_3)\cr
~~~Q_3~\to&~q_2'-q_2 ~=~(0,0,q_2' -q_2, 0)}
\auto\label{npl6}
$$}
The behavior of invariants is now
$$
\eqalign{&s_{12}~\st{\to} \infty~,~~~~
s_{23}~\to~2p_2p_3~,~~~~
s_{31}~\to ~2p_3p_1~,~~\cr
&s_{122'}~\st{\to} \infty~,~~~~
s_{233'}~\st{\to} \infty~,~~~~
s_{311'}~\to ~-2p_3q_3~ }
\auto\label{npl7}
$$
Comparing with (\ref{hp1}) and (\ref{ap1})-(\ref{ap4}), we see that 
the $L_3$ limit is the simple helicity-pole ``non-flip limit'' (\ref{hp1}) 
with
$$
p_1~\sim~u_1~,~~ p_2~\sim~u_2~,~~p_3~\sim~z_3
\auto\label{npl8}.
$$
We can also take $P_3$ to be in 
the same plane as $P_1$ and $P_2$ but with opposite space component. 
In this case $Q_1$ and $Q_2$ again aquire finite light-like components out 
of the transverse plane.
We define 

\noindent {\large $L_3')$} \newline
\noindent \parbox{3in}{ 
$$
\eqalign{ P_1~\to&~ P_1^+~= ~(p_1,p_1,0,0)~,~ p_1 \to \infty\cr
P_2~\to&~ P_2^+~= ~(p_2,p_2,0,,0)~,~p_2 \to \infty   \cr
P_3~\to&~ P_3^-~= ~(p_3,-p_3,0,0)~,~ p_3 \to \infty  }
$$}
\parbox{3in}{
$$ \eqalign{
~Q_1~\to&~q_2 -q_3 - q_3^+ ~=~ (-q^+_3,-q^+_3,q_2,0)\cr
~Q_2~\to&~~q_3 -q^+_3 -q_2'~=~(q^+_3,q^+_3,-q_2',q_3)\cr
~Q_3~\to&~q_2'-q_2~~= ~~(0,0,q_2' -q_2, 0)}
\auto\label{npl61}
$$}
The behavior of invariants is essentially the same as in (\ref{npl7}).
However, in contrast to $L_2'$, the light-like component $q^+_3$ can be
chosen independently of $q_3$ and so does not have to vanish if $q_3$
vanishes. 

From (\ref{np3}) and (\ref{npl6}) we see that the ``helicity-flip'' and 
``non-flip'' limits, $L_2$ and $L_3$ 
can respectively be distinguished by whether $p_1$ and $p_2$ are in 
opposite directions or the same direction on one light-cone. From 
(\ref{npl61}) it is also clear that the non-flip limit is truly a ``planar 
limit''. (\ref{np31}) differs from (\ref{npl61}) in that $P_2$ lies out of 
the plane.

\subhead{4.3 The S-W Representation and Regge Behavior}

As we outlined in the previous Section, the SW representation is obtained by
writing  appropriate partial-wave expansions for each set of hexagraphs
related by twisting. In particular, for the set of all hexagraphs related to
Fig.~4.4 by twisting we write 
$$ 
A^H_6 (z_1,z_2,z_3,u_1,u_2) = 
\sum  d^{J_1}_{-n_1,0} (z_1)\, d^{J_2}_{-n_2,0} (z_2)\, 
d^{J_3}_{n_1+n_2,0} (z_3)\,
u^{n_1}_1 u^{n_2}_2 a_{\til{\scriptstyle J}, \til{n}}
\auto\label{3pw}
$$
(As in our discussion of nonsense states in the previous Section, 
$-n_2$ is the $t_3$-channel center-of-mass helicity. Again 
we remark that we choose the present symmetric notation and language to make
direct contact with $s$-channel helicity amplitudes). The S-W transform is
obtained by converting the sums over $n_1, n_2, $ and $J_3$ to integrals. To
illustrate the general formalism more simply we again (temporarily) ignore
signature. In this case we can write 
$$ 
\eqalign{~~~~~~~~~~
A^H_6 &= \int_{>~ +~ <} ~{dn_1 ~(u_1)^{n_1}\over \sin \pi n_1}
\int_{>~ +~ <} ~{dn_2 ~(u_2)^{n_2} \over \sin\pi n_2} 
\int {dJ_3 ~d^{J_3}_{0,n_1+n_2} (z_3) \over \sin\pi(J_3-n_1-n_2)} 
\cr
&\times \sum^\infty_{J_1-|n_1|=0}\ d^{J_1}_{-n_1,0}(z_1)\ 
\sum^\infty_{J_2 - |n_2|=0}\ 
d^{J_2}_{-n_2,0}(z_2)\,
a {\centerunder{\raisebox{3.5mm}{$ \scriptstyle H_6,\til{\scriptstyle >}$}} 
{$\til{\scriptstyle J}\til{n}$}}
}
\auto\label{sw1}
$$
where the ${\raisebox{1mm}{\centerunder{$\scriptscriptstyle 
>$}{$\scriptscriptstyle <$}}}$ labels indicate the presence of separate 
integrals to reproduce the positive and negative helicity sums.

The triple-regge limits and helicity-pole limits can be studied by 
pulling the contours in (\ref{sw1}) to the left in the complex plane. (Again
we do not discuss the subtleties  of introducing second-type
representation functions etc. that are necessary to obtain a true asymptotic 
expansion). In the triple-regge limit, Regge poles at 
$l_1 = \alpha_1$, $l_2 = \alpha_2$, and $l_3 = \alpha_3$, give contributions
to each of the terms in the double sum in (\ref{sw1}) and we obtain a 
result very similar to (\ref{mrll}) (for simplicity we omit
the denominator sine factors)           
$$
\eqalign{A^H_6~~
\centerunder{$\large\sim$}{\raisebox{-4mm} 
{\centerunder{$z_1,z_2,$}{\raisebox{-4mm} 
{$ z_3, \rightarrow\infty$}}}}
~~&z_1^{\alpha_1}z_2^{\alpha_2}z_3^{\alpha_3}\sum^{\infty}_{N_1=0}
\sum^{\infty}_{N_2=0}\biggl[u_1^{\alpha_1 - N_1}u_2^{\alpha_2 - 
N_2}\beta_{\alpha_1, \alpha_2, \alpha_3, N_1, N_2}~+ \cr
&~
u_1^{-\alpha_1 + N_1}u_2^{\alpha_2 - 
N_2}\beta_{-\alpha_1, \alpha_2, \alpha_3, N_1, N_2}
~+~ u_1^{\alpha_1 - N_1}u_2^{-\alpha_2 + N_2}\beta_{\alpha_1, 
-\alpha_2, \alpha_3, N_1, N_2}\cr
&~~~
~~~~~~~~~+ u_1^{-\alpha_1 + N_1}u_2^{-\alpha_2 +  N_2}\beta_{-\alpha_1,
-\alpha_2, \alpha_3, N_1, N_2}\biggr]}
\auto\label{trp}
$$
where $\beta_{\alpha_1, \alpha_2, \alpha_3, N_1, N_2}$ is the Regge-pole 
residue of the F-G (analytically continued) ``non-flip helicity-amplitude'' 
$a \centerunder{\raisebox{3.5mm}{$\scriptstyle H_6,{\scriptstyle >>}~~~~~~~~$}}
{$ \scriptstyle J_1,J_2,J_3,n_1,n_2$}(t_1,t_2,t_3)$ at $J_i=\alpha_i, ~i=1,2,3$ and 
$n_i=J_i - N_i,~i=1,2$ and $\beta_{-\alpha_1, \alpha_2, \alpha_3, N_1, N_2}$
is the Regge-pole residue of the ``helicity-flip'' amplitude
 $a \centerunder{\raisebox{3.5mm}{$ \scriptstyle H_6,
{\scriptstyle ><}~~~~~~~~$}}
{$ \scriptstyle J_1,J_2,J_3,n_1,n_2$}(t_1,t_2,t_3)$ at
$J_i=\alpha_i, ~i=1,2,3$ and $n_1=-J_1 + N_1,n_2=J_2 - N_2 $. The
$u_i^{\pm\alpha_1}$ contributions come repectively from the
${\raisebox{1mm}{\centerunder{$\scriptscriptstyle 
>$}{$\scriptstyle <$}}}$ integrals in (\ref{sw1}). (The symmetry under
$u_1 \to 1/u_1, u_2 \to 1/u_2$ implies that the first and last sums in 
(\ref{trp}) can be identified, as can the second and third. When the 
hexagraph of Fig.~4.3 is part of a larger hexagraph this symmetry is, in 
general, not present.) 

To obtain the complete behavior of $M_6$ in the 
triple-regge limit we must add contributions corresponding to the additional
hexagraphs illustrated in Fig.~3.2. These contributions will have the
same general form as (\ref{trp}) but with the indices $1,2$ and $3$ cyclically
rotated. We also add twisted graphs by incorporating signature factors 
properly.

In analogy with (\ref{5.55}), the helicity-pole limit (\ref{hp1}) picks out
the first term of the first sum in (\ref{trp}) i.e. 
$$
A_6 
\centerunder{$\large\sim$}{\raisebox{-4mm} 
{\centerunder{$u_1,u_2,$}{\raisebox{-4mm} 
{$ z_3, \rightarrow\infty$}}}}
 (z_1u_1)^{\alpha_1}\, (z_2u_2)^{\alpha_2}\, z_3^{\alpha_3}\, 
\beta_{\alpha_1,\alpha_2,\alpha_3,0,0}
\auto\label{hp1r}
$$
while the second limit picks out the first term of the second sum i.e. 
$$
A_6 
\centerunder{$\large\sim$}{\raisebox{-4mm} 
{\centerunder{$u_1,1/u_2,$}{\raisebox{-4mm} 
{$ z_3, \rightarrow\infty$}}}}
 (z_1u_1)^{\alpha_1}\, (z_2u_2^{-1})^{\alpha_2} z_1^{\alpha_3} 
\beta_{\alpha_1,-\alpha_2,\alpha_3,0,0}
\auto 
$$
and so distinct helicity amplitudes, i.e. non-flip and flip, contribute in
the distinct helicity-pole limits while both amplitudes contribute in the
full triple-regge limit. This is why we refer to (\ref{hp1}) and (\ref{hp2}) 
respectively as non-flip and heicity-flip limits. Note that, as we 
anticipated in the previous sub-section, in both limits the dependence on 
both $z_1$ and $z_2$ is determined by the $u_1$ and $u_2$ dependence. This is 
necessary for the amplitudes to be directly expressible in terms of 
invariants, as is done in the next sub-section.

\subhead{4.4 Asymptotic Analytic Structure}

Consider how the cuts of Fig.~4.3 are represented asymptotically.
From (\ref{ap1})-(\ref{ap3}), we can write 
$$
\eqalign{ (z_1u_1)^{\alpha_1} (z_2u_2)^{\alpha_2} z_3^{\alpha_3}~ 
&= ~(z_1z_3u_1)^{\alpha_1}(z_1z_3u_2)^{\alpha_2}~
(z_3)^{\alpha_3-\alpha_1-\alpha_2}\cr
&\sim ~(s_{13})^{\alpha_1}(s_{2'3'})^{\alpha_2}(s_{11'3})^{ \alpha_3 - 
\alpha_1 - \alpha_2} }
\auto\label{hp1i} 
$$
showing how the hexagraph cuts of Fig.3.3(a) are represented in the limit
(\ref{hp1}). Similarly for the limit (\ref{hp2}), we can write 
$$
\eqalign{ (z_1u_1)^{\alpha_1} (z_2u_2^{-1})^{\alpha_2} z_3^{\alpha_3}~ 
&= ~(z_1z_3u_1)^{\alpha_1}\biggl({z_2z_3 \over u_2}\biggr)^{\alpha_2}~
(z_3)^{(\alpha_3-\alpha_1-\alpha_2)}\cr
&\sim ~(s_{13})^{\alpha_1}(s_{2'3'})^{\alpha_2}(s_{11'3})^{ \alpha_3 - 
\alpha_1 - \alpha_2}}
\auto\label{hp2i} 
$$
and so the cuts of Fig.~4.3(a) contribute similarly to both the non-flip
and helicity-flip limits. However, for the limit 
(\ref{hp2}) we can also write 
$$
\eqalign{  (z_1u_1)^{\alpha_1} (z_2u_2^{-1})^{\alpha_2} z_3^{\alpha_3} 
&\sim (z_1z_3u_1)^{(\alpha_1+\alpha_3-\alpha_2)/2}
\biggl({z_2z_3 \over u_2}\biggr)^{(\alpha_2+\alpha_3-\alpha_1)/2}
\biggl({z_1z_2 u_1 \over u_2}\biggr)^{(\alpha_1+\alpha_2-\alpha_3)/2}\cr
&\sim ~(s_{31})^{(\alpha_1+\alpha_3-\alpha_2)/2}
(s_{23})^{(\alpha_2+\alpha_3-\alpha_1)/2}
(s_{12})^{(\alpha_1+\alpha_2
-\alpha_3)/2}}
\auto\label{hp2i1}
$$
showing that the cuts of Fig.~4.3(b) are also present. Both sets of cuts are
represented simultaneously by the same asymptotic expression , which is 
equivalent to saying that asymptotically the two sets of cuts can not be 
distinguished. It is crucial that there is no expression corresponding to
(\ref{hp2i1}) for the limit (\ref{hp1}). As a result 
we conclude that both sets of cuts in Fig.~4.3 are present in the 
helicity-flip amplitude appearing in the limit (\ref{hp2}) while only those
of Fig.~4.3(a) appear in the non-flip amplitude. Conversely we expect 
amplitudes with both sets of cuts to appear in the helicity-flip amplitude 
and not in the non-flip amplitude.

The importance of this last discussion is as follows. The conventional
``triple-regge'' limit of the one-particle inclusive cross-section 
has been studied\cite{bw} in some detail in QCD. As we noted above, it is
in fact the non-flip helicity-pole limit (\ref{hp1}) that is involved. In
this limit only triple-regge behavior associated with the inclusive
cross-section discontinuities of Fig.~4.3(a) appears. The helicity amplitude
that appears is the same amplitude that appears in the reggeon unitarity
formula for the two-reggeon cut discontinuity. Consequently the 
triple-pomeron vertex that appears in the inclusive cross-section can be
identified with the vertex, discussed in the next 
Section, that appears in elastic scattering pomeron diagrams and in RFT. 
However, there are additional ``helicity-flip'' triple-regge vertices associated
with the helicity-flip amplitude appearing in the limit (\ref{hp2}), and
more generally with the full set of helicity flip amplitudes appearing in
the full triple-regge limit. These vertices appear in amplitudes containing
both the usual inclusive cross-section cuts and the second set of cuts
illustrated in Fig.~4.3(b). Such amplitudes have not been discussed within
QCD. We will discuss some of the simplest contributing Feynman diagrams in
Section 7. As we discuss in the next Section, the additional vertices make
very important contributions to the general solution of reggeon unitarity
for multiparticle amplitudes and, as a result, will play a crucial role in
our general construction of hadrons and the pomeron in QCD. 

\newpage

\mainhead{5. POMERON AND REGGEON DIAGRAM SOLUTIONS OF REGGEON UNITARITY }

In Section 3, we generically described a Regge pole participating in the 
generation of a Regge cut as a reggeon and gave the controlling ``reggeon 
unitarity'' equations. In this Section we discuss the solution of these
unitarity equations in terms of ``reggeon diagrams'', in analogy
with the Feynman diagram solution of conventional momentum space unitarity.
Historically such diagrams were first introduced\cite{gr} to describe the 
interactions of an even-signature pomeron Regge pole. Later they appeared as
describing\cite{bfkl,bs} the interactions of reggeized gluons in leading (and
next-to-leading) log calculations in massive gauge theories. Both pomeron
and reggeized gluon diagrams are often referred to generically as ``reggeon 
diagrams''. In this Section, for simplicity, we will use ``reggeon'' to 
refer exclusively to an odd signature (``reggeized gluon'') Regge pole, with
intercept close to unity. Therefore ``reggeon diagrams'' involve reggeized
gluons and ``pomeron diagrams'' involve pomerons. We will also use ``reggeon
unitarity'' exclusively for the unitarity condition on reggeons and use
``pomeron unitarity'' to describe the unitarity condition for pomerons. 
This will cause no confusion in this Section since we will not consider 
diagrams containing both reggeons and pomerons. A-priori they can certainly
appear simultaneously in diagrams. Indeed, our ultimate aim is to first
construct a reggeon diagram description of QCD amplitudes and then via the
analysis of infra-red divergences and the use of pomeron ``phase-transition
theory'' convert to a pomeron diagram description. At an intermediate stage 
there will in fact be diagrams containing both reggeons and pomerons. 

Pomeron and reggon unitarity equations differ only in the structure of
signature factors and one purpose of this Section is to describe the
diagrams for both cases in the same formalism. The most important new 
result will be the extension of the diagram formalism to
solve the unitarity equation (\ref{run}) for a large class of multiparticle
F-G amplitudes. For the reggeon case, there is 
a vector particle (the gluon) which becomes massless as the intercept 
of the reggeon goes to one. Massless particle states give rise to infra-red 
divergences of reggeon interactions which are very important in our later
discussion of QCD. In this Section we will consider only massive reggeons 
and will only briefly discuss the specific form of reggeon and pomeron
interactions. We begin with the simplest, and historically oldest, diagrams. 

\subhead{5.1 Pomeron Diagrams for Elastic Scattering}

We emphasize from the outset that we expect to use pomeron (or reggeon) 
diagrams to discuss infra-red phenomena only. Therefore we will be
interested in small $t_i$'s (and small $\kbar$'s) only. We denote the
pomeron trajectory by $j = \alpha_{\spom}(t)$, with ${\alpha_{\spom}(0) \sim
1}$. Since the pomeron has even signature, all multi-pomeron cuts are also
even signature and so signature factors can effectively be neglected. That
is, for small values of all the $t_i$ we can take, in (\ref{run}), 
$$ 
\sin{\pi\over 2}(\alpha_1 - \tau_1^{\prime}) \sim \cdots \sim \sin {\pi\over 2}
(\alpha_M - \tau_M^{\prime}) \sim 1  . 
\auto\label{sgf}
$$
$\xi_M$ simply gives a factor of $-1$ for each additional pomeron in the 
state and so for an $M$-pomeron state 
$$
\xi_{M} ~~\sim~~ (-1)^{M-1}
\auto\label{sgf1}
$$

We introduce the usual RFT variables, that is energies $E_i$ and
two-dimensional momenta $k_i$ as follows. We write 
$$
J_i = 1 - E_i \quad {\rm and} \quad t_i = k_i^2\qquad \forall ~i
\auto\label{rvr}.
$$                                     
so that (with $\Delta_k = 1 - \alpha(t_k))$
$$
\delta\bigl(J - 1 - \sum_{k=1}^M \Delta_k \bigr) ~~
\leftrightarrow ~~\delta\bigl(E - \sum_{k=1}^M \Delta_k \bigr)
\auto\label{eco}
$$
which we can regard as ``energy conservation'' by pomeron intermediate 
states. We can also write
$$
\int {dt_jdt_k\over \lambda^{1/2}(t_i,t_j,t_k)}~~\leftrightarrow~~
2 \int d^2\kbar_j d^2\kbar_k ~\delta^2\left(\kbar_i-\kbar_j-\kbar_k\right)
\auto\label{mco}
$$
which is ``momentum conservation'' for pomerons. 
The pomeron unitarity equation is initially derived for positive $t$ and the 
change of variables (\ref{rvr}) can be made with the $k_i$ taken to be 
two-dimensional Minkowski momenta. However, the continuation to negative $t$ 
is most simply done by rotating the plane of the $k_i$ so that they become 
spacelike and can be straightforwardly identified with the transverse 
momenta of $s$-channel Feynman diagram or unitarity calculations. The full
continuation of pomeron unitarity from the positive $t$ region, where it is 
first derived, is actually quite complicated\cite{siw} and it is
non-trivial, and very important, that the only $J$-plane singularities that
survive at negative $t$ are those due to Regge cuts. This implies that a
solution of pomeron unitarity in the small $t$ region should be sufficient
to satisfy full multiparticle $t$-channel unitarity equations. 

Note that since the amplitudes involved will be functions of the $t_i$
invariants, the ``transverse plane'' involved in (\ref{mco}) can be shifted
by the addition of an orthogonal light-like vector without changing the
resulting integrals. That is, the relation of the transverse plane to
four-dimensional momenta is ambiguous up to an orthogonal light-like vector.
This point will be important later in the Section. 

For elastic (particle) scattering the negative $t$ unitarity equation 
(\ref{run}), with the approximations (\ref{sgf}) and (\ref{sgf1}), 
is solved by pomeron diagrams as follows. The even-signature 
F-G amplitude $a^+(J,t)$ is written in the form 
$$
a^+(J,t)\equiv F\left(E,\kbar^2\right)=\sum^\infty_{m,n=1}
F_{mn}(E,\kbar^2)~,\auto\label{elu}
$$
where (omitting all factors of $(2\pi)^3$)
$$
\eqalign{
F_{mn}\left(E,\kbar^2\right)=~
	\int\prod_{i,j}& d^2\kbar_i d^2\underline{k'}_j~
	\delta^2 \biggl[\kbar-\sum^m_{i=1} \kbar_i\biggr]
	~\delta^2 \biggl[\kbar-\sum^n_{i=1} \underline{k'}_i\biggr]\cr
&
{ g_m ~g_n~
A_{mn}\left( E, \kbar_1,\ldots \kbar_m, \kbar'_1,\ldots
	\kbar'_n\right) \over 
\bigl(E-\sum^m_{i=1}\Delta(\kbar_i)\bigr)
\bigl(E-\sum^n_{j=1}\Delta(\underline{k'}_j\bigr)} ~,
}
\auto\label{elu1}
$$
The $g_m$ are couplings of pomerons to the external particles
which, in general, will be functions of the transverse momenta. In the 
approximation which gives (\ref{sgf}) and (\ref{sgf1}) we should take 
$$
g_m~ \sim~(i)^m~, 
\auto\label{sgf2}
$$
The $A_{mn}$ are pomeron scattering amplitudes. To include the simplest 
diagrams (without pomeron interactions) in (\ref{elu1}), the $A_{mn}$ should
include disconnected amplitudes, e.g. the completely disconnected 
amplitude
$$
A_{mn}(E,\til{k},\til{k}')~=~\delta_{mn}~\delta^2(\sum \kbar_i -\sum 
\kbar_i')~\Gamma^{-1}_m(E, \kbar_1,\ldots \kbar_m)
\auto\label{ue}
$$
where $\Gamma_m$ is the m-pomeron propagator
$$
\Gamma_m(E, \kbar_1,\ldots \kbar_m)~=~{1 \over \left[E- 
\sum_{r=1}^m\Delta_r(\kbar_r)\right] }
\auto\label{pro}
$$
(\ref{elu1}) is illustrated in Fig.~5.1

\begin{center}
\leavevmode
\epsfxsize=4in
\epsffile{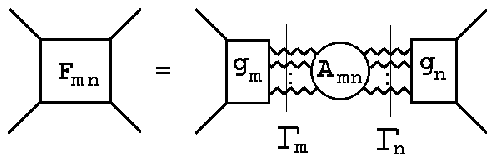}

Fig.~5.1 Multi-Pomeron Contribution to Elastic Scattering.

\end{center}
The amplitudes $A_{mn}$ must satisfy the pomeron
unitarity equation 
$$
\eqalign{
A_{mn}\biggl(E+i\epsilon,\til{k},\til{k'}\biggr)  
	- &A_{mn}\biggl(E-i\epsilon,\til{k},\til{k'}\biggr)=
\sum_r(-1)^r ~i \int \prod_s  d^2\kbar_s''
~ \delta [E- \sum^r_{s=1} \Delta_s ]\cr
& \times \delta^2[\kbar-\sum^r_{s=1}\kbar_s'']
~ A_{mr}\biggl(E+i\epsilon,\til{k},\til{k''}\biggr) 
A_{rn}\biggl(E-i\epsilon,\til{k''},\til{k'}\biggr).}
\auto\label{elu2}
$$
This equation is illustrated in Fig.~5.2.
 
\begin{center}
\leavevmode
\epsfxsize=4in
\epsffile{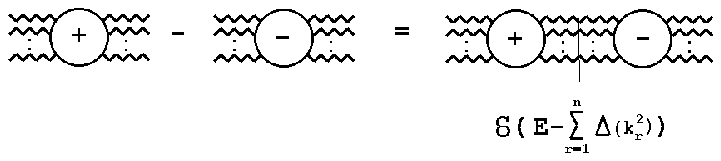}

Fig.~5.2 Unitarity for Pomeron Amplitudes

\end{center}

It is straightforward to write a general solution to (\ref{elu2}) in terms
of a (non-relativistic) graphical expansion involving arbitrary
(non-singular) vertices and propagators for states 
containing any number of pomerons. (That interactions are non-singular is 
assumed because of the absence of massless particles in the strong 
interaction). 

\begin{center}
\leavevmode
\epsfxsize=5in
\epsffile{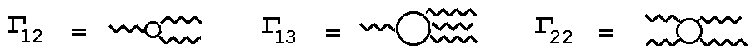}

Fig.~5.3 Pomeron Vertices

\end{center}
In the notation illustrated in Fig.~5.3, we take as interaction vertices 
$$
\Gamma_{12}~(=~\Gamma^*_{21})~=~ir_0~+ \cdots ~,~~~
\Gamma_{13}~=~\lambda_0~+ \cdots ~, ~~~
\Gamma_{22}~=~\lambda'_0~+ \cdots~,
\auto\label{pv}
$$
etc. The dots indicate that we could add transverse momentum 
dependence to the interaction vertices, but this would actually be 
inconsistent with making the approximations (\ref{sgf}), (\ref{sgf1}) and 
(\ref{sgf2}). It is important that all of these approximations are ultimately 
justified when the critical pomeron solution of RFT is formulated\cite{cri}.
It can be shown that all the neglected terms correspond to irrelevant
operators in the renormalization group scaling introduced at the critical
point. 

A general solution to (\ref{elu2}) is then given by the
complete set of diagrams having the general form illustrated in Fig.~5.4.

\begin{center}
\leavevmode
\epsfxsize=5in
\epsffile{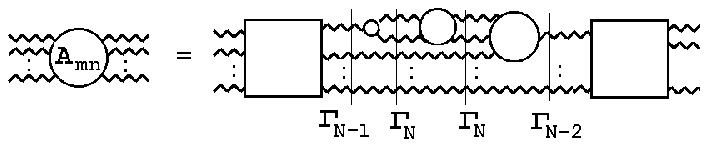}

Fig.~5.4 The General Form of Pomeron Diagrams 

\end{center}
These diagrams involve all possible 
combinations of propagators $\Gamma_m$, given by 
(\ref{pro}), coupled by the interaction vertices $\Gamma_{mn}$ given by 
(\ref{pv}).  There is an integration $\int d^2\kbar$ for each loop and 
momentum conservation is imposed at each vertex. 
The factor of $i$ in front of $r_0$ in (\ref{pv}), 
and all vertices for odd numbers of pomerons, reproduces the $(-1)^r$ factor
in (\ref{elu2}) when the usual graph cutting rules are applied. 

\subhead{5.2 Reggeon Diagrams for Elastic Scattering}

We consider next the modification of the diagrams of the last subsection 
when the pomeron is replaced by an odd-signature reggeon with trajectory 
$j = \alpha_R(t)$ such that ${\alpha_R(M^2) = 1}$, where $M^2 \sim 0$. 
The product signature rule says that odd number reggeon states appear in 
the odd-signature amplitude and even number states appear in the even 
signature amplitude. In a (spontaneously-broken) gauge theory 
the color quantum numbers break the signatured amplitudes up 
into sub-amplitudes.

For small values of all the $t_i$ we now take, in (\ref{run}), 
$$ 
\sin{\pi\over 2}(\alpha_i - \tau_1^{\prime}) ~\sim~ {\pi \over 2}~\alpha'
(t_i - M^2)  
\auto\label{sgfr}
$$
In the same approximation $\xi_M$ gives a factor of $+1$ when two 
odd-signature states are combined and a factor of $-1$ when an odd signature 
and even signature state are combined or when two even signature states are
combined. Instead of (\ref{elu}) we write 
$$
a^{\pm}(J,t)\equiv F^{\pm}\left(E,\kbar^2\right)=\sum^\infty_{n,m~ = ~even/odd}
F^{\pm}_{nm}(E,\kbar^2)~,
\auto\label{elur}
$$
where even/odd summations are respectively associated with the +/- sign
and (now omitting, in addition to the factors of $(2\pi)^3$, the factors of 
${\pi \over 2} \alpha'$ which compensate for the change in dimensions 
produced by the particle poles)
$$
\eqalign{ F^-_{m,n~=~odd}\left(E,\kbar^2\right)=~
	\int\prod_{i,j}& {d^2\kbar_i \over ({\kbar_i}^2 + M^2)}
 {d^2\underline{k'}_j \over ({\underline{k'}_j}^2 + M^2)}~
	\delta^2 \biggl[\kbar-\sum^m_{i=1} \kbar_i\biggr]
	~\delta^2 \biggl[\kbar-\sum^n_{i=1} \underline{k'}_i\biggr]\cr
& { G_m~G_n~
A^-_{mn}\left( E, \kbar_1,\ldots \kbar_m, \kbar'_1,\ldots
	\kbar'_n\right) \over 
\bigl(E-\sum^m_{i=1}\Delta(\kbar_i)\bigr)
\bigl(E-\sum^n_{j=1}\Delta(\underline{k'}_j\bigr)} ~,
}
\auto\label{unr}
$$
The $G_m$ are couplings of reggeons to external particles
and  the $A^-_{mn}$ are odd-signature reggeon scattering amplitudes. 
The scattering amplitudes $A^-_{mn}$ 
satisfy the reggeon unitarity equation 
$$
\eqalign{
A^-_{mn}\left(E+i\epsilon,\kbar\right)  
	-A^-_{mn}\left(E-i\epsilon,\kbar\right)=&
\sum_{r=odd}~(-1)^{(r-1)/2}\int \prod_s  
{d^2\kbar_s \over ({\kbar_s}^2 + M^2)}
~\delta^2[\kbar-\sum^r_{s=1}\kbar_s]\cr
& \delta [E- \sum^r_{s=1} \Delta_s ]
~A^-_{mr}\left(E+i\epsilon,\kbar\right) A^-_{rn}\left(E-i\epsilon,
	\kbar\right).}
\auto\label{unr1}
$$
$F^+_{mn}$ is similarly defined in terms of amplitudes $A^+_{mn}$ satisfying 
the analagous equation. 

The reggeon unitarity equations can again be solved by reggeon diagrams. We
can introduce general reggeon interaction 
vertices in the same way as we did 
for the pomeron. Because of signature conservation there is no 
$\Gamma_{12}$ vertex, only $\Gamma_{22}$ and $\Gamma_{13}$ vertices. 
For the m-reggeon propagator $\Gamma_m$ we take
$$
\Gamma_m(E, \kbar_1,\ldots \kbar_m)~=~{1 \over \prod (\kbar_r^2 + M^2)
\left[E- \sum_{r=1}^m\Delta_r(\kbar_r)\right] }
\auto\label{rpro}
$$

It is well-known that a reggeon diagram formalism is exactly what
emerges\cite{arw2,bs} from leading and next-to-leading log calculations in
gauge theories. This is a very non-trivial result. Indeed, as is explicitly
shown in \cite{bs}, matching sixth-order calculations with reggeon diagrams
allows $\Gamma_{22}$ to be extracted. The existing higher-order (eighth and
tenth-order) results are then predicted completely by iterating the reggeon 
interaction. This is consistent with the requirement that,
once the form of the reggeon
interactions is known, the structure of the full set of reggeon diagrams 
is determined by reggeon unitarity. However, the reggeon interaction 
obtained is quite complicated and so in the next sub-section we
digress from our general formalism to briefly summarize some of the results
obtained in massive
(i.e. spontaneously-broken) gauge theories. 

\subhead{5.3 Reggeon Diagrams in Gauge Theories}

Because of the presence of (close to) massless particles, the reggeon
interaction vertices of a gauge theory (unlike the pomeron vertices
discussed above) contain transverse momentum singularities and can not be
approximated as regular. For simplicity we assume in this Section that the 
gauge symmetry breaking has provided all gluons with the same mass. In 
Section 8 we will consider a more complicated situation.

In lowest order perturbation theory the trajectory
function is given by 
$$
\eqalign{ \alpha(q^2)~&= ~1~+~\Delta(q^2) \cr
&=~1~+~g^2 C ~(q^2 + M^2)~J_1(q^2)}
\auto\label{traj1}
$$
where $C$ is a color factor that we give below and 
$$
J_1(q^2)  \sim 
\int {d^2\underline{k}_1d^2\underline{k}_2   
\over (\underline{k}^2_1 + M^2)( \underline{k}^2_2  + M^2)}
\delta^2
[ \underline{q}-\underline{k}_1-\underline{k}_2 ] ~.
\auto\label{J1} 
$$

Introducing transverse momenta $\kbar_1, \kbar_2, \kbar_1', \kbar_2'$ 
that satisfy 
momentum conservation (i.e. $\kbar_1 + \kbar_2 = \kbar_1'+ \kbar_2'$)
we can write\cite{bs}
$$
\Gamma_{22}(\kbar_1,\kbar_2,\kbar_1',\kbar_2' )= 
a~(\kbar_1+\kbar_2)^2+b~M^2 - c~R_{22}\left(\kbar_1,
\kbar_2,\kbar_1',\kbar_2'\right)~,
\auto\label{6.9}
$$
where $a, b$ and $c$ are color factors that we discuss below 
and $R_{22}$ has the complicated structure 
$$                        
\eqalign{
R_{22}(\kbar_1,\kbar_2,\kbar_1',\kbar_2')= ~~
&~{{\left(\kbar^2_1+M^2\right)\left(
{\kbar^2_2}'+M^2\right)+\left(\kbar^2_2+M^2\right)\left(
{\kbar^2_1}'+M^2\right)}\over
{\left(\kbar_1-\kbar_1'\right)^2+M^2}}\cr
+ &~{{\left( \kbar^2_1+M^2\right)\left( {\kbar^2_1}'+M^2\right)+
\left( \kbar^2_2 +M^2\right)\left({\kbar^2_2}'+M^2\right)}
\over {\left( \kbar_1 -
\kbar_2'\right)^2+M^2}}~. }
\auto\label{6.10}
$$
The (massive) BFKL equation\cite{bfkl} is simply the color zero
reggeon ``Bethe-Salpeter'' 
equation obtained by iterating the reggeon interaction $\Gamma_{22}$ in 
reggeon diagrams. 

In other papers\cite{cw} we 
have outlined a program for constructing reggeon interactions by beginning 
with a $\Gamma_{12}$ vertex which contains a nonsense zero that ensures it 
does not participate directly as a reggeon vertex. The singular part of
reggeon interactions (including the massless limit of 
(\ref{6.10}) giving the BFKL kernel) can then be
constructed from $t$-channel particle discontinuities and the Reggeon Ward
identities discussed in the next Section. 
This construction implies that we can 
simultaneously discuss the color structure and the singularities of 
reggeon interactions due to particle (gluon) poles, by 
introducing the transverse momentum diagram notation illustrated in Fig.~5.5.
(Transverse momentum diagrams are essentially reggeon diagrams without 
reggeon propagators.)

\begin{center}
\leavevmode
\epsfxsize=4in
\epsffile{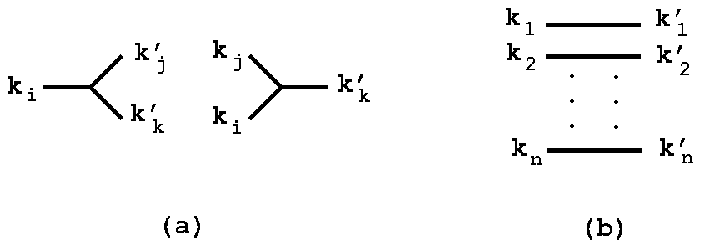}

Fig.~5.5 (a)Vertices and (b) Intermediate States in Transverse Momentum
Diagrams.
\end{center}

Amplitudes are obtained by combining vertices and intermediate states 
according to the following rules.

\begin{itemize}

\item{For each three-point vertex, illustrated in Fig.~5.5(a), we write a factor
$$
16\pi^3~ f_{ijk}~\delta^2 
(\sum k_i~  - \sum k_i')\left( (\sum k_i~)^2 + M^2 \right)
$$
where $f_{ijk}$ is the usual antisymmetric group tensor.}
\item{For each intermediate state, illustrated in Fig.~5.5(b), we write a factor
$$
(16\pi^3)^{-n}\int { d^2k_1...d^2k_n~ \over (k_1^2 + M^2) ... (k_n^2 +M^2) }
$$
}
\end{itemize}

The trajectory function (\ref{traj1}), with the color factor included, 
is now given by the simple 
transverse momentum diagram shown in Fig.~5.6.

\begin{center}
\leavevmode
\epsfxsize=1.5in
\epsffile{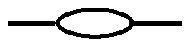}

Fig.~5.6 The Trajectory Function.
\end{center}
The interaction term $c  R_{22}$ is given by the sum of diagrams in
Fig.~5.7.
\begin{center}
\leavevmode
\epsfxsize=4.5in
\epsffile{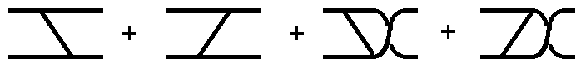}

Fig.~5.7 The Reggeon Interaction $R_{22}$
\end{center}
We have used a thick line in the above transverse momentum diagrams to 
specifically indicate that color factors are included in the same notation. 
Note that the interaction of 
Fig.~5.7 is not projected on a particular color channel in the $t$-channel.

The regular part of the reggeon 
interaction $\Gamma_{22}$ is more complicated to include in the diagram 
formalism. The zero mass part (i.e. the $(\kbar_1 + \kbar_2)^2$ term) is 
determined, from the singular part, by the reggeon Ward identities that we
discuss in the next Section. In the color channel with gluon quantum numbers 
the mass term can be included diagrammatically as shown 
in Fig.~5.8.
\begin{center}
\leavevmode
\epsfxsize=3in
\epsffile{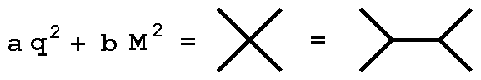}

Fig.~5.8 The Regular Interaction in the Reggeon Channel.
\end{center}

We also introduce a diagrammatic notation for color factors only 
that will be useful in the remainder of the paper. This is illustrated in 
Fig.~5.9 (since only color factors are involved we use thinner lines.)
\begin{center}
\leavevmode
\epsfxsize=4.5in
\epsffile{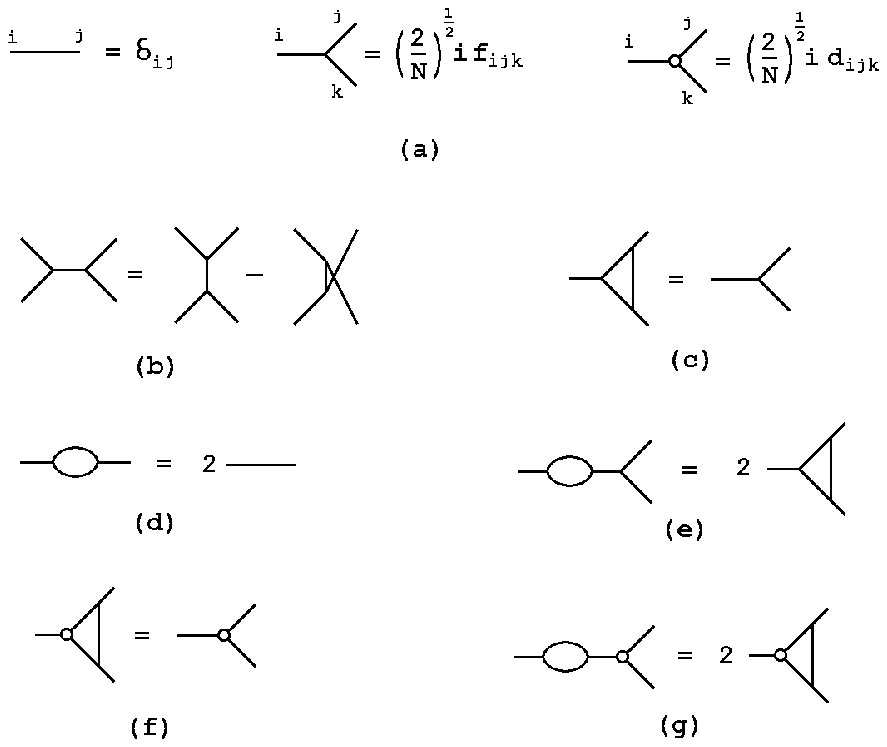}

Fig.~5.9 (a) Color Tensors, (b) The Jacobii Identity, (c) - (g) Relations 
Between Tensors.
\end{center}
We have included the 
symmetric $d$-tensor that exists in SU(N) for $N \geq 3$ and 
expressed a number of useful identities, not all of which are independent,
in the same notation. 

The reggeization of the gluon implies that in the gluon 
quantum number channel, the leading higher-order interactions give only 
simple Regge pole exchange. 
The necessary condition for reggeization is\cite{bfkl} the 
``bootstrap cancelation'' that is expressed in terms of transverse momentum
diagrams in Fig.~5.10. 
\begin{center}
\leavevmode
\epsfxsize=3in
\epsffile{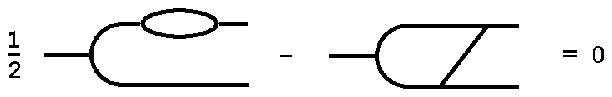}

Fig.~5.10 The Bootstrap Condition for Reggeization.
\end{center}
The momentum part of this equation is trivial, given the structure of the 
vertices. The color part follows from Fig.~5.9(e). 
The cancelation of Fig.~5.10 ensures that when the reggeon interaction
$\Gamma_{22}$ is included in the triple reggeon interaction,
only Fig.~5.8 survives and this
simply iterates the reggeization. 

It is interesting to note that, because of 
Fig.~5.9(f), the cancelation of Fig.~5.10 holds also if the left-hand vertex 
in each diagram is replaced by a vertex containing a $d$-tensor. This 
implies that in QCD there is an additional ``bound-state'' reggeon\cite{bw}
(or colored pomeron) in the symmetric octet channel with a trajectory that
is exchange degenerate with the reggeized gluon. We will refer frequently to
this feature in later Sections. 

It is clear from (\ref{traj1}) and (\ref{6.10}) that
both the trajectory function and the reggeon 
interaction are singular as the mass $M \to 0$. We will discuss the
significance of this singularity structure in detail in Section 8. In the 
next subsection we return to our abstract discussion and consider the 
extension of the elastic scattering formalism to
multiparticle amplitudes. We continue to illustrate most of our discussion 
with pomeron diagrams because specific examples are simpler to write down.
However, we will constantly emphasize the close similarity of pomeron and 
reggeon diagrams.
 
\subhead{5.4 Helicity Amplitude Pomeron Diagrams and Helicity-Flip Vertices}

We begin our discussion of multiparticle amplitudes by considering the 
implications of pomeron unitarity for the helicity-pole
limits (\ref{hp1}) and (\ref{hp2}) discussed in the previous Section. In 
both cases the leading behavior is described by a single (analytically 
continued) helicity amplitude which satisfies (\ref{run}) in a 
straightforward manner. 

Pomeron diagrams describing the non-flip limit which, as we noted, is the 
usual inclusive cross-section triple-Regge limit, were studied many years 
ago. The structure of the diagrams was derived directly from 
pomeron unitarity\cite{csw}, as we now describe, and also from hybrid Feynman 
diagram calculations\cite{abbs}. The results were the same. 
For simplicity we omit signature labels as in
Section 4, and again introduce Reggeon Field Theory notation by writing
for the non-flip amplitude, introduced via (\ref{sw1}), as 
$$
A^{{\cal N }} (E_1,E_2,E_3,q_1^2,q_2^2,q_3^2) ~\equiv~
a \centerunder{\raisebox{3.5mm}{$\scriptstyle H_6{\scriptstyle >>}~~~~~~~~$}}
{$\scriptstyle J_1,J_2,J_3,n_1,n_2$}(t_1,t_2,t_3)
\auto\label{ha1}
$$
with $ n_1 =J_1, ~n_2 =J_2, ~J_i =1-E_i, ~~ t_i ~=~q_i^2, ~i=1,2,3$. For the 
helicity-flip amplitude we similarly write
$$
A^{{\cal F}} (E_1,E_2,E_3,q_1^2,q_2^2,q_3^2) ~\equiv~
a \centerunder{\raisebox{3.5mm}{$\scriptstyle H_6{\scriptstyle ><}~~~~~~~~$}}
{$\scriptstyle J_1,J_2,J_3,n_1,n_2$}(t_1,t_2,t_3)
\auto\label{ha2}
$$
where, in this case, $n_1 =J_1, ~n_2 = - J_2$.
 
The crucial property of $A^{{\cal N}}$ and $A^{{\cal F}}$ is that 
they each satisfy a pomeron unitarity equation in all three of the
$E_i$-channels which is essentially the same as the unitarity equation for
elastic amplitudes. As a result we can write 
$$
 A^{\gamma } (E_1,E_2,E_3) ~
=~\sum_{m,n,r}~F^{\gamma}_{mnr}(\til{E},\til{t}) ~~~~~~~{\gamma} = {\cal 
N,F}
\auto\label{nfu}
$$
where $F^{{\cal N,F}}_{mnr}$ is constructed from pomeron diagrams as
illustrated in Fig.~5.11. 

\begin{center}
\leavevmode
\epsfxsize=5in
\epsffile{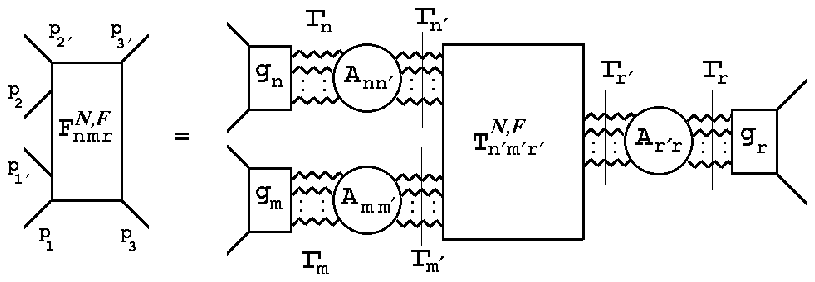}

Fig.~5.11 Triple-Regge Pomeron Diagrams for $F^{{\cal N,F}}_{mnr}$. 

\end{center}
The notation is the same as in Fig.~5.1 and the $A_{nn'}$ are again 
given by Fig.~5.4. 
The new element in Fig.~5.11 is the central vertex $T^{{\cal N,F}}_{m'n'r'}$ 
coupling the pomerons in each $E_i$-channel. The pomeron unitarity equation 
forces the diagrams to have the essentially factorized form of Fig.~5.11
where, apart from the $T^{{\cal N,F}}$, all the couplings and interactions
are identical to those appearing in elastic scattering. Indeed, the 
flip/non-flip distinction between the amplitudes is carried only by
the $T^{{\cal N,F}}$ vertices. 

If the $T^{{\cal N,F}}$ are connected amplitudes which can, like the $g_n$,  be 
treated as constants independent of all the reggeon
transverse momenta then the unitarity condition is clearly
satisfied. Each $E_i$-channel will have a separate transverse momentum plane 
and will be completely separate dynamically. Also, when the pomeron
intercept is close to unity and all transverse momenta are small, we have 
$$
E_1 \sim \sum \Delta (k^2_{m'}) \sim E_2 \sim \sum \Delta (k^2_{n'}) \sim 
\sum \Delta (k^2_{r'}) \sim E_3 \sim 0
\auto\label{trr}
$$ 
and so $T^{{\cal N}}_{m'n'r'}$ will coincide with the corresponding elastic
scattering pomeron vertex, in first approximation. However, as we emphasized 
in Section 3, ``helicity-flip vertices'' do not appear internally within the 
reggeon unitarity equation. It is important for the dynamical role of the 
anomaly that we discuss in later Sections that there are no vertices 
corresponding to the $T^{{\cal F}}_{m'n'r'}$ in elastic scattering pomeron
diagrams. These vertices appear only in the role of joining scattering 
channels, as in Fig.~5.11.

It is not necessary that $T^{{\cal N}}_{m'n'r'}$ or $T^{{\cal F}}_{m'n'r'}$
be connected. In fact diagrams involving disconnected $T^{{\cal N,F}}$
vertices are the most interesting dynamically since they couple the
transverse momentum dependence in the three channels. Such diagrams 
play a crucial role in our analysis. Therefore we want to be sure 
we fully understand their construction and their dynamical origin. As
the following discussion shows, there are various
subtleties when disconnected $T^{{\cal N,F}}$ vertices are involved. The 
relative definition of the transverse momentum planes becomes an issue, 
as well as the ordering of different disconnected interactions.

The simplest diagram with a disconnected vertex is that shown 
in Fig.~5.12.
\begin{center}
\leavevmode
\epsfxsize=2.5in
\epsffile{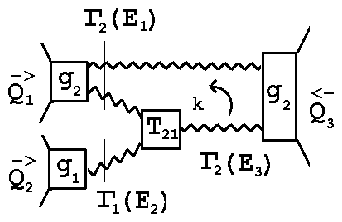}

Fig.~5.12 The Simplest Disconnected Triple-Regge Pomeron Diagram. 
\end{center}
This has the disconnected vertex illustrated in Fig.~5.13, i.e. one
disconnected pomeron line, together with a $T_{21}$ vertex. 
\begin{center}
\leavevmode
\epsfxsize=2in
\epsffile{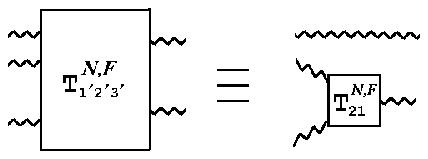}

Fig.~5.13 The Disconnected Vertex of $T_D$
\end{center}
We have used 
a square, and the $T_{21}$ notation, in order to
emphasize that the $T^{\cal F}_{21}$ vertex is distinct from the
$\Gamma_{21}$ vertex appearing in elastic scattering pomeron diagrams. 

The diagram of Fig.~5.12 is particularly simple since there is only one
transverse momentum integral. The diagram is written explicitly as 
$$
\eqalign{
F^{\gamma}_{122} ( E_1,E_2,&E_3,Q_1^2,Q_2^2, Q^2_3 )\cr
&=~  ~g_1~g_2^2~  
\Gamma_1(E_2)\int d^2\kbar ~~~~\Gamma_2(E_1)~
\Gamma_2(E_3)
~T^{\gamma}_{21}\left((Q_1 
+k)^2,Q_2^2,(Q_3-k)^2\right)\cr
&=~ ~{g_1~g_2^2~   \over 
[E_2 - \Delta(Q_2^2)]}
\int{ d^2\kbar ~~~~ T_{21}\left((Q_1 + k)^2,Q_2^2,(Q_3-k)^2\right) 
\over [E_1 - \Delta(\kbar^2) - \Delta((Q_1-\kbar)^2)]
[E_3  -  \Delta(\kbar^2) - \Delta((Q_3-\kbar)^2) ]} } 
\auto\label{TDS}
$$
where $\gamma = {\cal N, F}$. It is important to have a consistent physical
interpretation of the ``transverse momentum'' in this diagram. 

As we stated in the previous Section, we will always define the 
``transverse plane'' to be the 2-3 plane. An immediate question is
whether the integration in (\ref{TDS}) can be taken to be in the transverse 
plane. For the reggeon cuts of Fig.~5.12 to
be generated correctly, both $Q_1$ and $Q_3$ must either lie in the
$\kbar$-plane or lie outside of it only by an orthogonal light-like vector.
Having in mind the underlying Feynman diagram origin of Regge behavior, we
also expect that the presence of the pomeron connecting
the $Q_1$ and $Q_3$ external vertices requires the integration to be 
transverse to large light-cone momenta at these vertices. In principle, this
also defines the plane for the integration. 

At this point it becomes crucial that we are considering a helicity-pole
limit, rather than a triple-Regge limit. The helicity-flip helicity-pole limit 
is the more complicated case. Consider the particular kinematics of the
limit $L_2$ defined by (\ref{np3}). In this case the ``transverse plane'' is
indeed the the $Q_i$-plane but it is not orthogonal to $P_3^+$. However,
if we take the realization $L_2'$ given by (\ref{np31}), then the transverse
plane is orthogonal to both $P_1$ and $P_3$. Also $Q_3$ lies in the
transverse plane and $Q_1$ is obtained by adding an orthogonal light-like
vector to a vector in the transverse plane. Therefore, by using $L_2'$ it is
clear that the integration in Fig.~5.12 can indeed be taken to be the
transverse plane (provided we utilize the transverse plane ambiguity). 

For the non-flip limit, the $L_3'$ description given in
(\ref{npl61}) shows immediately that the transverse momentum integral can be 
defined to be the transverse plane, once the (light-like vector) ambiguity 
for $Q_1$ is exploited. There is one important difference 
between the contribution of Fig.~5.12 in the flip and non-flip limits, apart 
from the different $T_{21}$ vertices. This is in the implicit light-like 
components carried by the $Q_i$. Using the $L_2'$ and $L_3'$ limits 
to justify writing the diagram as an integral in the transverse plane 
implies that if there is an infra-red divergence as $Q_1 \to 0$ 
then in the flip limit this is associated also with a vanishing longitudinal 
momentum, whereas for the non-flip limit this is not the case. This 
is important if infra-red divergences of this kind are ultimately to be 
interpreted as related to wee partons and the ambiguities of light-cone
quantization at zero longitudinal momentum. 

Clearly it is also important that in a helicity-pole
limit the full six-point amplitude becomes dependent on only six of the
eight independent variables i.e. three invariants  conjugate to $E_1$, $E_2$
and $E_3$ and the three $t_i$. The transverse integrals we are describing
are able to represent the full amplitude only when it is independent of the
remaining angles. 

Consider next the diagram of Fig.~5.14. 
As indicated, there are now two transverse momentum integrals. 
The $\kbar_1$-integration should be be orthogonal to the light-cone momenta
at the $Q_1$ and $Q_3$ vertices while the $\kbar_2$-plane should be
orthogonal to the light-cone momenta at the $Q_2$ and $Q_3$ vertices. 
However, to construct the diagram we can construct the $\kbar_1$ loop first 
using the $L_2'$ limit for the helicity-flip limit, or the $L_3'$ limit
for the non-flip limit. Then, having the invariant amplitude expressed
as an integral in the transverse plane, we can smoothly interchange the form 
of $P_1$ and $P_2$ and similarly construct the $\kbar_2$ loop. The 
conclusion is that both integrations can be taken to be in the transverse
plane. 
\begin{center}

\leavevmode
\epsfxsize=2.5in
\epsffile{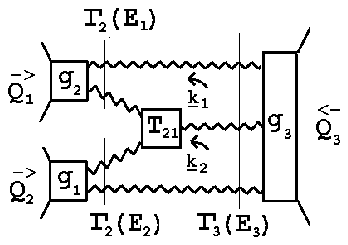}

Fig.~5.14 Another Disconnected Triple-Regge Pomeron Diagram. 
\end{center}

In the notation illustrated, we can write Fig.~5.14 (as another relatively
simple explicit example) in the form 
$$
\eqalign{
T^{\gamma}& ( E_1,E_2,E_3,Q^2_1,Q^2_2, Q^2_3 )~=~ 
\int d^2\kbar_1 d^2\kbar_2 ~g_2^2~ g_3~ \Gamma_2(E_1)
\Gamma_2(E_2)\Gamma_3(E_3)\cr
&~~~~~~~~~~~~~~~~~~~~~~~~~~~~~~~~~~~~~~~~~~~~~
\times T^{\gamma}_{21}\left((Q_1 +k_1)^2,(Q_2 -k_2)^2,(k_1-k_2)^2\right)\cr
&=~ 
\int d^2\kbar_1 d^2\kbar_2 ~
{g_2^2~ g_3~ 
 \over [E_1 - \Delta(\kbar_1^2) - \Delta((Q_1-\kbar_1)^2)]
 [E_2 - \Delta(\kbar_2^2) - \Delta((Q_2-\kbar_2)^2)]} \cr
&~~~~~\times 
{T^{\gamma}_{21}\left((Q_1 +k_1)^2,(Q_2 -k_2)^2,(k_1-k_2)^2\right)
 \over [E_3 - \Delta((Q_1 - \kbar_1)^2) - \Delta((Q_2-\kbar_2)^2) 
- \Delta((\kbar_1 +\kbar_2)^2)]}
}
\auto\label{TD}
$$
An extension of the above discussion shows that there is no difficulty in
constructing transverse momentum integrals for general diagrams of the form
of Fig.~5.11 and Fig.~5.14 in which multiple pomerons are exchanged, provided
there is only a single disconnected vertex. It is important to remember, 
however, that the ``physical'' transverse momenta involved, in general 
contain light-like momenta orthogonal to the transverse plane that we 
integrate over. For helicity-flip limits the presence of the light-like 
components has a special infra-red significance.

Next we consider Fig.~5.15 as an example of a diagram of the form of 
Fig.~5.11
in which there are apparently two disconnected central vertices.
Diagrams of this kind are particularly relevant for the arguments of later 
Sections.
\begin{center}
\leavevmode
\epsfxsize=2.4in
\epsffile{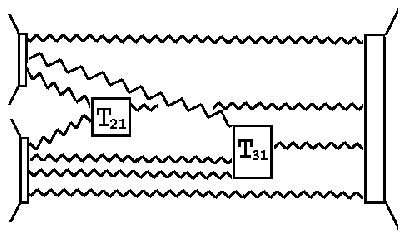}

Fig.~5.15 A Triple-Regge Pomeron Diagram with Two Central Vertices.

\end{center}
In this diagram there are five transverse momentum 
integrations and four reggeon propagators. From the above discussion, all of 
the transverse integrations can be taken to be in the same plane.
A-priori it is not clear, however, which of the $T_{21}$ and $T_{31}$
vertices is ``energy non-conserving''. By starting at each of the external 
particle couplings and considering the unitarity condition for each possible 
cut of the diagram it is straightforward to show\cite{csw} that the diagram
must contain only one unique energy non-conserving vertex. (The same result
was obtained for $A^{\cal N}$ amplitudes by direct calculation of hybrid
Feynman diagrams\cite{abbs}). The vertex occurs where there is a transition 
from $E_1$ and $E_2$ propagators to $E_3$ propagators. In particular, if we
insert propagators as shown in Fig.~5.16 we determine that, 
as indicated by the notation,
\begin{center}
\leavevmode
\epsfxsize=2.5in
\epsffile{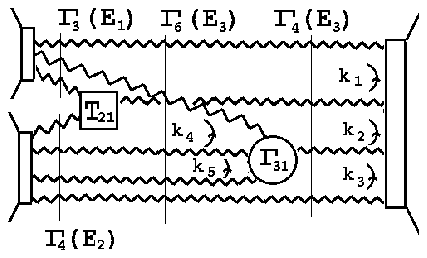}

Fig.~5.16 The Diagram of Fig.~5.15 with a Particular Non-Conserving Vertex. 
\end{center}
the $T_{21}$ vertex is the energy non-conserving vertex. 
Alternatively, as illustrated 
in Fig.~5.17, we can insert propagators in the same diagram in such a 
manner that the $T_{31}$ vertex is non-conserving. 
In the inclusive cross-section this freedom of choice is the freedom to 
choose the rapidity-ordering of the two vertices. From the present 
perspective it is the topological ambiguity in the insertion of propagators
which gives the freedom of choice.

\begin{center}
\leavevmode
\epsfxsize=2.5in
\epsffile{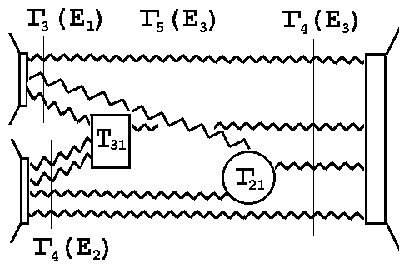}

Fig.~5.17 The Diagram of Fig.~5.15 with an Alternative Non-Conserving Vertex. 
\end{center}

The most general set of diagrams for both $A^{{\cal N}}$ and $A^{{\cal F}}$ 
involves all possible connected and disconnected $T^{{\cal N}}_{m'n'r'}$ 
and $T^{{\cal F}}_{m'n'r'}$ respectively. As in the last example, one 
diagram topology will often generate a number of distinct diagrams which 
differ only by which reggeon propagators are inserted. All such diagrams are 
considered as distinct. 

\subhead{5.5 Helicity Amplitude Reggeon Diagrams}

From our discussion of reggeon and pomeron diagrams for elastic scattering
it is clear that we can construct helicity amplitude reggeon diagrams in
close parallel with the construction of pomeron diagrams. For reggeon
diagrams signature plays an important role and so the new vertices $T^{{\cal
N,F}}_{m'n'r'}$ carry signature labels for each $E_i$ channel. Signature
is not conserved by the new vertices (in addition to energy) although 
$T^{{\cal N}}_{m'n'r'}$ will carry a nonsense zero at $E_3 - E_2 - E_1 = 0$
when signature is not conserved. The signature non-conserving 
$T^{{\cal F}}_{m'n'r'}$, i.e. the helicity-flip vertices,
need not contain such a factor. 

A particularly interesting situation occurs when a nonsense zero appears in
one (or more) reggeon vertices involved in an ordering ambiguity of the
kind discussed for pomeron vertices in the previous sub-section. As discussed 
in \cite{cw}, reggeon interactions involving nonsense zero vertices can be 
constructed by simply allowing the zeroes to cancel a corresponding reggeon 
propagator. The logic behind this is that the zero will not appear in 
unsignatured amplitudes and that in such amplitudes the corresponding 
reggeon diagram can be constructed with the reggeon propagator present. When 
the signatured amplitude is formed the cancelation of the reggeon propagator 
by the nonsense zero will occur. For example, if we consider Fig.~5.17 
to be a reggeon diagram then $\Gamma_{12}$ will be a signature 
non-conserving, energy conserving, vertex with a nonsense zero. This 
nonsense zero will effectively cancel the $\Gamma_5(E_3)$ propagator and so 
the $\Gamma_{12}$ and $T_{31}$ vertices should simply be combined to obtain
a single disconnected, energy non-conserving, vertex as illustrated in 
Fig.~5.18.
\begin{center}
\leavevmode
\epsfxsize=3in
\epsffile{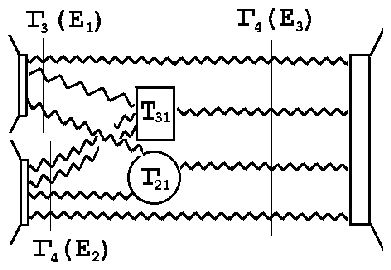}

Fig.~5.18 A Reggeon Diagram with a Disconnected Non-Conserving Vertex. 
\end{center}

We must determine the new $T^{{\cal N,F}}_{m'n'r'}$ vertices, scattering, by
direct calculation in QCD. We will construct important massless quark
components of these new vertices in Section 7. They play a crucial role in
our infra-red analysis. 

\subhead{5.6 Higher-Order Amplitudes}

Consider next the hexagraph amplitude $H_8$ shown in Fig.~5.19.

\begin{center} 
\leavevmode 
\epsfxsize=5.5in 
\epsffile{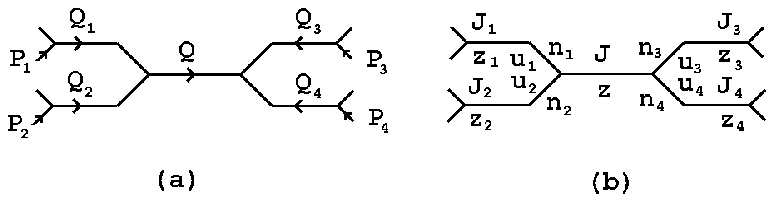} 

Fig.~5.19 A Hexagraph $H_8$ for $M_8$, (a) momenta, (b) angular variables, 
angular momenta and helicities. 

\end{center}
We consider both non-flip and helicity-flip limits at both vertices.
A sufficient description of the behavior of invariants in both limits is 
$$
\eqalign{&P_1.P_2 \sim \left( {u_1 \over u_2} + {u_2 \over u_1} \right)~, 
~~~P_1.P_3 \sim z\left(u_1u_3 + {1 \over u_1u_3}\right)~, 
~~~P_3.P_4 \sim \left( {u_3 \over u_4} + { u_4 \over u_3}\right)~, \cr
&P_1.Q_3 \sim z \left(u_1 + {1 \over u_1}\right)~, ~~~Q_1.Q_3 \sim z~, 
~~~P_4.Q_1 \sim z\left(u_4 + {1 \over u_4}\right) 
~ ~~ \cdots \cr
&~ P_1.Q,~P_2.Q,~ P_3.Q,~P_4.Q  ~~~\hbox{ finite } }
\auto\label{hp31}
$$
The double non-flip limit is 
$$
u_1,~u_2,~z, ~u_3, u_4 ~\to ~\infty 
\auto\label{hp32}
$$
while if the left-vertex, say, is helicity flip the limit is 
$$
u_1,~u^{-1}_2,~z, ~u_3, u_4 ~\to ~\infty 
\auto\label{hp33}
$$
If $u_1 \sim u_2$ and $u_3 \sim u_4$ then we see from (\ref{hp31}) that 
in the double non-flip limit (\ref{hp32}) both 
$P_1.P_2$ and $P_3.P_4$  are finite, whereas in (\ref{hp33}) $P_1.P_2 \to 
\infty$. 

Both limits are maximal helicity-pole limits and so the S-W representation 
shows that only a single helicity amplitude is involved. We can write the 
amplitude that appears in the double non-flip
limit (\ref{hp32}) as 
$$
A^{{\cal N_LN_R}}(E_1,E_2,E,E_3,E_4)~,
\auto\label{hp3}
$$
where $ J_i=n_i= 1-E_i, ~~i=1,2,3,4$ and $J=1-E$. Similarly we can write 
the flip/non-flip amplitude appearing in the limit (\ref{hp33}) as
$$
A^{{\cal F_LN_R}}(E_1,E_2,E,E_3,E_4)~, 
\auto\label{hp4}
$$
where now $J_i=n_i= 1-E_i, ~~i=1,3,4 $ and $J_2=-n_2=1-E_2,~J=1-E$, etc.
We have used an obvious 
generalization of notation in which, for example, ${\cal F_LN_R}$ denotes 
non-flip at the right vertex and helicity-flip at the left vertex. 

To understand how two-dimensional transverse momentum diagrams describe the 
limit, we discuss the realization of the limits (\ref{hp32}) and
(\ref{hp33}) in terms of light-cone momenta as
follows. For the double non-flip limit 
(\ref{hp3}) we take as external light-cone momenta
$$
\eqalign{ P_1~\to&~ P_1^+~= ~(p_1,p_1,0,0)~~~~
P_2~\to~ P_2^+~= ~(p_2,p_2,0,0)  \cr
P_3~\to&~ P_3^-~= ~(p_3,-p_3,0,0)~~
P_4~\to~ P_4^-~= ~(p_4,-p_4,0,0) }~~~\eqalign{ p_i \to \infty ~~\forall~ i }
\auto\label{8pl}
$$
It is clear from (\ref{hp31}) that to realize the internal $z \to \infty$
limit the $Q_i$ must also carry light-cone momenta, i.e. 
$$
\eqalign{ ~~~~~~Q_1~\to&~q_1^+ ~+~ Q_1^{\perp} 
~~~~~~~~~Q_2~\to~- q_1^+ ~+~ Q_2^{\perp}\cr 
Q_3~\to&~q_3^- ~+~ Q_3^{\perp}
~~~~~~~~~Q_4~\to~-q_3^- ~+~ Q_4^{\perp} }
\auto\label{8pl0}
$$
where the $q_i^{\pm}$ lie in the plane of the light-cone momenta 
(\ref{8pl}). The $q_i^{\pm}$ are large, but not as large as the $p_i$. As we 
discussed after defining $L_3'$ in the previous Section, when the limit is 
non-flip (at both vertices) there is no problem in choosing the light-cone 
momenta independently from the transverese momenta. 
The $Q_i^{\perp}$ are orthogonal to the light-cone momenta and lie in the
transverse plane. Momentum conservation gives
$$ 
Q~=~~Q_1^{\perp} ~+~ Q_2^{\perp} ~=~Q_3^{\perp}~+~ Q_4^{\perp} 
\auto\label{8pl1}
$$

For the non-flip/flip limit of (\ref{hp4}), one possibility is to utilise 
$L_2'$ and take 
$$
\eqalign{~~~~~~ P_1~\to&~ P_1^+~= ~(p_1,p_1,0,0)~ ~~~~~~~~~~~
P_2~\to~ P_2^+~= ~(p_2,0,0,p_2)  \cr
P_3~\to&~ P_3^-~= ~(p_3,-p_3,0,0)~~~~~~~~~~~
P_4~\to~ P_4^-~= ~(p_4,-p_4,0,0)}
\auto\label{8pl2}
$$
while for the internal momenta we take $Q_3$ and $Q_4$ as above
except that now we require specifically that 
$$
Q_3^{\perp} ~+~ Q_4^{\perp} ~~= ~~Q~~=~(0,0,q_2,0)
\auto\label{8pl4}
$$
so that $Q$ is still orthogonal to all four of the $P_i$ (this condition 
determines that we are considering a helicity-pole limit).  For $Q_1$ and 
$Q_2$ we take 
$$
Q_1~\to~\tilde{q}_1^- ~+~ Q_1^{\perp}~~~~~~~~~~~~~
Q_2~\to~- \tilde{q}_1^+ ~+~ Q_2^{\perp} 
\auto\label{8pl5}
$$
where $Q_1^{\perp}$ and $Q_2^{\perp}$ again lie in the transverse 
plane but $\tilde{q}_1^+$ is chosen to ensure 
orthogonality to both $P_1$and $P_2$ i.e. if 
$$
Q_1^{\perp}~=~(0,0,q_{12},q_{13}) ~~~~~~and ~~~~~~ 
Q_2^{\perp}~=~(0,0,q_2 - q_{12}, - q_{13}) 
\auto\label{8pl6}
$$
then
$$
\tilde{q}_1^+ ~=~(q_{13},q_{13},0,0)
\auto\label{8pl61}
$$

Taking a helicity-flip limit at a vertex again requires the introduction of 
light-like components determined by the spacelike components, 
for the corresponding $Q_i$. 
To realize the internal $z \to \infty$ limit it would suffice to take only 
$q_3^-$ large. We can not take $\tilde{q}_1^+$ large, i.e. take
$q_{13} \to \infty$ since with the 
definitions (\ref{8pl6}) and (\ref{8pl61}) this would imply $Q_1^2, Q_2^2 \to 
\infty$. To contribute to, or to realize the $z \to \infty$ limit with 
$\tilde{q}_1^+$ we must instead apply a Lorentz boost simultaneously to 
$P_2$ and $\tilde{q}_1^+$ that preserves their orthogonality. We write
$$
P_2 \to (p_2 C, p_2 S, 0, p_2)~, ~~\tilde{q}_1^+ \to
(q_{13}(C+S), q_{13}(C + S), 0, 0)
\auto\label{8pl8}
$$
where $C = cosh \zeta$ and $S= sinh \zeta$. We can then take 
$\zeta \to \infty$ as (all or) part of the limit $z \to \infty$. 
We notice that relative to $P_2$, the light-like component of 
$\tilde{q}_1^+$ continues to vanish as $q_{13} \to 0$.

The double flip limit 
$$
u_1,~u^{-1}_2,~z, ~u_3, u^{-1}_4 ~\to ~\infty 
\auto\label{hp35}
$$
introduces the amplitude $A^{{\cal F_LF_R}}(E_1,E_2,E,E_3,E_4)$.
To introduce a light-cone realization we proceed similarly. However,
we now have the extra subtlety that $\tilde{q}$
momenta have to be introduced for both vertices and a Lorentz boost $\zeta$ 
is essential at one (or both) of the vertices to realize the internal  
$z \to \infty$ limit. 

It seems that in a general helicity-pole limit we can 
always find a kinematic representation in which each of the internal $Q_i$ 
momenta is out of the transverse plane only by an orthogonal light-like 
vector. As the foregoing and following discussion shows, this feature
underlies the fact that helicity-pole limits can be described by 
helicity-amplitudes that satisfy pomeron and reggeon unitarity via
transverse momentum integrals. We repeat that, for the QCD physics of 
divergences associated with the anomaly that we discuss in later 
Sections, it is important to remember that in helicity-flip limits
the physical ``transverse momenta'' involve a closely related
light-like longitudinal component. 
 
\subhead{5.7 Pomeron and Reggeon Diagrams for Higher-Order Amplitudes}

The general form of the diagrams for each of the helicity amplitudes
corresponding to Fig.~5.19 is illustrated in Fig.~5.20.
(For simplicity we have not explicitly included propagators.) 

\begin{center} 
\leavevmode 
\epsfxsize=3in 
\epsffile{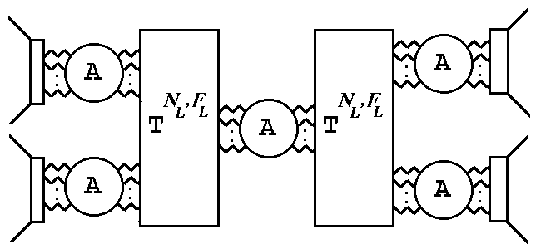} 

Fig.~5.20 The Structure of Pomeron and Reggeon Diagrams for 
$A^{{\cal N_LN_R}}, A^{{\cal F_LN_R}} ... $
\end{center}
As implied by the notation, the $T^{{\cal N,F}}$ vertices are the same as those 
that appear in the $A^{{\cal N,F}}$ discussed above - including disconnected 
vertices. A-priori it is not obvious that the resulting diagrams involving 
disconnected vertices coupling disconnected interactions actually make sense.
To see that this is the case it will be helpful to consider further 
specific examples.

Consider next the diagram of Fig.~5.21, which involves both disconnected 
vertices and and a disconnected amplitude, first as a contribution to
$A^{{\cal N_LN_R}}$ then as a contribution to $A^{{\cal F_LN_R}}$.
To make sense within our formalism it must be possible to write this diagram 
as a single integral in the transverse plane. According to our previous
discussion, the $\kbar$-integration should be orthogonal to the large
momenta at the $Q_1$ and $Q_3$ vertices.  Also for the Regge cuts in each of
the $Q_1$, $Q$ and $Q_3$ channels to have the correct discontinuity, each of
these momenta should either lie in the plane or be outside only by a
light-like vector. 
\begin{center} 
\leavevmode 
\epsfxsize=4in 
\epsffile{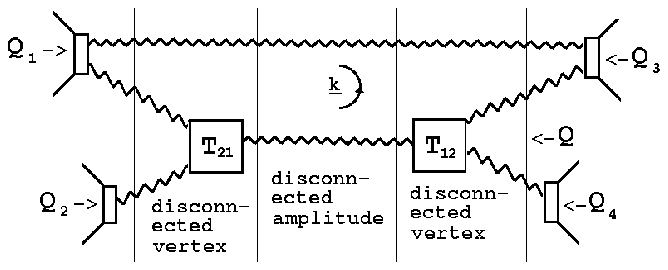} 

Fig.~5.21 A Diagram with Disconnected Components. 
\end{center}

To discuss $A^{{\cal N_LN_R}}$ we use the kinematics of 
(\ref{8pl})-(\ref{8pl1}). It is then clear that all of the requirements we 
have just listed are straightforwardly satisfied if we indeed take the 
$\kbar$-integration to be in the transverse plane. 
(Note that if we remove the external vertices, the same reggeon amplitude 
appears within elastic scattering pomeron diagrams except 
that the rapidities of the $T_{21}$ and $T_{12}$ vertices are integrated 
over to produce energy conservation.)
To consider $A^{{\cal F_LN_R}}$ we instead use the kinematics of 
(\ref{8pl2})-(\ref{8pl61}). Again the necessary requirements are satisfied
if the integration is in the transverse plane. 
We conclude that Fig.~5.21 gives a well-defined contribution 
to each of $A^{{\cal N_LN_R}}$, $A^{{\cal F_LN_R}}$, $A^{{\cal N_LF_R}}$
and $A^{{\cal F_LF_R}}$. As we have emphasized, whether the amplitude is 
flip or non-flip at each vertex is determined by whether the $T_{21}$ and 
$T_{12}$ vertices are flip or non-flip. When helicity-flip
vertices are involved, the amplitude has no relationship to elastic
scattering amplitudes. 

As an example with an important new feature we consider contributions to the 
hexagraph of Fig.~5.22.
We consider the helicity-pole limit in which all the vertices are non-flip.
\begin{center} 
\leavevmode 
\epsfxsize=3in 
\epsffile{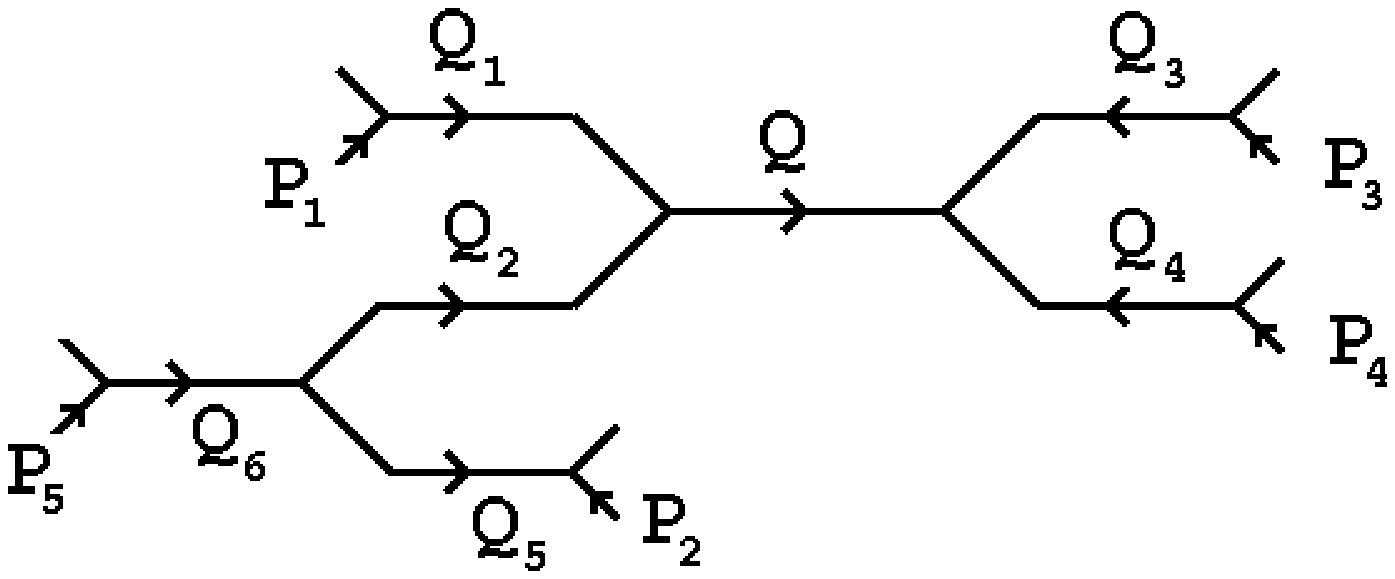} 

Fig.~5.22 A Higher-Order Hexagraph.
\end{center}
The general form of pomeron and reggeon diagrams contributing in this limit
is shown in Fig.~5.23.
The internal box couplings once again indicate either $T^{{\cal N}}$ or
$T^{{\cal F}}$ vertices which are both connected and disconnected. 
\begin{center} 
\leavevmode 
\epsfxsize=2.7in 
\epsffile{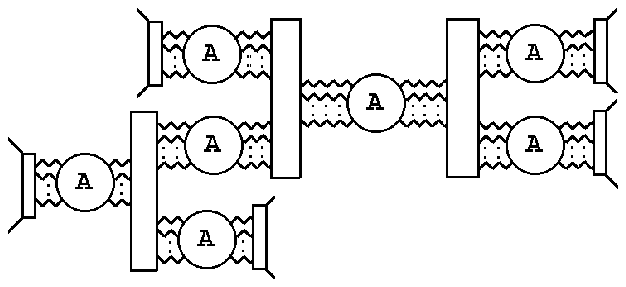} 

Fig.~5.23 The Form of Reggeon/Pomeron Diagrams for Fig.~5.21
\end{center}

We can set up a light-cone kinematic realization of the full non-flip limit by 
extending the discussion of the double non-flip limit of Fig.~5.19. We take 
$P_1$, . , $P_4$ to have the same form as in (\ref{8pl}) and in addition 
take 
$$
P_5~\to~ P_5^+~= ~(p_5,0,0,p_5)~. 
\auto\label{10pl}
$$
We also take $Q$, $Q_1$, . , $Q_4$ as in (\ref{8pl0}) and (\ref{8pl4}). In
addition to $Q_1$and $Q_2$, $Q_5$ and $Q_6$ must also be orthogonal to $P_1$
and $P_2$. $Q_6$ must be orthogonal to $P_5$ while $Q_2$ and $Q_5$ should 
not be. We therefore take $Q_5$ and $Q_6$ to have the form
$$
\eqalign{ &Q_5~\to~~ Q_5^{\perp}~ + ~q_1^{+}~ - ~\tilde{q}_6^- \cr
& Q_6~\to ~~ Q_6^{\perp} ~ + \tilde{q}_6^- }
\auto\label{10pl1}
$$
where $Q_5^{\perp}$ and $Q_6^{\perp}$ lie in the transverse 
plane but $\tilde{q}_6^-$ is chosen to ensure 
orthogonality of $Q_6$ to $P_5$ i.e. if 
$$
Q_6^{\perp}~=~(0,0,q_{62},q_{63}) 
\auto\label{10pl2}
$$
then
$$
\tilde{Q}_6^- ~=~(q_{63},q_{63},0,0)
\auto\label{8pl7}
$$

We see from (\ref{10pl2}) and (\ref{8pl7}) that to realize a sufficiently 
complicated non-flip limit we have had to introduce a 
light-like component for some of the $Q_i$ which are correlated with the 
transverse plane component. Previously this was only necessary to realize 
helicity-flip limits. The internal Regge and helicity-pole limits,
associated with the $Q$ and $Q_2$ lines respectively, can be realized by
taking $q_1^+$ and $q_3^-$ large appropriately. Alternatively a Lorentz
boost $ \zeta$ could be applied as in (\ref{8pl8}). To preserve the
orthogonality conditions the boost has to be applied simultaneously to all
of $P_2, P_5, Q_5$ and $Q_6$. 

Now consider the contribution of the pomeron diagram of Fig.~5.24  
to the limit under discussion. 
\begin{center} 
\leavevmode 
\epsfxsize=4in 
\epsffile{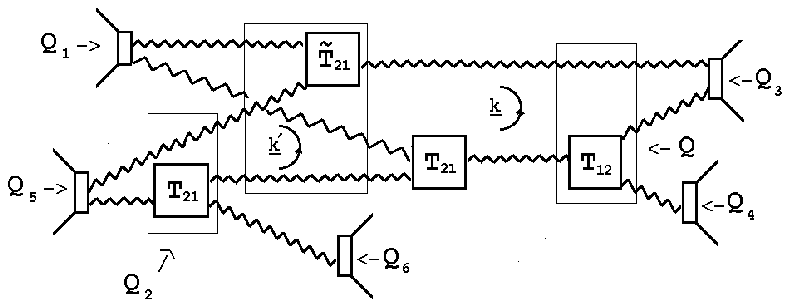} 

Fig.~5.24 A Pomeron Diagram Having the Form of Fig.~5.23.

\end{center}
With the above kinematics, both the $\kbar$ and $\kbar'$ integrations can
be taken to be in the transverse plane.
The internal boxes of Fig.~5.23 are indicated as thin-line boxes in 
Fig.~5.24. 
We now observe that, while the overall helicity-pole limit is entirely
non-flip, the $\tilde{T}_{21}$ vertex in Fig.~5.24 must actually be a
helicity-flip vertex. Although not directly 
coupled in the hexagraph of Fig.~5.22, $P_1$ and $P_5$ are in a 
relative helicity-flip limit. To see this we simply compare the form we have 
given for $P_1, P_5$ and $P_3$ with $P_1, P_2$ and $P_3$ in the $L_2'$ limit 
(\ref{np31}). Therefore if we introduce an internal  vertex coupling the
corresponding external vertices it must be a helicity-flip vertex. 
Comparing with Fig.~5.21 we see that the addition of the additional $P_5$ 
momentum, in a new plane, has produced a helicity-flip interaction 
accompanying
a helicity non-flip interaction (i.e. the $T_{21}$ vertex to the right 
of the $\tilde{T}_{21}$ vertex in Fig.~5.24). 

The pomeron diagram of Fig.~5.24 and the hexagraph of Fig.~5.22 have the
general form illustrated in Fig.~5.25. 
\begin{center} 
\leavevmode 
\epsfxsize=4.5in 
\epsffile{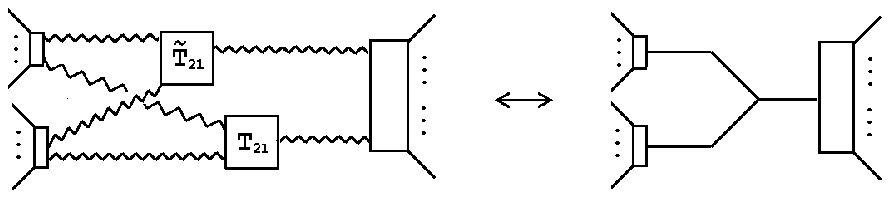} 

Fig.~5.25 A Pomeron Diagram and Corresponding Hexagraph. 
\end{center}
The point made in our discussion of Fig.~5.24 extends to general diagrams 
having the form of Fig.~5.25. That is, in a non-flip helicity-pole limit,
corresponding to the exposed vertex of the hexagraph of Fig.~5.25, a
helicity-flip vertex can appear as an energy non-conserving vertex
accompanying an energy conserving non-flip vertex, provided the left-hand
external couplings have sufficient structure. 
Since the $\tilde{T}_{21}$
vertex is the only one enclosed 
by a box in Fig.~5.24, this appears to violate our rule that the non-flip 
nature of the limit is correlated with that of the vertex. However, the two 
vertices picked out in Fig.~5.25 have (at first sight) an ordering ambiguity
and should be thought of as an overall disconnected 
vertex. That the vertex is non-flip is then determined by the presence 
of the $T_{21}$ vertex. 

The ordering ambiguity in the pomeron diagram of Fig.~5.25 is of the
kind we have discussed earlier. Apparently, the $\tilde{T}_{21}$
vertex can appear to the left or to the right
of the $T_{21}$ vertex. However, the helicity-flip vertex 
$\tilde{T}_{21}$ must be energy non-conserving for the diagram to be
consistent with pomeron unitarity. This is not the case when the
$\tilde{T}_{21}$ vertex is to the right of the ${T}_{21}$ vertex. Therefore
there is no diagram correponding to this possibility. In general we need
not distinguish the ordering of the vertices in Fig.~5.25 if we select 
specific pomeron states in each of the hexagraph channels and 
regard the combination of disconnected vertices as a single 
pomeron interaction. For example, if we 
consider the two-pomeron state in each of the channels in Fig.~5.25, we can
regard the $\tilde{T}_{21}$ and $T_{21}$ vertices as combining to produce a 
single disconnected vertex coupling the three two-pomeron states. If the 
pomerons are replaced by reggeons then, as we discussed in subsection 5.5,
$T_{21}$ contains a nonsense zero, and the ordering is similarly irrelevant.

The importance of our discussion
of Figs.~5.24 and 5.25 will become apparent in our QCD analysis 
when we are looking for bound-state amplitudes in Section 8. We will be 
looking for non-flip amplitudes within reggeon diagrams which also have 
infra-red divergences associated with helicity-flip vertices.
The crucial dynamics will be produced by accompanying 
helicity-flip processes that occur as we have just discussed. 

In Section 3.7 we 
noted that internal particle poles occur only in association with internal 
V subgraphs. The simplest hexagraph that contains an internal 
scattering amplitude associated entirely with internal 
Regge pole particle poles is that illustrated in Fig.~5.26.

\begin{center} 
\leavevmode 
\epsfxsize=3in 
\epsffile{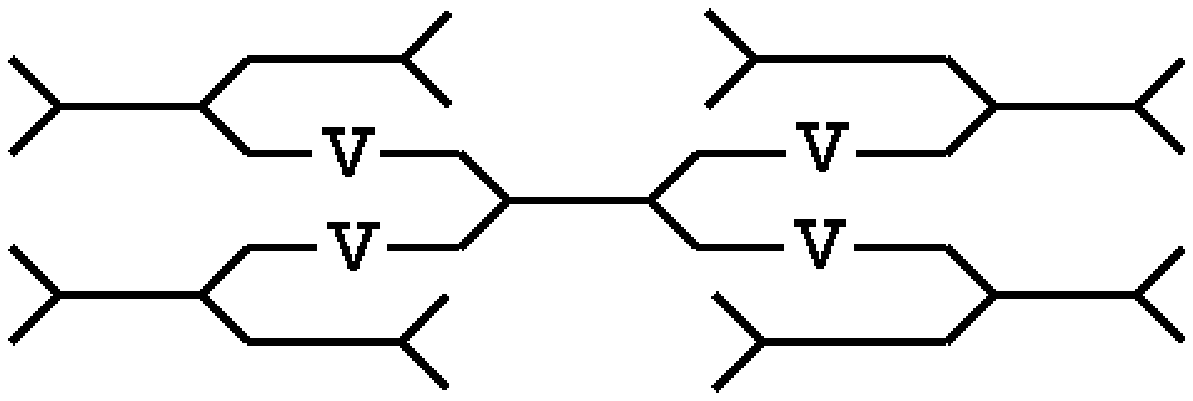} 

Fig.~5.26 A Hexagraph Containing Four V Subgraphs

\end{center}
In this hexagraph we have added, to each of the $Q_i$ lines of Fig.~5.19, the 
same additional vertices that we added to the $Q_2$ line to obtain the 
hexagraph of Fig.~5.22. When Regge poles (with trajectories close to
particle poles) are inserted for each of the V lines, the four-particle
amplitude enclosed in the thin-line box can 
be factorized off, first as a four-reggeon amplitude, and then as a
four-particle amplitude as the reggeons generate particle poles.
In our QCD analysis the Regge poles we will be looking for will (eventually) 
be those of bound-state hadrons and the amplitude will be that for pomeron 
exachange. The general form of pomeron and reggeon diagrams for the
hexagraph of Fig.~5.26 is illustrated in Fig.~5.27. 

\begin{center} 

\leavevmode 
\epsfxsize=3in 
\epsffile{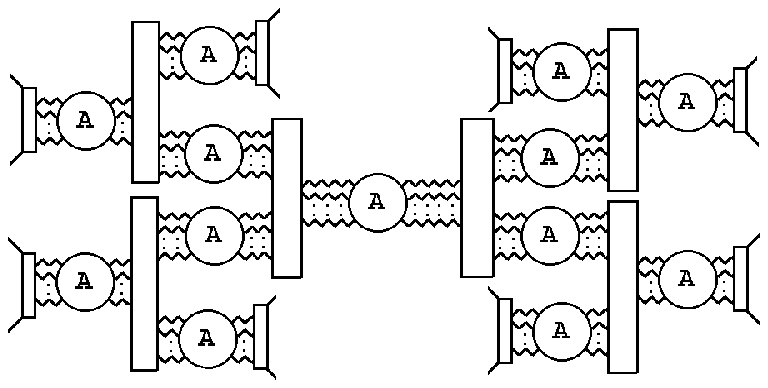} 

Fig.~5.27 The Structure of Pomeron and Reggeon Diagrams for the Hexagraph of
Fig.~5.26. 

\end{center}
The internal boxes once again contain $T^{{\cal N}}$ and $T^{{\cal F}}$
vertices. 

\subhead{5.8 General Helicity Amplitudes}

It should now be clear how our discussion generalises to any hexagraph. We 
isolate a single helicity-amplitude by an appropriate helicity-pole limit
(which in general will involve a combination of non-flip and flip limits for
the relevant $u_{ij}$ variables). Given the $T^{{\cal N}}$
and $T^{{\cal F}}$ vertices, the associated pomeron and reggeon
diagrams can then be constructed. A relatively simple example of the more 
complicated graphs that we will discuss in Section 8 is shown in Fig.~5.28. 
\begin{center} 

\leavevmode 
\epsfxsize=2.5in 
\epsffile{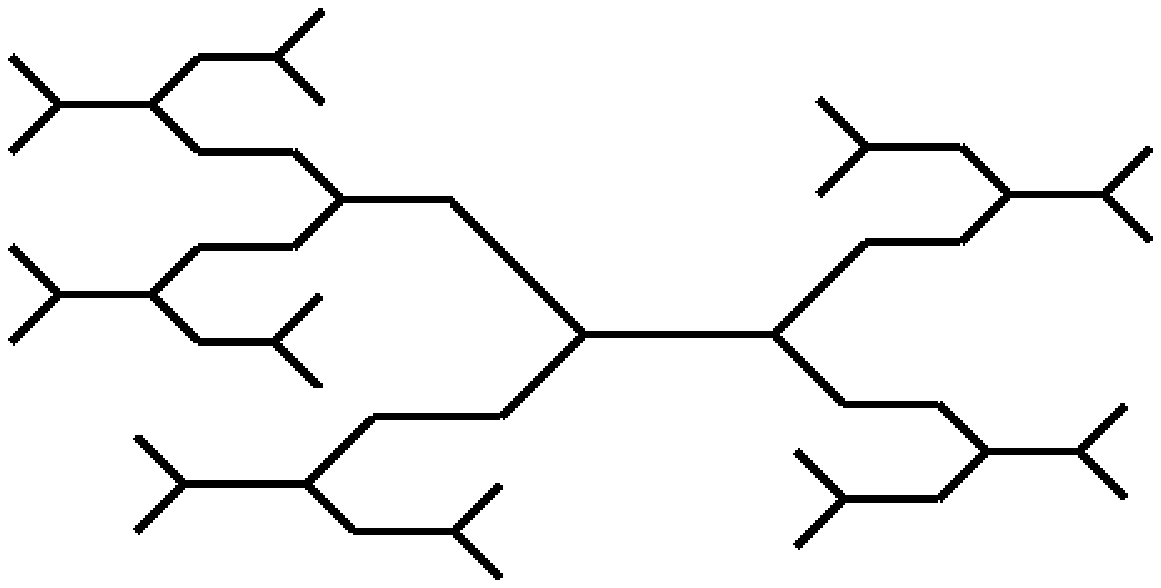} 

Fig.~5.28 A Relatively Simple Example of a Class of Hexagraphs.

\end{center}
We again emphasize 
that while the diagrams are constructed as 
two-dimensional integrals in a single transverse plane, when a helicity-flip 
vertex is involved, a correlated light-like vector is implicitly added to 
this plane to obtain the ``physical'' transverse momentum.
This is presumably deeply connected with the relationship between the 
QCD infra-red divergence results we will obtain and the zero-mode longitudinal 
momentum ambiguities of light-cone quantization.

\newpage

\mainhead{6. QUARK-REGGEON COUPLINGS AND REGGEON WARD IDENTITIES }

In this and the following Sections we will be concerned exclusively with 
QCD. The reggeons we consider are
specifically the reggeized gluons of QCD. In the infra-red analysis of 
Section 8 we will discuss setting the gluon mass to zero in some 
detail. In this Section we will simply omit 
the mass because we want to discuss some of the simplest infra-red divergences 
that occur when quarks are involved. We particularly focus on the 
inter-relation of such divergences with ``reggeon Ward identities''. 

We begin by constructing the lowest-order ``quark-reggeon'' couplings, i.e. the 
couplings for multi-reggeon exchange in on-shell quark scattering. Since a 
reggeon reduces to a gluon at $\underline{k}^2 = 0$, multi-reggeon amplitudes 
are, in general, necessarily given by corresponding (on-shell) gluon
amplitudes at zero transverse momentum. It follows from the formula
for F-G amplitudes\cite{arw1} that the particular (nonsense) reggeon 
amplitudes which provide the couplings for 
Regge cuts can be expressed as integrals of discontinuities, i.e. in terms
of on-shell $s$-channel intermediate states. We have not given this formula
here because, for multireggeon couplings, it is quite cumbersome. Here we 
will simply utilise the outcome. That is, the lowest-order contribution of a
particular multi-reggeon exchange to a scattering amplitude is 
given by that part of the corresponding high-energy multigluon exchange
amplitude having the appropriate (Regge cut) signature and in which all
intermediate $s$-channel states are put on-shell, i.e. no logarithms (of the
energy) are generated. This is what we will exploit to calculate reggeon
couplings. We will also note the even signature color octet case 
discussed in sub-section 5.3. In this case there is
effectively an ``AFS cancellation'' and the anticipated two-reggeon cut
contribution is replaced by a new Regge pole.

\subhead{6.1 Elementary Reggeon Couplings}

Consider the coupling $G_1$ of a fast (massive) quark to a single gluon -
temporarily ignoring color factors. The quark propagator gives 
$$
\eqalign{
{{\gamma\cdot p+m}\over{p^2-m^2}}
\quad&\centerunder{$\large\sim$}{\raisebox{-4mm} 
{$p_+ \rightarrow\infty$}}\quad
{{\gamma_-p_+ + \st{p}_\perp +\cdots} \over {p^2-m^2}}\cr
&~~~~\equiv {{\gamma_-+\gamma_\perp\cdot(p_\perp/p_+)+ 0(1/p^2_+)} 
\over {\left[ p_- - {{p^2_\perp - m^2} \over {p_+}}\right]}}
}
\auto\label{faq}
$$
For a quark initially and finally on-shell, we remove the 
$(p^{2} - m^{2})^{-1}$ factor and so, in lowest-order 
perturbation theory, 
$$
G_{1\mu}\sim g\gamma_-p_+\gamma_\mu\gamma_-p_+ ~ \sim ~\gamma_-p^2_+ 
\hspace{0.5in} {\rm if}\ 
\gamma_\mu = \gamma_+ ~.
\auto\label{faq2} 
$$
Choosing the frame in which the initial quark has $p_\perp = 0$ we have 
$\gamma_-p_+ = m$ and obtain
$$
G_{1\mu}\sim gmp_+\delta_{ - \mu} ~~\equiv  ~G_1p_+
\auto\label{faq3}
$$
Therefore we anticipate that, in a scattering process, the leading power
behavior (for $p_+\to\infty$) will be obtained if the spin of the scattering
quark is conserved, that is there is helicity 
conservation. In particular, for the scattering via single gluon exchange
of a fast quark with momentum $p_1$ off a quark with momentum $p_2$, we
obtain the helicity-conserving amplitude 
$$
g^2m^2 p_{1+}\delta_{-\mu} 
\left[ {{g_{\mu\nu}} \over {q^2_\perp}}\right] \delta_{\nu +}p_{2-}
\sim {{g^2m^2s} \over {q^2_\perp}}~~=G_1^2~{s \over t}
\auto\label{faq4}
$$
Lorentz invariance requires, of course, that this result hold independently
of whether $p_{2-}$ is large or not. If $p_{2-}$ is not large, the spin 
structure for the fast quark simply picks out, via gluon exchange, the
relevant spin component of the slow quark. 

Next we look for the lowest-order 2-reggeon coupling within the amplitude 
for a fast quark to exchange two gluons. As we described above, we ignore 
logarithms and place each intermediate state propagator on-shell (via $k_-$
and $k_+$ integrations). The denominator is thus removed from (\ref{faq})
also for intermediate states and, in analogy with (\ref{faq3}), we obtain 
$$
G_{2\mu_1\mu_2} \sim g^2\gamma_-p_+\gamma_{\mu_1}\gamma_-p_+ 
\gamma_{\mu_2}\gamma_-p_+ 
\sim g^2\gamma_-p_+\ \hspace{0.5in}{\rm if}\ \mu_1 = \mu_2 = + ~. 
\auto\label{faq5}
$$
giving the coupling illustrated in Fig.~6.1(a).

\begin{center}
\leavevmode
\epsfxsize=4in
\epsffile{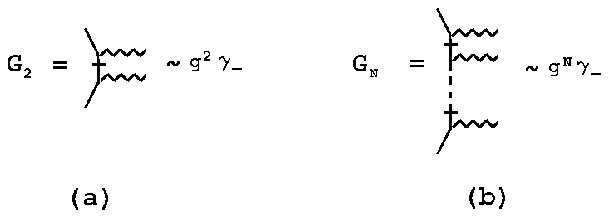}

Fig.~6.1 Quark-reggeon Couplings.

\end{center}
So the quark spin structure is again preserved and the unsignatured (helicity 
conserving) amplitude for 2-reggeon exchange has the lowest-order form
$$
A_2~\sim ~i ~G_2^2 ~s
\int {d^2\underline{k}_1d^2\underline{k}_2   
\over \underline{k}^2_1 \underline{k}^2_2 }
\delta^2
[\underline{q}-\underline{k}_1-\underline{k}_2 ]
\auto\label{faq6} 
$$
(Note that we should not cross the gluon lines in obtaining (\ref{faq6})
- the corresponding Feynman diagram gives only a real logarithm that we are not
interested in.) 

In the $J$-plane (\ref{faq6}) gives (writing $E=1-J$)
$$
A_2(E,q^2)~\sim ~{G_2^2  \over E}
\int {d^2\underline{k}_1 ~d^2\underline{k}_2   
\over \underline{k}^2_1 ~ \underline{k}^2_2 }
\delta^2
[ \underline{q}-\underline{k}_1-\underline{k}_2 ]
\auto\label{faq6E} 
$$
Higher-order contributions convert $E^{-1}$ to a two-reggeon propagator and
(\ref{faq6E}) takes the usual two-reggeon form 
$$
A_2(E,q^2)~\sim ~G_2^2 
\int {d^2\underline{k}_1~d^2\underline{k}_2   
\over \underline{k}^2_1 ~\underline{k}^2_2 }
{\delta^2
[ \underline{q}-\underline{k}_1-\underline{k}_2 ]
 \over [E -\Delta(\underline{k}^2_1) - \Delta(\underline{k}^2_1]}
\auto\label{faq6R} 
$$
The reggeon interactions described in sub-section 5.3 (in particular the 
full BFKL kernel) also appear as higher-order contributions.

Proceeding in the same way, we obtain the N-reggeon coupling 
illustrated in Fig.~6.1(b)
$$                                 
\eqalign{
G_{N\mu_1\cdots \mu_N}~ &\sim~ g^N\gamma_-p_+\gamma_{\mu_1}\gamma_-p_+ 
\cdots \gamma_-p_+\gamma_{\mu_N}\gamma_-p_+ \cr
&\sim g^N\gamma_-p_+\ \hspace{0.5in}{\rm if}\ 
\mu_1 = \mu_2 = \cdots = \mu_N = + } 
\auto\label{faq7}
$$
and for the unsignatured N-reggeon amplitude
$$
\eqalign{
A_N~\sim ~(i)^{N-1}m^2 G_N^2~s 
\int d^2\underline{k}_1\cdots& d^2\underline{k}_N\delta^2
[\underline{q}-\underline{k}_1-\underline{k}_2 
\cdots -\underline{k}_N] \cr
&\times {{1} \over {\underline{k}^2_1 }}\cdots
{{1} \over {\underline{k}^2_N }}
}
\auto\label{faq8} 
$$
so that helicity remains conserved. Again (\ref{faq8}) is the lowest-order 
component of the $E$-plane amplitude
$$
A_N(E,q^2)~\sim ~ G_N^2~
\int { d^2\underline{k}_1\cdots d^2\underline{k}_N
\over \underline{k}^2_1 \cdots
\underline{k}^2_N }
{\delta^2 [\underline{q}-\underline{k}_1-\underline{k}_2 
\cdots -\underline{k}_N] \over
[E -\Delta(\underline{k}^2_1) \cdots - \Delta(\underline{k}^2_N]}
\auto\label{faq8E} 
$$
Note that, once an overall factor 
of $m^2$ is absorbed by the normalization of the scattering states, the 
reggeon-couplings are independent of the quark mass $m$. It is also 
important for the discussion in the next Section that we need take only one 
of the scattering quarks to be fast in order to derive the kinematic 
structure of the lowest-order multi-reggeon exchange diagrams. The kinematic 
structure of the fast quark coupling to the exchanged gluons always imposes the 
same kinematic structure on the slow quark couplings. 

Positive (or negative) signatured amplitudes are obtained 
by adding (or subtracting) the corresponding TCP twisted  amplitude. That is 
we make a TCP transformation on one vertex or the other to which the 
multi-reggeon state is coupled and add (or subtract) the amplitude obtained.  
For the two-reggeon state we replace the fast quark coupling $G_2$ of a 
quark with a particular helicity by the coupling of a fast antiquark with 
the opposite helicity. Helicity conservation makes the parity part of the
twist trivial since parity conservation implies that the vertices for both
helicities are equal. Consequently the only effect of the TCP transformation 
is to replace the color factor of the quark by that for the antiquark.
For an abelian theory (QED) this simply changes the sign of the charge.
As a result the exchange of an even(odd) number of
reggeons contributes to the even(odd) signatured amplitude. This is the
normal signature rule for Regge cuts. (Of course, the photon is not actually 
reggeized in QED). When color factors are introduced, the TCP twist also
involves (color) charge conjugation. In this case, provided helicity is
conserved, signature can be identified with color charge parity. 

\subhead{6.2 Color Factors}

We define the color charge conjugation operation on the color matrix of the
gluon field by 
$$
A^i_{\alpha \beta} ~\to ~ -~ A^i_{\beta \alpha } ~
\auto\label{ccp}
$$
The minus sign indicates an intrinsic odd color parity for the gluon.
Quarks are transformed to antiquarks. We will discuss color parity for 
quarks in more detail in our second paper. For the purposes of this paper it 
is sufficient that color charge conjugation simply reverses the order of
multiplication of color matrices in internal quark loops. 

For SU(2) color, quark-quark scattering (via two gluon exchange and higher)
contains two $t$-channel color representations, i.e. in the $t$-channel 
$$
2~\otimes~2~\to ~1~\oplus~3
\auto\label{col}
$$
The singlet representation is symmetric (even color parity), while the triplet 
is antisymmetric (odd color parity). It is well-known\cite{arw2,bs} that at 
next-to-leading log the singlet amplitude contains the two-reggeon
cut while the triplet contains only the reggeizing gluon. It follows from
the bootstrap cancelation of Fig.~5.10. For gluon-gluon 
scattering we can have 
$$
3~\otimes~3~\to ~1~\oplus~3~\oplus~5
\auto\label{col1}
$$
and the $I=2$, symmetric, five dimensional representation also gives a 
two-reggeon cut. For three gluon exchange and higher, helicity conservation
implies that in quark-quark scattering, the odd number 
reggeon exchanges appear in the color triplet channel while the even number 
exchanges appear in the color singlet channel.

For SU(3) color, quark-quark scattering contains three $t$-channel 
representations.
$$ 
3~\otimes~\bar{3}~\to ~1~\oplus~{1 \over 2}(8_a~\oplus~8_s)
\auto\label{col2}
$$
Again (at next-to-leading log) the symmetric singlet gives a two-reggeon cut 
and the antisymmetric octet gives the reggeized gluon. However, as we 
noted in our discussion following Fig.~5.10, in the 
symmetric octet channel the kernel is identical with the reggeizing 
antisymmetric kernel and there is an ``AFS cancellation''. That is the 
two-reggeon cut is replaced by an even signature Regge pole\cite{bw}. The lowest
order amplitude is still (\ref{faq6E}) but in higher-orders it is converted
to the form 
$$
A_2(E,q^2)~\sim ~{J_1(q^2) \over E - g^2 q^2 J_1(q^2)}
\auto\label{8rp}
$$
where $J_1(q^2)$ is given by (\ref{J1}).
Ultimately, this will be very important for our construction of the QCD
Pomeron. It will also be important that, when helicities are not conserved,
the TCP twisting process involves both parity and color charge parity. In
general, helicity conservation 
implies that even-signature combinations of odd-signature and even-signature
reggeons will appear in both the singlet and $8_s$ channels while the odd
signature combinations will appear only in the $8_a$ channel. 

It is clear from (\ref{faq6}) and (\ref{faq8}) that reggeon diagrams
involving the scattering of on-shell quarks are infra-red divergent with the 
divergence arising from the integral over (gluon) transverse momenta. This
divergence is present even when the reggeon state carries zero color.
It is important to understand the origin of this divergence and how it 
relates to gauge-invariance. For this purpose we now discuss the  
``reggeon Ward identities" that, for reggeon amplitudes, are a direct 
requirement of gauge invariance.

\subhead{6.3 Reggeon Ward Identities}

Reggeon amplitudes can be defined directly in terms of 
analytically-continued partial-wave 
amplitudes or by the relevant multi-regge or helicity pole limit. In terms 
of multiparticle partial-wave amplitudes, it is straightforward to write
$$
a_{J_1,J_2,J_3,J_4,J_5,...} ~~\centerunder{$\longrightarrow$}{\raisebox{-3mm} 
{$\scriptstyle J_i \to \alpha_i, i=1,..,4$}}~~ \Pi^4_{i=1}~{\beta_i \over
(j_i-\alpha_i)} ~A_{\alpha_1,\alpha_2,\alpha_3,\alpha_4}(J_5,...)
\auto\label{rea}
$$
and to define $A_{\alpha_1,\alpha_2,\alpha_3,\alpha_4}(j_5,...)$ as a 
multi-reggeon amplitude. (For simplicity we omit the labels $N_i$ 
which give the differences between angular momenta and helicity labels in 
the F-G continuation involved). Multi-reggeon scattering amplitudes can be
defined in momentum space by writing a S-W representation involving the 
remaining $J_i$ or by simply taking a multi-Regge limit in which 
the Regge poles involved are exchanged. As we have illustrated in previous 
Sections, we can define such limits in terms of invariants and also
in terms of light-cone momenta in a particular Lorentz frame. 

To make our general discussion specific, we consider an 
eight-point function, as in Fig.~6.2, and suppose that the multi-regge or 
helicity pole limit considered involves $s_i \to \infty$ i=1,..,4, where
each $s_i$ is associated with a particular reggeon as illustrated.

\parbox{3in}{\begin{center}
\leavevmode
\epsfxsize=2.5in
\epsffile{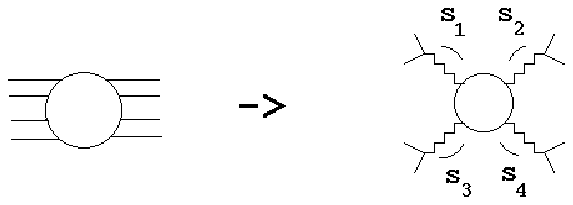} 
\end{center}}  $ \equiv~~\Pi^4_{i=1}
~s_i^{\alpha_i}~A_{\alpha_1,\alpha_2,\alpha_3,\alpha_4}$
\begin{center}
 
Fig.~6.2 A Reggeon Amplitude Extracted from the Eight-Particle Amplitude.

\end{center}
Consider the reggeon associated with $s_1$. We can choose a 
Lorentz frame in which the limit $s_1 \to \infty$ is defined by 
$p_+ \to \infty, k \to \kbar$ where $p$ and $k$ are the momenta labelled 
in Fig.~6.3 and $\kbar$ is the transverse momentum carried by the 
reggeon.

\begin{center}
\leavevmode
\epsfxsize=4in
\epsffile{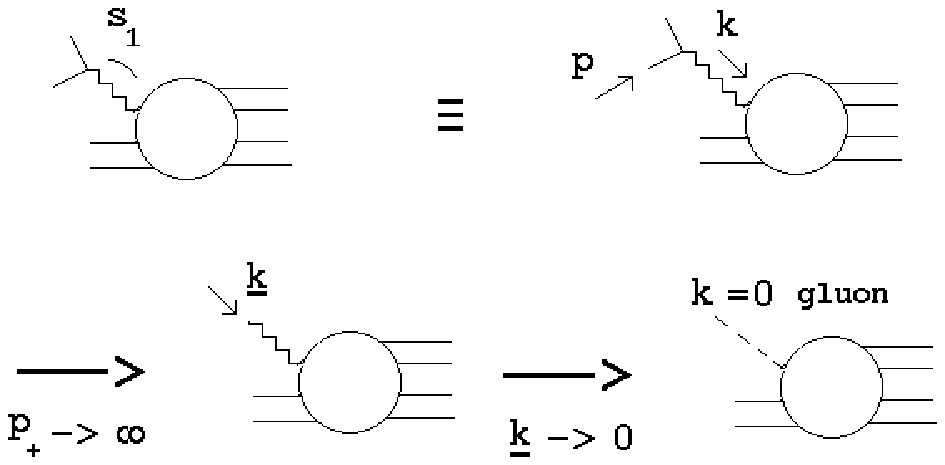}

Fig.~6.3 Reduction of a Reggeon Amplitude to a Gluon Amplitude.

\end{center}
Since the four-momentum $k$ is reduced to a 
transverse momentum $\kbar$ by the Regge limit, the further limit 
$\kbar \to 0$ is equivalent to setting $k=0$. Because of reggeization,
the reggeon amplitude must, as illustrated, give a $k=0$ gluon amplitude. 
Since the reggeon amplitude is embedded in an on-shell 
S-Matrix amplitude, we obtain the zero momentum limit of the amplitude
($\VEV{A_{\mu}(k)~...~}$) for an off-shell gluon to couple to an S-Matrix 
element. 

Gauge invariance implies directly that the gluon amplitude 
$\VEV{A_{\mu}(k)~...~}$ satisfies the simple Ward identity\cite{gth}
$$
k_{\mu}~\VEV{A_{\mu}(k)~...~}~=~0
\auto\label{wd2}
$$
Differentiating this identity (treating each component of $k$ as 
independent) we obtain 
$$
\eqalign{ &\VEV{A_{\mu}~...~}~+~\Biggl[{\partial \VEV{A_{\nu}~...~}\over 
\partial k_{\mu}}\Biggr]_{k=0}~k_{\nu}~=0 \cr
=>~&\VEV{A_{\mu}~...~}~~\centerunder{$\to$} {\raisebox{-5mm} 
{$k_{\mu}\to  0$}}0~~~~~~~if~~ ~\Biggl[{\partial \VEV{A_{\nu}~...~}\over 
\partial k_{\mu}}\Biggr]_{k=0} ~
~\st{\to}~\infty}
\auto\label{wd3}
$$
implying that the gluon amplitude and also (if there is no subtlety with the 
Regge limit) the reggeon amplitude should vanish at zero transverse 
momentum. This is what we refer to as a reggeon Ward identity. By similarly 
defining the additional $s_i \to \infty$ limits as light-cone 
limits, the argument can (a-priori) be extended to an arbitrary number of
reggeon transverse momenta vanishing. 

In general, therefore, (massless) reggeon amplitudes vanish linearly in 
$\kbar$ when any transverse momentum $\kbar \to 0$. This is a
direct consequence of gauge invariance. It is straightforward to check that 
$\Gamma_{22}$ defined by (\ref{6.9}) and (\ref{6.10}) has this property when
$M=0$. However, if the quark-reggeon
couplings discussed above had this property, the infra-red divergences of
(\ref{faq6}) and (\ref{faq8E}) would not occur. So why do the quark-reggeon
couplings not satisfy reggeon Ward identities ?

\subhead{6.4 On-shell Quarks}

In parallel with our discussion of fast quarks above, we consider the
coupling of a gluon to on-shell quarks in the form 
$$ 
\Gamma_{\mu}(p,p')~=~(\gamma\cdot p+m)~\gamma_{\mu}~(\gamma\cdot p'+ m)
\auto\label{ons}
$$
The Ward identity (\ref{wd2}) is easily shown to hold. 
$$
\eqalign{(p-p')_{\mu}\Gamma_{\mu}(p,p`)~&=~(\gamma\cdot p+m)(\gamma\cdot p - m
- \gamma\cdot p + m )({\gamma\cdot p'+m})\cr
&=~(p^2 - m^2)(\gamma\cdot p'+m) ~-~(\gamma\cdot p+m)({p'}^2 - 
m^2)\cr
&=~0 }
\auto\label{ons1}
$$
after applying the on-shell condition for the initial and the final quark.

To compare with the argument of the previous sub-section we should 
evaluate the reggeon coupling $G_1$ by calculating quark-quark scattering
in a frame in which one quark has infinite momentum but the momentum of the
quark we are considering has finite momentum. The fast quark can then be 
identified with the line carrying momentum $p$ in Fig.~6.3 and the finite 
momentum quark vertex identified with the remaining amplitude that satisfies 
the reggeon Ward identity. Therefore, we identify the quark momentum $p$ in
(\ref{ons}) with $p_2$ in (\ref{faq7}) and take
$$
p \equiv p_2 ,~~~ p'\equiv p_2 +k~, ~~~p_2 =( p_{2+},p_{2-},p_{22},0)~,
~~~k=(0,k_-,k_2,k_3) ~.
\auto\label{kin}
$$
The remnant of
the fast quark Regge limit is that $k_- \to 0$.
A-priori, since all the momenta involved are finite, (\ref{wd3}) goes through
straightforwardly. However, since both $p_2$ and $p'$ are on mass-shell 
$$
p_2^2=m^2, ~~~(p_2+k)^2=m^2
 ~~\rightarrow~~~2 p_{2+}k_-= 
2 p_{22} k_2  + k_3^2
\auto\label{ons2}
$$
Therefore, if we keep $p_2$ finite, we can not treat $k_-$ and the 
components of 
$\underline{k}$ as independent variables. In particular, 
$$
\left[{\partial k_3 \over \partial k_-}\right]_{k=0} ~~\sim  ~~~
(k_-)^{-1/2}
\auto\label{ons22}
$$
so that 
$$
\left[{\partial \over \partial k_-} \Gamma_3 \right]_{k = 0}
~\sim ~
{\partial k_3 \over \partial k_-} 
{\partial \over \partial k_3 } \Gamma_3
 ~\sim ~{k_-}^{-1/2}~{\partial \over \partial k_3 } \Gamma_3
\auto\label{ons3}
$$
implying that 
$$
G_{1}~ \equiv ~\Gamma_+ ~\sim ~k_3 {\partial \over \partial k_-} \Gamma_3~
\sim~ {\partial \over \partial k_3 } \Gamma_3 ~~
\centerunder{$\st{\to}$}{\raisebox{-5mm} {$k \to 0$}}~0
\auto\label{ons4}
$$
Since (\ref{ons2}) also requires $k_- \sim k_2$, the transverse component 
$\Gamma_2$ similarly satisfies
$$
\Gamma_2 ~\sim ~k_3 {\partial \over \partial k_2} \Gamma_3~
\sim~ {\partial \over \partial k_3 } \Gamma_3 ~~
\sim ~ \Gamma_+
\auto\label{ons44}
$$
In the gluon Ward identity the contributions of $\Gamma_+ $ and $\Gamma_2$ 
cancel, while the Regge limit picks out just $\Gamma_+$.

Clearly the mass-shell constraint conflicts with the derivation of 
the reggeon Ward identity. Note that, since $G_{N}$ is given by a sequence
of on-shell quark scatterings, this coupling also need not vanish when any,
or all, of the $\underline{k}_i \to 0$. We conclude that the reggeon Ward
identity does not hold for reggeons coupling directly to on-shell quarks  -
even though the related gluon Ward identity implied by gauge invariance
still holds. Conversely, when reggeons couple through off-shell quarks or
gluons, as is in general the case, the reggeon Ward identities follow
directly from gluon Ward identities. (Note that all of the above discussion
goes through straightforwardly when the quark mass $m$ is set to zero.) 

\subhead{6.5  Reggeon Ward Identities in Reggeon Diagrams}

The vanishing of massless reggeon interactions at $\kbar = 0$, due to the
reggeon Ward identities, is crucial for the infra-red properties of reggeon
diagrams when the gluon is massless. As elaborated in \cite{cw} the
infra-red 
finiteness of the BFKL kernel, as well as  next-to-leading order corrections, 
is a direct consequence of this property. Explicit next-to-leading order
calculations have verified\cite{fl} that the reggeon Ward identities hold 
also for the quark production amplitudes that produce next-to-leading order
quark loop interactions in the BFKL kernel. From the above discussion it is
clear, however, that we could expect a violation of the reggeon Ward
identities (but not the gluon Ward identities), if there is an
infra-red divergence within a reggeon interaction due directly to a 
loop of on-shell quarks. The reggeon interaction would then involve 
the on-shell quark couplings discussed above. 

Note that a violation of the reggeon Ward identities can not be produced
by a loop of on-shell (massless) gluons. This is because we can use 
$t$-channel helicities to describe the polarizations of the on-shell gluons. 
Since a reggeon, at zero $\kbar$, is also a $t$-channel gluon it follows from 
helicity conservation that the reggeon can not couple to a pair of on-shell
gluons in the loop. Hence the reggeon must decouple from the gluon loop at 
$\kbar = 0$. Consequently, any divergence due to an internal
loop of on-shell quarks can not be cancelled by an internal gluon loop. Not
surprisingly perhaps, a quark loop divergence occurs only in very special
situations (related to the infra-red triangle anomaly) and is a subtle
phenomenon to isolate. The purpose of the remaining Sections is to establish
that such a phenomenon can indeed occur. 

We can describe how the reggeon Ward identities are normally satisfied
diagrammatically (for quark-loop interactions of the kind that we are
interested in) as follows. It is well known that to obtain the gluon Ward
identity (\ref{wd2}) for a multi-gluon amplitude it is necessary, at the
Feynman diagram level, to sum diagrams 
in which the gluon involved is attached in all possible ways to the remainder 
of the diagram.  This is illustrated for a class of diagrams containing a
quark-loop in Fig.~6.4. (Diagrams of this kind will be of particular
interest to us in the next Section.) 

\begin{center}
\leavevmode
\epsfxsize=5in
\epsffile{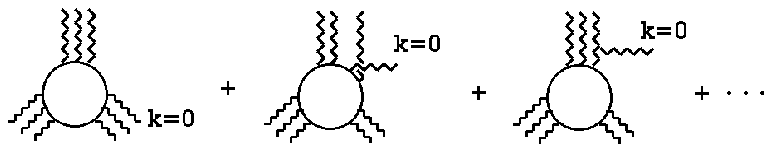}

Fig.~6.4 A Ward Identity Diagram Sum

\end{center}
If some or all of the gluons are replaced by reggeons then, in general, a 
similar sum over all related reggeon/Feynman diagrams gives the reggeon Ward 
identity. The number of diagrams involved is much smaller if we 
generalize the argument we gave above for putting intermediate state 
particles on-shell to obtain particle-reggeon couplings. To obtain a 
multi-reggeon coupling from diagrams such as those of Fig.~6.4,  we 
first consider which hexagraph is involved and then put corresponding quark 
lines on-shell to obtain the relevant multiple discontinuity. We will not 
elaborate the argument for this procedure - which we follow through in more 
detail in the next Section, but note only that it is directly due to the
fact that multiparticle F-G amplitudes are expressed in terms of the
multiple discontinuities of the hexagraph involved. (From the discussion of
Sections 3 and 4, we have seen how multi-regge behavior explicitly reflects
the hexagraph cut structure of amplitudes.) 

As an important example, suppose we replace all the gluons in the first
diagram of Fig.~6.4 by reggeons and embed the diagram in a six-quark
amplitude, as illustrated in Fig.~6.5. 

\begin{center}
\leavevmode
\epsfxsize=2.5in
\epsffile{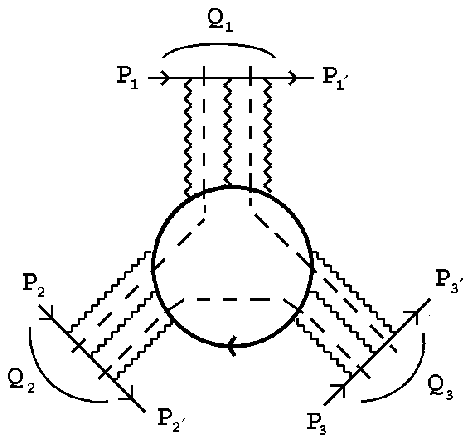}

Fig.~6.5 A Triple Discontinuity.

\end{center}
(We evaluate this diagram explicitly in the next Section.) If we 
associate this diagram with the hexagraph of Fig.~4.4, the cuts shown as 
dashed lines in Fig.~6.5 correspond to the triple discontinuity of Fig.~4.3(b).
Since some quarks remain off-shell, after the triple discontinuity is taken, 
reggeon Ward identities should hold after we sum over all related
diagrams having the same triple discontinuity. The most direct way to show
this is to follow Fadin and Lipatov\cite{fl} and introduce
reggeon/reggeon/gluon effective vertices in addition to the quark-reggeon
couplings we have already introduced. The results of \cite{fl} 
can then be applied to show that, provided the quark loop integration 
introduces no problems, the 
diagrams of Fig.~6.6 combine to give a reggeon Ward identity zero as $\kbar
\to 0$. ($\kbar$ is the transverse momentum carried by a single reggeon.)

\begin{center}
\leavevmode
\epsfxsize=4.5in
\epsffile{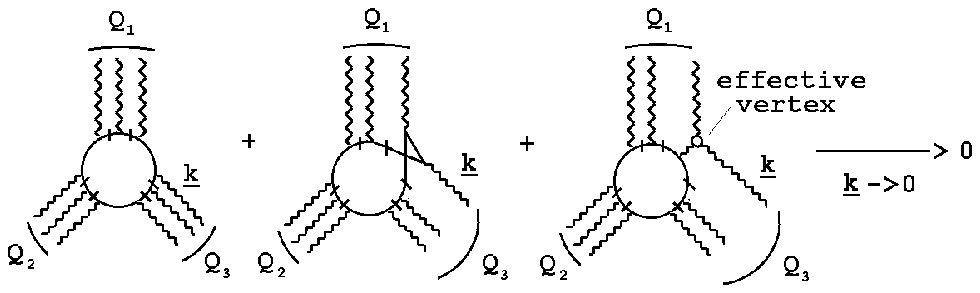}

Fig.~6.6 Quark Loop Couplings Giving a Reggeon Ward Identity. 

\end{center}
In the first two diagrams all reggeons couple directly to the quark loop. In
the third diagram there is a gluon line coupling a two-reggeon/gluon
``effective vertex'' to the quark loop. 

\subhead{6.6 Pauli-Villars Regulator Quarks}

There is a very important difference between the quark-loop reggeon interaction 
vertices appearing in Figs.~6.5 and 6.6 and those appearing in elastic
scattering\cite{fl}. In both cases the process of obtaining reggeon vertices
from Feynman diagrams involves putting quark lines on-shell. However, for a
quark loop contributing as an elastic scattering interaction there 
is always a sufficient number of discontinuities taken through the loop
to effectively reduce the dimension of the loop integration. In contrast, in 
the example of Fig.~6.5 the quark lines can be put on shell by using only the 
longitudinal momentum integrations for the other 
loops involving reggeons/gluons.
Consequently the quark loop remains as a four-dimensional integration. This
feature is associated with the fact that the multi-regge limit of
interest can be defined with the complete quark loop at rest (as we will
explicitly do in the next Section). 

As the quark lines are put on-shell, the ultra-violet convergence of the
quark loop is significantly reduced. In the first two diagrams of Fig.~6.6
there are three quark propagators remaining off-shell while in the third
diagram only two quark propagators remain off-shell. 
Therefore in all three diagrams the quark loops are power divergent with the 
third diagram being particularly badly divergent. Although, higher-order
diagrams may provide additional convergence there is no a-priori reason why 
this should be the case. Because there is no loss of dimension in the loop 
integration, in general we can expect that the reduced quark loops 
(produced by the multi-regge kinematics we discuss) are no more convergent
than the quark loops encountered in the original definition of the theory.
This implies that a regulator is necessary to define these loops. While a
regulator can straightforwardly be applied in the definition of the theory,
we can not do this here. In our case, the need for a regulator implies that
the multi-regge behaviour of the underlying Feynman diagrams is not
correctly given by the reduction to reggeon diagrams that we are implicitly
assuming. 

If the reduction of Feynman diagrams to a reggeon diagram gives infinite
coefficients involving power divergent subdiagrams, then the multi-regge
behaviour of the underlying diagrams must be larger by a power than that of
the reggeon diagram. This phenomenon provides a real threat to the unitarity
boundedness of the theory. (We will return to the significance of this in
our second paper.) As will become clear from our discussion in the next two
Sections, it is the infra-red contribution of the triangle diagram which
will eventually dominate the dynamical picture that we develop. However, we
would like a starting point in which we have both gauge invariance and
a finite reggeon diagram formalism. This requires a definition of the 
contribution of quark loops to reggeon interactions which, when the quarks 
are massive, is finite and satisfies the reggeon Ward identities. To achieve 
this we introduce large mass Pauli-Villars regulator ``quarks'',
in addition to the light quarks that we ultimately take to
be massless. The regulator quark loops have the opposite sign to the 
physical light quark loops. To ensure there are no reggeon diagram 
ultra-violet divergences, the safest procedure is to keep the regulator mass 
finite. In the following we will make only occasional reference to the
regulator quark mass $m_{\lambda}$ which will provide a finite  ultra-violet
cut-off in the quark sector.  It's presence means that, in the quark sector,
the theory is not unitary at this mass-scale. We will ultimately remove
$m_{\lambda}$ after we have extracted infra-red divergences associated with
the massless quarks. 

With the Pauli-Villars
cut-off, the reggeon Ward identities will be satisfied straightforwardly, as
illustrated in Fig.~6.6, as long as the light quark mass is non-zero.  
When the quarks are massless, an infra-red divergence problem arises
which leads to another important difference between the 
diagrams of Fig.~6.6. The three off-shell propagators in the first two 
diagrams will generate a
triangle Landau singularity enhancing zero transverse momentum quark threshold 
singularities. In the first diagram this singularity occurs when 
$$
Q_1, ~~Q_2,~~Q_3 ~~\longrightarrow~~0
\auto\label{tris}
$$
In the third diagram there is no triangle singularity. In the next two
Sections we will see how the presence of the triangle singularity produces a
violation of the reggeon Ward identities when the zero quark mass limit is
taken. We will also see that the presence of the ultra-violet regulator
sector plays an important role in the way the limit is realized. 

\newpage

\mainhead{7. TRIPLE-REGGE HELICITY-FLIP VERTICES}

In this Section we study Feynman and reggeon diagrams of the kind discussed
at the end of the last Section, all of which involve a quark loop. We will
study such diagrams in the variety of triple-Regge limits 
discussed in Section 4. Our aim is to extract parts of the helicity-flip
reggeon vertices $T^{{\cal F}}_{m'n'r'}$ discussed in Section 5 which have
special (singular) zero quark mass properties with respect to the reggeon
Ward identities. As anticipated in the last Section, we initially consider
particular Feynman diagram contributions involving on-shell quarks and then
deduce the structure of corresponding reggeon couplings. We will build up 
to diagrams with the complexity of Fig.~6.5. We begin, however, with the
diagram of Fig.~7.1 involving single gluon exchange. 

\begin{center}
\leavevmode
\epsfxsize=2in
\epsffile{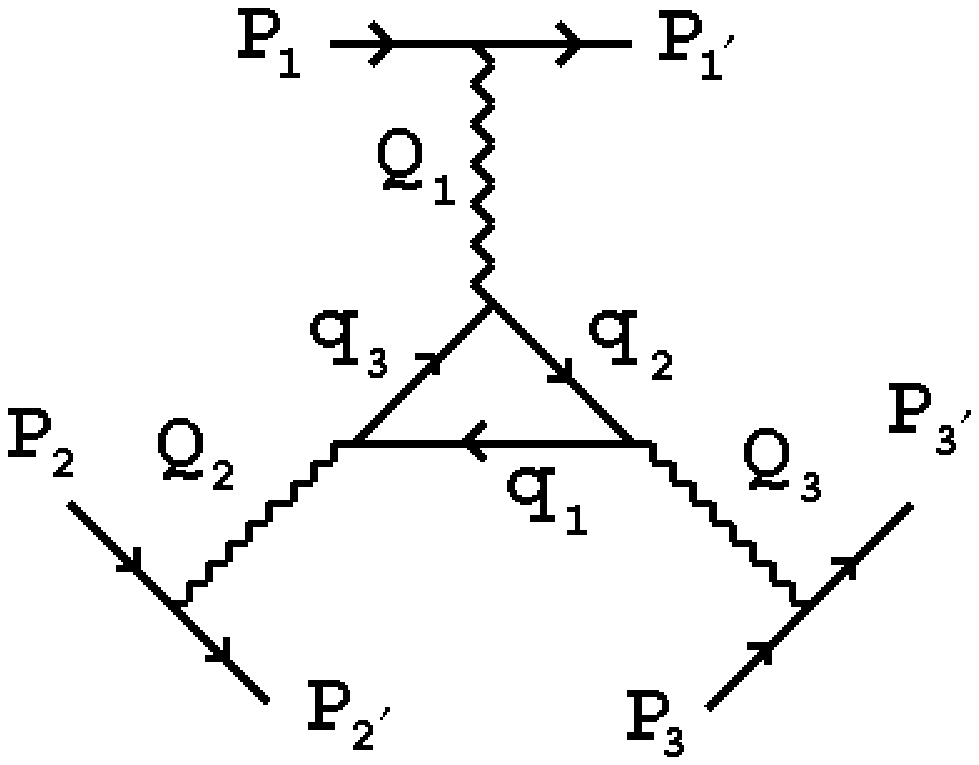}

Fig.~7.1 A triple gluon vertex. 

\end{center}

\subhead{7.1 Feynman Diagram Limits}

Consider the behavior of Fig.~7.1 
in the limits defined in sub-Section 4.2. Since each limit is 
defined in terms of fast external quarks we can simply apply
(\ref{faq}) to these quarks and leave the quark loop to be evaluated
at finite momentum. Initially we omit color factors and take the quark mass
$m \neq 0$. In this case the quark loop gives (apart from a normalization 
factor) the usual vertex function 
$$
\Gamma_{\mu_1 \mu_2 \mu_3}(q_1,q_2,q_3,m) = i\int {  d^4 k~ Tr \{ \gamma_{\mu_1}
(\st{q}_3 + \st{k} + m ) \gamma_{\mu_2} (\st{q_1} + \st{k} + m ) 
\gamma_{\mu_3} (\st{q}_2 + \st{k} + m) \} 
\over [ (q_1 + k)^2 - m^2 ][ (q_2 + k)^2 - m^2 ]
[ (q_3 + k)^2 - m^2 ]}
\auto\label{gam}
$$
(Since we implicitly consider Pauli-Villars regulator quarks to be present
as we discussed in the last Section, we ignore ultra-violet divergence
problems.) Denoting the full amplitude corresponding to Fig.~7.1 by
$T^{111}$ and using (\ref{faq}) we obtain the analogous result to
(\ref{faq4}) for the limit $L_1$, i.e. 
$$ 
T^{111}~\to~T^{111}_{L_1}~\sim  
~ g^6~ { p_1p_2p_3 \over t_1 t_2 t_3 } ~\Gamma_{1^+2^+3^+}(q_1,q_2,q_3)
\auto\label{tl1}
$$ 
where $t_1 = Q_1^2$ etc., and $\Gamma_{1^+2^+3^+}$ is defined by 
$\gamma_{i^+} = \gamma_0 + \gamma_i,~i=1,2,3$. In this Section, for 
simplicity, we continue to omit the gluon mass. For the limit $L_2$
we similarly obtain 
$$
T^{111}~\to~T^{111}_{L_2}~\sim  
~ g^6~ { p_1p_2 p_3 \over t_1 t_2 t_3 } ~\Gamma_{1^+1^-3^+}(q_2',q_2,q_3)
\auto\label{tl2}
$$ 
and for the limit $L_3$
$$
T^{111}~\to~T^{111}_{L_2}~\sim  ~ g^6 ~ { p_3 \over t_1 t_2 t_3 } 
~\Gamma_{1^+1^+3^+}(q_2',q_2,q_3)~.
\auto\label{tl3}
$$ 

Our further discussion of infra-red divergences and reggeon Ward identities 
in the next Section will center on that part of the vertex functions
(\ref{tl1})~-~(\ref{tl3}) that behaves non-uniformly with respect to the two
further limits 
$$
i)~~ q_1 \sim q_2 \sim q_3 \sim Q \to 0 ~~~~~~~ii)~ ~m \to 0 
\auto\label{tl4}
$$
We will be studying effects that are closely related to the infra-red triangle 
anomaly\cite{cg}. At first sight it might seem that we should not encounter
such behavior. Firstly, (\ref{tl1}) - (\ref{tl3}) involve 
$\Gamma_{\mu_1 \mu_2 \mu_3}$ evaluated with (transverse) momenta orthogonal
to the appropriate (light-cone) Lorentz indices. Therefore 
(\ref{tl1}) - (\ref{tl3}) do not contribute to the divergence of the 
triangle graph in which the anomaly resides. However, as we discussed in 
sub-section 6.4, if on-shell quarks are involved, transverse and 
longitudinal components of vertex functions are linked by the underlying 
gluon Ward identities, even though the Regge limit picks out just the 
longitudinal component. Consequently if the transverse component
contains an infra-red divergence of the triangle graph, associated with the
anomaly, in which the quarks are placed on-shell, this will also appear in the 
longitudinal component. Even so, since only vector (rather than axial vector)
couplings appear in $\Gamma_{\mu_1 \mu_2 \mu_3}$, we again would not expect the 
anomaly to appear. In fact, as we build up multi-reggeon interactions in the
following, we will consider (originally) non-local couplings to the triangle
graph that are ``axial-vector like'' in the Regge limit, infra-red, region
of interest. 

The most singular behavior in the 
the limit $i)$ involves all three denominator poles and 
the minimum internal momentum dependence from the 
numerator. Since the trace of an odd number of $\gamma$-matrices 
vanishes, the only $m$ dependence of the numerator of 
$\Gamma_{\mu_1 \mu_2 \mu_3}$ comes from the terms containing a factor of 
$m^2$. Denoting this ``$m^2$ part'' by ${\Gamma}_{\mu_1 \mu_2 \mu_3,m^2}$ we 
have  
$$
\eqalign{{\Gamma}_{\mu_1 \mu_2 \mu_3,m^2}~ 
{\centerunder{$\longrightarrow$}{\raisebox{-4mm} 
{$Q \to 0$} }}&~ i~ m^2
 \int { d^4 k \over [ k^2 - m^2 ]^3 ~+~ O(Q)} \cr
& \times { Tr \{ \gamma_{\mu_1}
(\st{q}_3 + \st{k}) \gamma_{\mu_2} \gamma_{\mu_3} + \gamma_{\mu_1}
\gamma_{\mu_2} (\st{q}_1 + \st{k}) \gamma_{\mu_3} + \gamma_{\mu_1}
\gamma_{\mu_2} \gamma_{\mu_3} (\st{q}_2 + \st{k}) } 
}
\auto\label{tl0}
$$
In the leading term, the numerator terms that are odd in $k$ vanish after
integration and so 
$$
\eqalign{ {\Gamma}_{\mu_1 \mu_2 \mu_3,m^2}~ &
{\centerunder{$\longrightarrow$}{\raisebox{-4mm} 
{$Q \to 0$} }}~ {\Gamma}_{\mu_1 \mu_2 \mu_3,0}(q_1,q_2,q_3) \cr
&~~~\equiv ~ i~ R ~ Tr \{ \gamma_{\mu_1}
\st{q}_3  \gamma_{\mu_2} \gamma_{\mu_3} + \gamma_{\mu_1}
\gamma_{\mu_2} \st{q}_1\gamma_{\mu_3} + \gamma_{\mu_1}
\gamma_{\mu_2} \gamma_{\mu_3} \st{q}_2 \} }
\auto\label{tl5}
$$
where 
$$
R~=~ m^2 ~\int {d^4 k \over [ k^2 - m^2 ]^3 } ~~
=  ~\int {d^4 y \over [ y^2 - 1 ]^3 } 
\auto\label{tl51}
$$
Clearly
$$
{\Gamma}_{\mu_1 \mu_2 \mu_3,0}(q_1,q_2,q_3) 
{\centerunder{$\sim$}{\raisebox{-4mm} 
{$Q \to 0$} }} ~~Q
\auto\label{tl52}
$$
If we reverse the order of the limits $i)$ and $ii)$ we obtain, instead of
(\ref{tl5}), 
$$ 
{\Gamma}_{\mu_1 \mu_2 \mu_3,m^2}~ 
\centerunder{$\sim $}{\raisebox{-6mm} 
{$m \to 0$} }~m^2~~\longrightarrow ~0
\auto\label{tl6}
$$

If we consider $T^{111}$ as an isolated Feynman diagram, defined directly in 
the massless theory, (\ref{tl5}) will not be present. However, we will 
shortly consider reggeon interactions containing 
${\Gamma}_{\mu_1 \mu_2 \mu_3,m^2}$. In the next Section we will see
that the non-uniformity of (\ref{tl5}) and (\ref{tl6}) implies that if the 
reggeon Ward identities are satisfied for $m \neq 0$, then 
${\Gamma}_{\mu_1 \mu_2 \mu_3,0}$ is present in these interactions when
$m = 0$. Note that the presence of $m^2$ in the numerator of
${\Gamma}_{\mu_1 \mu_2 \mu_3,m^2}$ indicates two
helicity flips of the quarks in the loop. That the helicity-flip processes
do not decouple in the limiting process, where the limit i) is taken before
the limit ii), is clearly a consequence of the triangle singularity
infra-red divergence produced as all three internal quark propagators go
on-shell. The presence of this divergence is therefore crucial. 

Consider now the contribution of (\ref{tl5}) to (\ref{tl1})-(\ref{tl3}).
In (\ref{tl1}) we will have a contribution 
$$ 
T^{111}_{L_1,0}~\sim  
~ g^6~ { p_1p_2p_3 \over t_1 t_2 t_3 } ~\Gamma_{1^+2^+3^+,0}(q_1,q_2,q_3)
\auto\label{tl7}
$$
where
$$
\eqalign{ \Gamma_{1^+2^+3^+,0}
&\sim ~(Tr \{ \gamma_{1^+}\gamma_{2^+}\gamma_1\gamma_{3^+}\}q_1~ 
+~ Tr\{\gamma_{1^+}\gamma_{2^+}\gamma_{3^+}\gamma_2 \}q_2 ~+~
Tr\{ \gamma_{1^+}\gamma_3\gamma_{2^+}\gamma_{3^+}\} q_3) \cr
&\sim ~(q_1 + q_2 + q_3) }
\auto\label{tl8}
$$
Similarly, in (\ref{tl2}) we will have a contribution 
$$
\eqalign{ T^{111}_{L_2,0}~&\sim { p_1p_2p_3 \over t_1 t_2 t_3 } 
 ~\Gamma_{1^+1^-3^+,0}(q_2',q_2,q_3)\cr 
 &\sim ~{ p_1p_2p_3 \over t_1 t_2 t_3 } 
~ (Tr \{ \gamma_{1^+}\gamma_{1^-}\gamma_2\gamma_{3^+}\}q_2' ~+~
(Tr \{ \gamma_{1^+}\gamma_{1^-}\gamma_{3^+}\gamma_2\}q_2 ~+~
Tr\{ \gamma_{1^+}\gamma_3\gamma_{1^-}\gamma_{3^+}\}q_3 ) \cr
&\sim ~{ p_1p_2p_3 \over t_1 t_2 t_3 } ~ q_3  }
\auto\label{tl9}
$$
and in (\ref{tl3}) we will have 
$$
\eqalign{ T^{111}_{L_3,0}~&\sim  \Gamma_{1^+1^+3^+,0}(q_2',q_2,q_3)\cr
 &\sim  ~ 
Tr \{ \gamma_{1^+}\gamma_{1^+}\gamma_2\gamma_{3^+} \}q_2' ~+~
Tr \{ \gamma_{1^+}\gamma_{1^+}\gamma_{3^+}\gamma_2 \}q_2 ~+~
Tr \{ \gamma_{1^+}\gamma_3\gamma_{1^+}\gamma_{3^+}\} q_3  \cr
&= ~0
}
\auto\label{tl10}
$$
We conclude from (\ref{tl7}) that when the additional limits (\ref{tl4}) are 
taken after the triple-Regge limit, there is a non-zero contribution of 
the helicity flip process (\ref{tl5}). There are
three terms. (\ref{tl9}) suggests that just one of the three terms
appearing in the triple-Regge limit appears in the helicity-flip
helicity-pole limit. (We will see shortly that this is the case. It
can not be straightforwardly deduced from (\ref{tl7}) and (\ref{tl9}) since 
we redefined the momentum components of the
$Q_i$ in going from one limit to the other.) The result of (\ref{tl10})
shows that there is no contribution of the helicity flip process (\ref{tl5}) 
in the simple non-flip helicity-pole limit. 

Next we consider some higher-order Feynman diagrams in order to determine 
how they contribute to 
higher-order reggeon couplings. Suppose first that we replace one or more 
of the gluons in Fig.~7.1 by two gluon exchange, as illustrated in Fig.~7.2. 

\begin{center}
\leavevmode
\epsfxsize=3.5in
\epsffile{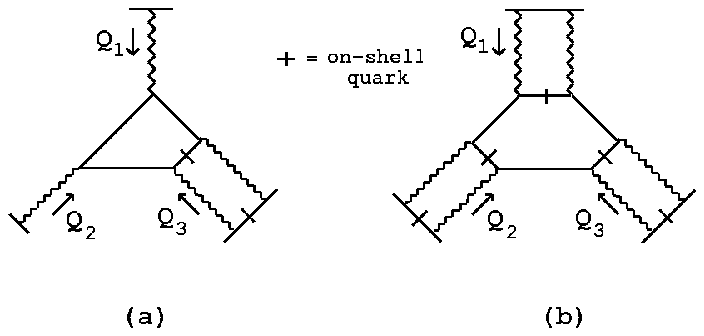}

Fig.~7.2 Triple Couplings Involving Two Gluon Exchange.

\end{center}
We again evaluate the quark loop at finite momentum. 
Our interest in two gluon states is to extract two-reggeon (Regge cut) 
couplings and so we calculate
the diagram with on-shell intermediate states as illustrated. (We can
justify this by evaluating the appropriate multiple discontinuity to 
calculate the relevant F-G amplitude or we can simply suppose that 
we have carried out the related longitudinal
integrations). Denoting now the full amplitude for Fig.~7.2(a) by 
$T^{112}$ we obtain for the limit $L_1$, in analogy with (\ref{faq6}), 
$$ 
T^{112}~\to~T^{112}_{L_1}~\sim  
~ i~ g^8~ {p_1p_2p_3 \over t_1 t_2} ~J_1(t_3)
~\Gamma_{1^+2^+3^+}(q_1,q_2,q_3) 
\auto\label{tl20}
$$
and so again, after the further limits (\ref{tl4}) are taken, 
there is a contribution of the form (\ref{tl5}), i.e.
$$
T^{112}_{L_i,0} ~ \sim~ -i~t_3 J_1(t_3)~ T^{111}_{L_i,0}
~~~~i=1,2,3
\auto\label{tl11}
$$
Denoting the full amplitude for Fig.~7.2(b) by $T^{222}$ we similarly obtain
$$ 
T^{222}~\to~T^{222}_{L_1}~\sim 
~ (i)^3 ~ g^{12}~ p_1p_2p_3~ J_1(t_1) J_1(t_2) J_1(t_3)
~\Gamma_{1^+2^+3^+}(q_1,q_2,q_3) 
\auto\label{tl12}
$$
and so
$$
T^{222}_{L_i,0} ~ \sim~ -i~t_1 t_2 t_3 J_1(t_1)J_1(t_2)J_1(t_3)
~ T^{111}_{L_i,0}
~~~~i=1,2,3
\auto\label{tl13}
$$

We can continue adding gluons (as we did for single quark couplings 
in the last Section) and obtain correponding results. The diagram of Fig.~7.3 
contains the triple coupling of three gluon states which we anticipate will
give the first multi-reggeon coupling appearing in Fig.~6.6. This coupling
will be very important in the following. 

\begin{center}
\leavevmode
\epsfxsize=2in
\epsffile{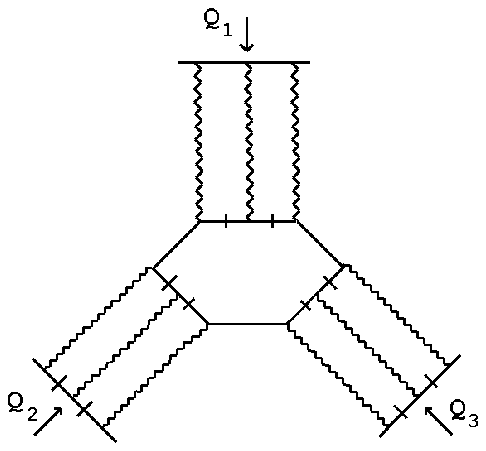}

Fig.~7.3 A Triple Vertex for Three Gluons. 

\end{center}
In this case we obtain, as above, for the limit $L_1$
$$ 
T^{333}~\to~T^{333}_{L_1}~\sim  
~ i^6~ g^{16}~ p_1p_2p_3 ~J_2(t_1)J_2(t_2)J_2(t_3)
~\Gamma_{1^+2^+3^+}(q_1,q_2,q_3)
\auto\label{tl14}
$$ 
where (continuing to omit normalization factors)
$$
J_2(q^2)~=~~\int {d^2 k \over (k-q)^2 }J_1(k^2)
\auto\label{J2}
$$
and in all the $L_i$ limits 
$$
T^{333}_{L_i,0} ~ \sim~ -~t_1 t_2 t_3 J_2(t_1)J_2(t_2)J_2(t_3)
~ T^{111}_{L_i,0}
~~~~i=1,2,3
\auto\label{tl15}
$$

Note that in extracting the $T^{333}_{L_i,0}$ from the diagram of Fig.~7.3, 
we have put on-shell (the denominators of) all those quark propagators that
we had not already put on-shell in converting the multi-gluon coupling to a
multi-reggeon coupling. Therefore the $T^{333}_{L_i,0}$ couplings actually
involve a loop of on-shell quark propagators and so, from the discussion of
the last Section, might be anticipated to be associated with a violation of
the reggeon Ward identities. To establish that such contributions actually
appear in multi-reggeon couplings we must first consider the color factors
involved. 

\subhead{7.2 Color Factors}

In this sub-section we discuss the color factors that should be added to the 
diagrams considered in the last sub-section. We use the tensor notation 
introduced in Fig.~5.9.
The quark relations shown in Fig.~7.4 are then
sufficient to evaluate the color factors for any number of gluons 
coupling to a single quark loop.

\begin{center}
\leavevmode
\epsfxsize=4.5in
\epsffile{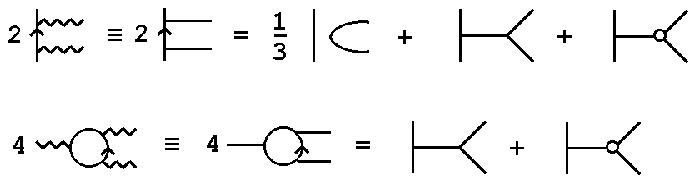}

Fig.~7.4 Color Factors For Quark-Gluon Couplings. 

\end{center}
We can form multi-gluon (multi-reggeon) states with color $1$, $8_a$ and $8_s$
by combining $\delta$,$~f$, and $d$-tensors appropriately with gluon fields.
The color parity of such a state will then be given by a product of factors
of (-1) for each gluon field and (-1) for each $f$-tensor. 

From the second relation of Fig.~7.4, the quark loop in Fig.~7.1 gives a
color factor proportional to 
$$ 
   d_{i_1i_2i_3} ~+~ i~ f_{i_1i_2i_3} 
\auto\label{cf}
$$
where $i_1$ is the color label for the gluon carrying momentum $Q_1$ etc. 
Consider next the addition of the diagram of Fig.~7.5 (which is the only 
other topologically distinct quark loop three gluon interaction).

\begin{center}

\leavevmode
\epsfxsize=4.5in
\epsffile{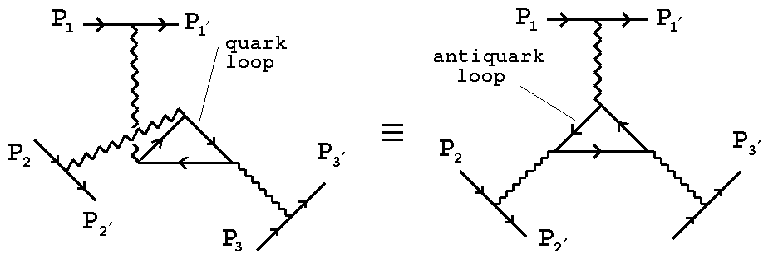}

Fig.~7.5 An Additional Quark Loop Interaction.

\end{center}
The diagram of Fig.~7.5 differs from that of Fig.~7.1 by permutation of 
the color matrices which (within the trace) is the same as reversal of the 
direction for multiplication. The result is complex conjugation of the color 
factor. Since the sign of the $q_i$ is also reversed Fig.~7.5 can, as 
illustrated, be obtained directly from Fig.~7.1 by replacing the quark loop
by an anti-quark loop. For the $O(m^2)$ part with which we are 
concerned, the two diagrams combine to give 
$$
\eqalign{ T^{111}_{L_1,0}~ &\sim ~g^6 {p_1p_2p_3 \over t_1t_2t_3}~ 
\biggl[ ( d_{i_1i_2i_3} + if_{i_1i_2i_3} )  
(q_1 + q_2 + q_3)~-~ 
( d_{i_1i_2i_3}- i f_{i_1i_2i_3} )(q_1 + q_2 + q_3) \biggr]\cr
&= ~2g^6 ~ i~f_{i_1i_2i_3}~{p_1p_2p_3 \over t_1t_2t_3}~ (q_1 + q_2 + q_3) 
}
\auto\label{cf1}
$$

The color factors for the diagrams of Fig.~7.2 are, of course, more 
complicated. The two gluons can 
form states with $t$-channel color $1$, $8_a$ and $8_s$. 
For Fig.~7.2(a) the color factor contains each of the color tensors 
illustrated in Fig.~7.6. 
To extract the full discontinuity giving the Regge cut coupling we must also 
add the contribution obtained by replacing the quark loop of Fig.~7.2(a)
with an antiquark loop. 
The factor of $i$ associated with the on-shell 
quark now also changes sign. As a result only the real part of the color factor
remains, i.e. the first three diagrams in Fig.~7.6, which contain an even
number of $f$-tensors. 
\begin{center}
\leavevmode
\epsfxsize=4.5in
\epsffile{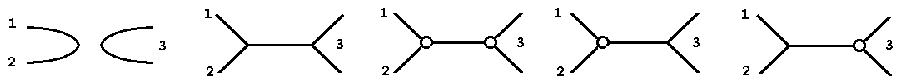}

Fig.~7.6 Color Factors For Fig.~7.2(a)

\end{center}
These color factors describe, successively, the
coupling of $1$, $8_a$ and $8s$ two-gluon states to the two single gluons.

Moving on to Fig.~7.2(b) we again add the corresponding diagram
with an antiquark loop and, because of the factor of $(i)^3$ for each 
on-shell antiquark, select the color 
diagrams containing an even 
number of $f$-tensors, i.e. the diagrams shown in Fig.~7.7. 

\begin{center}

\leavevmode
\epsfxsize=4.5in
\epsffile{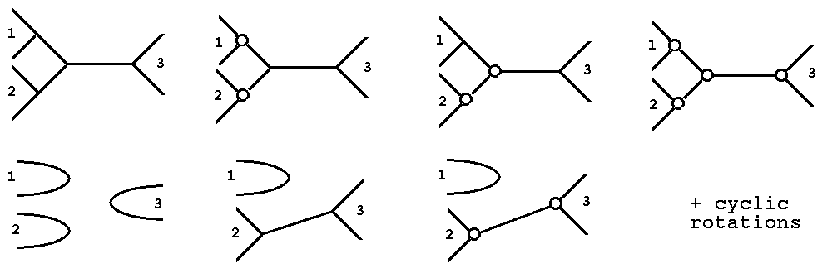}

Fig.~7.7 Color Factors for the Six Gluon Vertex. 

\end{center}
The first diagram in Fig.~7.7 gives an antisymmetric coupling of three
two-reggeon states, each carrying odd color parity. The second gives an
antisymmetric coupling of two even color parity states and one odd color parity
state, and so on. 

Finally we consider color factors for the triple coupling of 
three-gluon states shown in Fig.~7.3. Now we have an even number of factors 
of $i$ from on-shell quarks and so color diagrams with an odd
number of $f$-vertices survive when we add the antiquark loop. Three particular
color factors that we will be interested in are those of Fig.~7.8.

\begin{center}

\leavevmode
\epsfxsize=4in
\epsffile{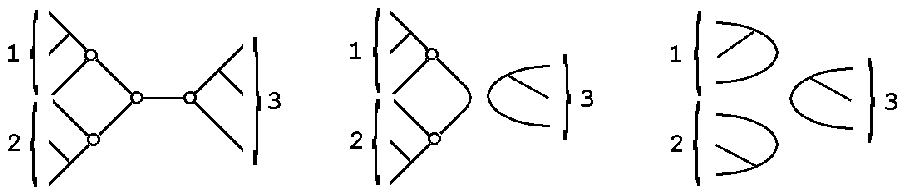}

Fig.~7.8 Color Factors for the Triple Coupling of Three Gluon States. 

\end{center}
These are couplings which contain an odd number of $f$-vertices but 
provide a symmetric triple coupling of three-gluon states which each carry 
even color charge parity.

\subhead{7.3 Reggeon Interaction Vertices - Kinematic structure}

We discuss now the implications of the results of the last two
sub-sections for reggeon interaction vertices. 
First we consider how the structure of the $T_{L_i,0}$ that we have
discussed relates to the general triple-regge analysis of Section 4. 

We can rewrite the above formulae in terms of invariants 
either by writing, for example,  
$$
\eqalign{p_1p_2p_3 q_1 ~&=~~
 (p_1p_3)(p_2p_3) (p_3q_3)^{-1} q_3^2 \cr
&\equiv ~~(s_{23})( s_{31})(s_{11'3})^{-1}~ q_3^2}
\auto\label{Linv}
$$
or we can instead write 
$$
\eqalign{ p_1p_2p_3 q_3 ~&=~~
 (p_1p_3)^{1/2}(p_2p_3)^{1/2} (p_1p_2)^{1/2} q_3 \cr
&\equiv ~~(s_{31})^{1/2}( s_{23})^{1/2}(s_{12})^{1/2}~ q_3 }
\auto\label{Linv1}
$$
It then remains to express $q_3$ directly in terms of invariants. For the 
special kinematics of the triple-Regge limit $L_1$ this is particularly
simple, i.e. we can write 
$$
q_3 = [Q_1.Q_2 ]^{1/2} ~.
\auto\label{L1q3}
$$ 

Comparing with (\ref{hp2i}) and (\ref{hp2i1}) we recognize (\ref{Linv}) and
(\ref{Linv1}) as having the form appropriate for a triple-Regge 
helicity-flip amplitude with 
$\alpha_1=\alpha_2=\alpha_3=1$. The two expressions, (\ref{Linv}) and
(\ref{Linv1}) correspond to the lowest-order contribution from the cuts of
Fig.~4.3 (a) and (b) if we suppose that the $\alpha_i$ can be expanded 
perturbatively around unity (as is the case for the trajectory of the 
reggeized gluon). Therefore we can potentially associate the $q_3$ term in
$T_{L_1,0}$ (see (\ref{tl3})) with the first of the three hexagraphs in
Fig.~4.2. Similarly the other two $q_i$ terms could be associated with the
other two hexagraphs. Of course, the Feynman diagram of Fig.~7.1 has no 
cuts. The cuts appear only as the gluons reggeize in higher-orders. The
higher-order loop diagrams of Figs.~7.2 and 7.3 do have cuts, and in particular
the diagram of Fig.~7.3 (with the quark lines initially uncut) clearly has
all the relevant cuts necessary to contribute to the helicity-flip limit.
(See the discussion in sub-section 4.4.) 
If the cuts through this diagram generate Regge cut couplings, as we are
anticipating, then we can directly associate the three $q_i$ terms 
in (\ref{tl14}) with the three hexagraphs of Fig.~4.2. 

In the $L_2$ limit we can proceed similarly and again use (\ref{Linv}) and 
(\ref{Linv1}) to argue that $T_{L_2,0}$ can be associated 
with the first of the three hexagraphs in Fig.~4.2. So, 
just as the general arguments imply, each of three terms appearing in the 
triple-Regge limit is separately picked out by the corresponding
helicity-flip limit. Again higher-order contributions can produce 
reggeization of the gluons and convert (\ref{Linv}) and (\ref{Linv1}) to the 
form (\ref{hp2i}) and (\ref{hp2i1}) respectively. 

In the $L_2$ limit $q_3$ has a slightly more complicated expression in terms 
of invariants, i.e.
$$
q_3~=~ \pm ~{[-\lambda(Q_1^2,Q_2^2,Q_3^2)]^{1/2} \over
2 [Q_3^2]^{1/2} }
\auto\label{q3}
$$
(We will discuss shortly the significance of the choice of sign in 
(\ref{q3}).) Note that (\ref{q3}) satisfies reggeon Ward identities in that it
vanishes linearly when either $Q_1 \to 0$ or when $Q_2 \to 0$. 
When the $Q_i$ are spacelike,   
$[- \lambda(Q_1^2,Q_2^2,Q_3^2)]^{1/2}$ is the area of the triangle formed by 
the three momenta and so it vanishes when any one of them 
vanishes. The denominator spoils the vanishing for $Q_3 \to 0$. In fact, if
$q_3 \neq 0$, the numerator of the corresponding quark propagator in 
(\ref{tl5}) is off-shell . This is what allows two of the reggeon Ward
identities to hold. In contrast, Both of the quark propagators which form
the $Q_3$ channel are strictly on-shell and so the $Q_3 \to 0 $ limit gives
the on-shell result. In anticipation of the next Section we note that if all
three reggeon Ward identities hold, we expect the vertex to have dimension
two in the $Q_i$ (as would be obtained, for example, by simply removing the
denominator in (\ref{q3})). 

\subhead{7.4 Reggeon Interaction Vertices - Signature}

The discussion of the previous subsection shows that, kinematically, each of
the $T_{L_2,0}$ that we have considered could appear in the corresponding
lowest-order multi-reggeon helicity-flip amplitude. However, we have not 
yet discussed color parity and signature. As we noted in Section 4,
signature is defined via a TCP twist that combines color (charge) parity and
space parity. Since we are discussing helicity-flip amplitudes, we 
expect that space parity plays a non-trivial role. The helicity-flip is
reflected in the presence of the $q_i$-factors and indeed the sign of
$q_3$, as given by (\ref{q3}), is changed under the parity transformation 
associated with signature. This change takes place for each of the three 
$t_i$-channels. 
 
Consider first $T^{111}_{L_2,0}$, with the color factor (\ref{cf1}) 
included. The denominator 
factors of $t_i$ are, of course, the usual gluon particle poles. We use 
(\ref{Linv1}) to extract the (potential) helicity-flip reggeon vertex
$$
 T_{111}^{ {\cal F},0}~=~i~f_{i_1i_2i_3} ~q_3 
\auto\label{tm}
$$ 
with $q_3$ given by (\ref{q3}). We keep the $0$ superscript to indicate both
that this is a particular contribution to the general vertex and that it is
defined at zero quark mass. Note that since (\ref{Linv1}) expresses the
triple-Regge behavior in terms of invariants that have no kinematic
singularities in the $Q_i~,$ it defines the appropriate vertex to extract if 
we wish to consider singular behavior as the $Q_i \to 0$. 

We introduce signature in the $t_i$-channel by making a TCP 
transformation of the corresponding initial and final scattering states 
together with the vertex involved. For the $t_3$-channel, therefore, we regard 
reggeon 1 as scattering into reggeon 2 by exchanging reggeon 3. 
Interchanging 1 and 2 gives a factor  
of -1 from the color parity of the $f$-tensor and a further factor
of -1 from the parity change of sign of $q_3$. 
Consequently, reggeon 3 should be even rather than odd signature if
$T_{111}^{ {\cal F},0}$ is to appear in the vertex. We conclude that
$T_{111}^{ {\cal F},0}$ does not contribute to the triple reggeon vertex.
Equivalently, when we sum over all the diagrams for quark and antiquark 
scattering necessary to define signatured amplitudes, the pieces we have
extracted are canceled. The combination of external quark and antiquark 
vertices requires odd signature for the (reggeized) gluons to couple while 
the central vertex requires even signature. 

Consider next $T_{112}^{ {\cal F},0}$ with the color factor given by the first
three diagrams of Fig.~7.6. To give a reggeon coupling the factor of $J_1(t_3)$ 
must be converted to a two reggeon propagator in higher-orders (or, for the 
$8_s$ Regge pole discussed in the last Section, $J_1(t_3)$ 
must contribute to reggeization).
In the first color diagram, there is no color factor and so the change of 
sign of the momentum factor is in direct conflict with the required even 
signature of the two-reggeon state. In the second diagram, one $f$-tensor 
forms an $8_a$ two-reggeon state which then couples to the two single
reggeons via a vertex of the same form as (\ref{tm}). 
In this case the color and momentum factors do combine to give even signature 
in the $t_3$-channel. In the $t_1$ (and $t_2$) channel, the situation is 
more complicated. Because the reggeon states in the $t_2$ and $t_3$ channels 
are distinct, there is no simple parity property for their interchange.
However, in the next Section we will be interested in the situation in which 
all reggeons in a reggeon state carry zero transverse momentum and the state 
itself produces a universal canonical transverse momemtum dependence. In 
this case we need not distinguish between distinct reggeon states when 
interchanging them to obtain signatured couplings. Consequently in 
discussing the signature effects of color factors we only need consider the 
symmetry of the color tensor in the reggeon vertex and not the tensors 
involved in forming the reggeon states. In particular in the $t_1$ and 
$t_2$ channels we only need consider the symmetry of the tensor in
the vertex (\ref{tm}). Combined with the negative parity of $q_3$ this gives 
even signature for the $t_1$ and $t_2$
channels, where odd signature is required. So again there is no vertex.
The third diagram of Fig.~7.6 replaces the $f$-tensor of (\ref{tm}) with a 
$d$-tensor and so
gives odd signature for the $t_3$-channel where even signature is required.

\subhead{7.5 Reggeon Interaction Vertices - Anomalous Reggeon States and the 
Anomalous Odderon } 

Now consider $T_{222}^{ {\cal F},0}$ with the color factors given by the
diagrams of Fig.~7.7. In this case the vertex color factor has to provide a
change of sign to compensate for the change of sign of the momentum factor
in order to to give even signature in each of the channels. The first
diagram of Fig.~7.7, which exists in both SU(2) and SU(3), achieves this by
coupling three $8_a$ two-reggeon states with a vertex of the form
(\ref{tm}). Therefore $T_{222}^{ {\cal F},0}$ can appear 
in a triple coupling of two-reggeon states that have ``anomalous
color parity'', i.e. the color parity is not equal to the signature. 
Normally (i.e. in next-to-leading log perturbation theory) because of 
helicity-conservation there is no $8_a$ two-reggeon state. As we noted in 
sub-section 6.2, the two-reggeon state has color parity +1 and signature +1. 
We refer to states with  anomalous color parity as ``anomalous reggeon'' 
states. Such reggeon states will  will not appear when
quarks scatter with their helicity conserved (as is the case for the
leading-order perturbative couplings discussed in the last Section and 
must be the case to all orders when the quarks are massless.) However, 
these states will couple between $T_{222}^{ {\cal F},0}$ vertices. 

The second diagram of Fig.~7.7 also provides an interesting coupling. It 
does not exist in SU(2), but in SU(3) it
gives a $T_{222}^{ {\cal F},0}$ vertex of the form (\ref{tm}) that couples
an anomalous $8_a$ state to two even signature $8_s$ states that are not
anomalous. As we discussed 
in subsections 5.3 and 6.2, the $8_s$ even signature channel contains a 
bound-state reggeon that is exchange-degenerate with the reggeized gluon. 
If we denote an anomalous reggeon state by 
``A'' and a normal reggeon state by ``N'', the first two diagrams of 
Fig.~7.7 respectively produce 
\newline ~
\newline \centerline{ ``~A~A~A~''~~~~and~~~~``~A~N~N~''}
\newline ~
\newline couplings. This is analagous to the well-known ``$~AAA + AVV~$''
structure of the triangle anomaly, where A denotes an axial vector coupling
and V denotes a normal vector coupling. All the remaining diagrams in
Fig.~7.7 contain a symmetric vertex color factor that can not offset the odd
parity property of $q_3$. 
 
Finally we consider $T_{333}^{ {\cal F},0}$. Again this has the kinematic
structure of (\ref{tm1}) but now with color factors such as the three diagrams 
shown in Fig.~7.8. These are the only diagrams giving a triple coupling of
anomalous three-reggeon states (i.e. AAA couplings). The anomalous color
parity three reggeon state will play an important role in the next Section.
We refer to it as the ``Anomalous Odderon'' state. In SU(3) we can 
form an anomalous odderon either as a color octet or a color singlet by 
using the tensors shown in Fig.~7.9.
\begin{center}

\leavevmode
\epsfxsize=2.5in
\epsffile{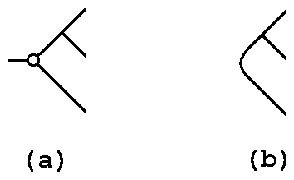}

Fig.~7.9 SU(3) Color Tensors for (a) the Octet Odderon (b) the Singlet 
Odderon.
\end{center}

The first diagram of Fig.~7.8 couples three color octet anomalous odderons. 
We obtain odd signature for each three-reggeon state by combining the even
color parity of the central $d$-tensor with the odd parity of the momentum
factor, i.e. the three-reggeon states couple with an effective triple vertex
$$
T_{333}^{ {\cal F},0}~=~
d_{i_1i_2i_3} q_3
\auto\label{tm1}
$$ 
where, again, $q_3$ is given by (\ref{q3}). The three-reggeon states have
even color parity since they are obtained by combining an odd $df$ factor
with an odd number of gluons (reggeons). 

The second diagram of Fig.~7.8 couples two color octet anomalous odderons
and one color singlet. The third diagram in Fig.~7.8 couples three color
singlet anomalous odderons and simply leads to a vertex 
$$
T_{333}^{ {\cal F},0}~=~q_3
\auto\label{tm2}
$$
with no color factor. (\ref{tm2}) exists in both SU(2) and SU(3). In SU(2) 
there is only a color singlet anomalous odderon. However, the SU(3) color octet 
anomalous odderon has a component that transforms as an SU(2) singlet with 
respect to an SU(2) subgroup. For this component the 
SU(2) version of the third diagram of Fig.~7.8 is obtained from the first
two SU(3) diagrams by projecting onto the SU(2) subgroup. Since the
three-reggeon states 
carry anomalous color parity they also will not couple to single quarks 
scattering with their helicity conserved. Again, these states will couple
between $T_{333}^{ {\cal F},0}$ vertices. 

The above arguments generalise to any number of gluons 
coupling via a single quark loop. It is straightforward to show that there 
are AAA $~T_{223}^{ {\cal F},0}$ 
vertices of the form (\ref{tm1}) and (\ref{tm2}) 
that couple anomalous color octet (triplet for SU(2)) two-reggeon states to 
color octet and color singlet 
anomalous odderons respectively. The first possibility exists only in SU(3),
of course. A $T_{233}^{ {\cal F},0}$ vertex of the form (\ref{tm}) 
exists in SU(3) with color octet anomalous odderon states. There is no 
corresponding vertex for color singlet anomalous odderons. 
Although we have discussed only the lowest-order couplings explicitly, it is 
clear that there is a large 
set of even and odd signature anomalous color parity multireggeon states that 
couple through helicity-flip vertices of the kind that we have 
isolated. Such vertices can appear in reggeon diagrams only within the 
$T^{{\cal F}}_{m'n'r'}$ vertices discussed in Section 5. 

In addition to the $T^{ {\cal F},0}$ AAA vertices there will also be
a corresponding variety of ANN vertices. In most of our discussion in the next
Section we will specifically consider only the AAA couplings of anomalous 
reggeon states. We will see that the dynamics is determined by the AAA
couplings, most importantly because the AAA coupling (\ref{tm})
provides the only $T^{ {\cal F},0}$ coupling (either AAA or ANN) of color
zero states within SU(2) and ultimately it is SU(2) color singlet couplings 
and infra-red divergences that will interest us. 

\subhead{7.6 General Couplings of Anomalous Reggeon States }

Note that while the anomalous reggeon states do not couple to
helicity-conserving elastic scattering states, they will couple in general
multiparticle amplitudes, provided only that the initial and final states 
have different parity properties. A general amplitude of this kind is
illustrated in Fig.~7.9.
\begin{center} 

\leavevmode
\epsfxsize=2in
\epsffile{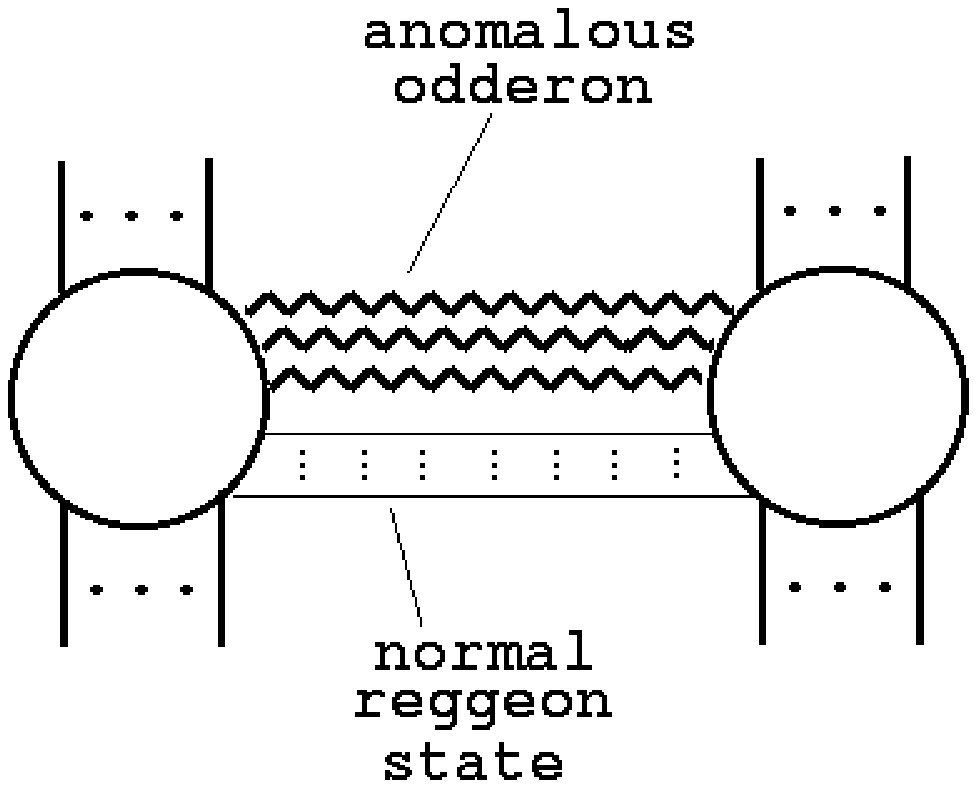}

Fig.~7.9 A General Amplitude Containing the Anomalous Color Parity Three 
Reggeon State - the ``Anomalous Odderon''
\end{center}
In such amplitudes the anomalous reggeon couplings will automatically 
satisfy the reggeon Ward
identities. The distinctive feature of the helicity-flip couplings we have 
discussed in this Section is that 
they are associated with a violation of these identities in the
massless quark theory. This is the subject of the next Section. 

\newpage 

\mainhead{8. INFRA-RED DIVERGENCES AND CONFINEMENT.} 

In the last Section we found that anomalous color parity reggeon states
couple through the special helicity-flip vertices that we isolated.
These vertices appear in 
massless quark Feynman diagrams only when the $m \to 0$ limit is taken after
a zero transverse momentum limit. In this Section we describe how, within
reggeon diagrams containing the relevant interactions, imposition of the
reggeon Ward identities with $m \neq 0$ implies 
these  vertices survive the $m \to 0$ limit.  We will then 
indicate how, in the particular circumstances that the SU(3) gauge
symmetry of QCD is broken to SU(2), infra-red divergences appear as $m \to
0$. These divergences produce what we call ``a confinement phenomenon''. By
``confinement'' we mean that a particular set of color-zero states is
selected that contains no massless multigluon states and has the necessary
completeness 
property to consistently define an S-Matrix. That is, if two or more of this
set of states initially scatter via QCD interactions, the final states
consist only of arbitrary numbers of the same set of states. Our discussion
is no more than an outline argument and certainly is not a rigorous proof 
that this form of confinement occurs. Nevertheless, we believe the argument
is straightforward and there is no reason to believe it can not 
be improved significantly. 

\subhead{8.1 Properties of Massless Reggeon Interactions}

Before discussing the effects of the helicity-flip quark loop interactions,
we first summarize what is known from existing calculations about 
the general properties of the elastic scattering reggeon amplitudes
$A^{\tau}_{mn}$ discussed in Section 5 when the gluon mass $M \to 0$. The 
best-known example of an elastic scattering reggeon amplitude is, of course,
the BFKL kernel\cite{bfkl}. We first recall the infra-red properties of this
kernel. 

Taking the massless limit in (\ref{6.10}) and including the trajectory 
contribution (\ref{traj1}) as part of the interaction, we obtain the 
leading-order singular part of the color zero kernel. This
can be written in terms of transverse momentum diagrams as in Fig.~8.1. 
\begin{center}
\leavevmode
\epsfxsize=3in
\epsffile{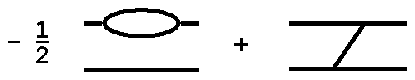}

Fig.~8.1 The Singular Part of the BFKL Kernel
\end{center}
(The full kernel is obtained by adding the diagrams with the initial states 
interchanged.) We have not shown the regular part of the kernel. As we 
remarked earlier, the regular part is uniquely determined\cite{cw} 
from the singular part by the requirement
that the full kernel satisfy the reggeon Ward identities. 
Since the notation includes momemtum-conserving  $\delta$-functions, the
diagrams are formally 
scale-invariant (even though potentially infra-red divergent). The
infra-red cancelation that provides the finiteness of the kernel 
is illustrated diagrammatically in 
Fig.~8.2. The dashed line carries zero transverse momentum. 
\begin{center}
\leavevmode
\epsfxsize=2.5in
\epsffile{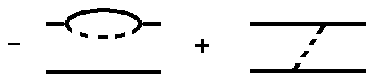}

Fig.~8.2 Infra-red Finiteness of the BFKL Kernel.
\end{center}
This cancelation is present only in the color zero channel.
When higher-order corrections to the kernel are calculated, the infra-red 
finiteness and
reggeon Ward identities persist\cite{fl}. Therefore for
our purposes, it is sufficient to frame our discussion in terms of the
leading-order diagrams.

As we have emphasized in previous Sections, the helicity-flip interactions
do not appear in elastic scattering reggeon diagrams. As a consequence, when
$M \to 0$, gauge invariance implies that the reggeon Ward 
indentities hold for all the $A^{\tau}_{mn}$. For $t = Q^2 \neq 0$, the
resulting zeroes are sufficient to compensate for any internal infra-red
divergences of the $A_{mn}$ due to the reggeon propagators (i.e. due to
the particle pole factors of $(\kbar_r^2 +M^2)^{-1}$) that we have 
included in (\ref{rpro}) as defining a reggeon propagator). Therefore, for
$Q^2 \neq 0$, all infra-red divergences arise only from particle
singularities within the reggeon interactions. 

We anticipate that the above features of the BFKL kernel generalise as 
follows. When reggeization effects are included as part of the
interaction, all color zero interactions are 
infra-red finite for $Q^2 \neq 0$. 
For non-zero color all interactions are
infra-red divergent, even when $Q^2 \neq 0$. As discussed in \cite{arw2},
reggeon unitarity implies that these divergences necessarily exponentiate
amplitudes to zero as $M \to 0$. Therefore only reggeon states with zero
$t$-channel color survive in the massless limit. Note that this is not
equivalent to confinement since the multi-reggeon states are still present
and produce a branch-point at $Q^2 = 0 $. Most important for our purposes, 
the infra-red 
finiteness of the interactions implies that the 
canonical divergence of the multi-reggeon state $Q^2 = 0 $, i.e.
$$
\int d^2\underline{k}_1\cdots  d^2\underline{k}_N\delta^2
\left(\underline{Q}-\underline{k}_1-\underline{k}_2 
\cdots -\underline{k}_N\right) \times {{1} \over {\underline{k}^2_1 }}\cdots
{{1} \over {\underline{k}^2_N }}
~~~~\sim ~~~{1 \over Q^2 }
\auto\label{mrd} 
$$
persists in the presence of interactions. Normally this divergence is
eliminated (e.g. in discussions of the BFKL equation) by using gauge
invariant couplings (the external particle-reggeon couplings $G_m$ in
(\ref{unr})) that have reggeon Ward identity zeroes. 

\subhead{8.2 Infra-red Scaling of Helicity-Flip Vertices}

We now begin our discussion of an infra-red phenomenon involving 
massless reggeons and massless quark helicity-flip vertices. A focal point
for most of the following discussion will be the reggeon diagram shown in
Fig.~8.3 in which anomalous odderon reggeon states containing three massless
reggeized gluons are coupled by two helicity-flip vertices. 
\begin{center} 

\leavevmode
\epsfxsize=2.5in
\epsffile{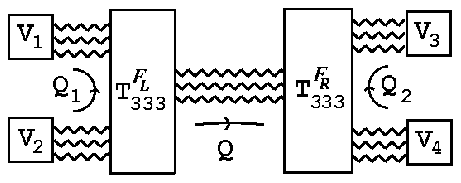}

Fig.~8.3 A Reggeon Diagram Involving Anomalous Odderon Reggeon States.

\end{center}
We suppose that this diagram is embedded in a larger diagram so that 
$Q, Q_1$ and $Q_2$ are each integrated over. The $V_i$ boxes
represent the remainder of the full diagram (in general they will be 
indirectly coupled by additional reggeons). An example of such an embedding 
is the diagram shown in Fig.~8.4. 
\begin{center}

\leavevmode
\epsfxsize=3in
\epsffile{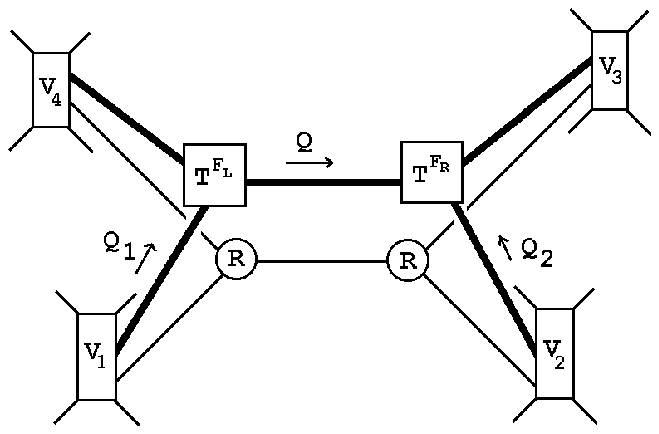}

Fig.~8.4 Embedding Fig.~8.3 in a Larger Diagram. 

\end{center}
In this diagram 
\raisebox{-0.7mm}{\epsfxsize=0.3in \epsffile{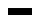}} 
$~$represents the anomalous 
odderon reggeon state and, for the moment, 
\raisebox{-0.3mm}{\epsfxsize=0.3in \epsffile{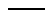}} $~$ represents any 
normal (i.e. non-anomalous) combination of reggeons. 
We take both 
\raisebox{-0.7mm}{\epsfxsize=0.3in \epsffile{dss15.ps}} and 
\raisebox{-0.3mm}{\epsfxsize=0.3in \epsffile{dss16.ps}} $~$ to be 
color singlets. For our initial discussion the gauge group could be either 
SU(3) or SU(2). Although we do not show ($A_{mn}$)
interactions within the reggeon states, they can be present within both the
odderon and the normal states without modifying our discussion. We will
discuss later interactions that link reggeons in the normal state with those
in the odderon state. Fig.~8.4 will correspond to the hexagraph in Fig.~5.26 
and will be of the form illustrated in Fig.~5.27 provided the $V_i$ have the
necessary structure. Comparing with the diagram of Fig.~5.24 it is then
clear that $T^{F_L}$ ($\equiv T_{333}^{ {\cal F}_L}$)
and $T^{F_R}$ ($\equiv T_{333}^{ {\cal F}_R}$) can contribute as 
helicity-flip vertices.
(These vertices must, of course, be energy non-conserving and couple distinct 
scattering channels as shown. In all the diagrams we discuss there will 
be a combination of a ``regular'' vertex $R$ and a $T^F$ vertex that appear
together as a single disconnected reggeon interaction. The regular vertex 
will be a non-flip, energy-conserving vertex that could appear in elastic 
scattering reggeon diagrams.) 

We suppose that $T_{333}^{ {\cal F}_L}$ and $T_{333}^{ {\cal F}_R}$ contain 
all the
diagrams analagous to those of Fig.~6.6, together with the corresponding 
Pauli-Villars regulator diagrams, that are 
needed to obtain the full range of reggeon Ward identities when 
the quark mass $m \neq 0$. Both vertices contain 
$\Gamma{\mu_1\mu_2\mu_3,m^2}$ contributions.
We concentrate on the infra-red region where we expect the presence
of the $T_{333}^{ {\cal F},0}$ vertices to be most significant, i.e. we 
consider the region 
$$ 
Q_1 ~\sim ~Q_2 ~\sim ~Q ~\to 0
\auto\label{sca}
$$ 
We also consider the internal phase-space region of the reggeon states where
each reggeon carries transverse momentum $k_i \sim Q$. In this region, as we
discussed above, color zero reggeon interactions can be present, but
because they are infra-red finite the full reggeon state scales 
canonically as ``$ 1/Q^2$ ''. Fig.~8.3 then gives 
$$
\eqalign{ \int \cdots {d^2Q_1 \over Q_1^2}{d^2Q_2 \over Q_2^2}& ~~
V_1(\cdots,Q_1)V_2(\cdots,Q - Q_1)
V_3(Q_2,\cdots) V_4(Q - Q_2,\cdots)\cr
& ~~\times \int {d^2Q \over Q^2 (Q-Q_1)^2 (Q - Q_2)^2 
}T_{333}^{ {\cal F}_L}(Q_1,Q)
T_{333}^{ {\cal F}_R}(Q,Q_2)\cr
& ~~\cr
&~~~~~~ ~~~~\times~~\hbox{[reggeon propagators]}
}
\auto\label{rwv}
$$

A vital property of the $T^{ {\cal F},0}$ vertices is that they have 
dimension one with respect to transverse momentum. This should be contrasted
with the dimension two of the elastic reggeon interaction vertices which 
appear in the $A_{mn}$, for example $\Gamma_{22}$ given by (\ref{6.9}) and 
(\ref{6.10}). When combined with the
momentum conserving $\delta$-function, dimension two interactions naturally
produce a scale-invariant massless reggeon theory in the infra-red region. 
As we observed following (\ref{q3}), the loss of
a dimension is coupled to the loss of a reggeon Ward identity. Since this 
identity is reinstated by the addition of the extra diagrams of Fig.~6.6, we 
expect the full $T_{333}^{ {\cal F}}$ to have the normal dimension two 
infra-red behavior. Therefore, when $m \neq 0$ the limit (\ref{sca}) will 
give 
$$
T_{333}^{ {\cal F}} ~~\sim ~~Q^2 
\auto\label{sca1}
$$
whereas
$$
T_{333}^{ {\cal F},0} ~~\sim ~~ C ~Q
\auto\label{sca2}
$$
where $C$ is a constant which depends on precisely how the limit (\ref{sca})
is defined in terms of the $Q_i$ and also contains a color factor. 

Let us first ignore the $Q_1$ and $Q_2$ dependence of the $V_i$ 
vertices, and consider the behavior of the remainder of (\ref{rwv}) in the 
region (\ref{sca}). If we insert (\ref{sca1}) for $T_{333}^{ {\cal F}}$ we 
obtain
$$
\int {d^2 Q  \over Q^2 }~~\biggl( \int {d^2 Q \over Q^4} 
T_{333}^{ {\cal F}_L}(Q,Q)\biggr)\biggl( \int {d^2 Q \over Q^4} 
T_{333}^{ {\cal F}_R}(Q,Q)\biggr)
~~ \sim ~~~
\int_0 ~{d^6 Q \over Q^6} 
\auto\label{rwv1}
$$
which is only logarithmically divergent and so any power convergence provided 
by the $V_i$ will be sufficient to give a finite integral. Since each
of the $V_i$ will, in general, satisfy a reggeon Ward identity giving 
$$
V_i(Q) \centerunder{$\sim$}{\raisebox{-4mm}{$Q \to 0$}}  ~~Q ~~~~\left( 
\equiv ~V(Q) \right)
\auto\label{sca3}
$$
we expect no infra-red divergence problem - provided (\ref{sca1}) holds. 
(We will use V(Q) generically in the following to indicate a coupling that
vanishes linearly in $Q$.)

If we instead insert the behavior (\ref{sca2}) for $T_{333}^{{\cal F}}$
and now include the $V_i~$, we find that (\ref{rwv1}) is replaced by 
$$ 
\int_0~ {d^6 Q \over Q^8}~ \prod_i~ V_i(Q)
\auto\label{rwv3}
$$ 
In this case at least three of the $V_i$ must satisfy (\ref{sca3}) to ensure 
convergence. If we choose, say, $V_1$ and $V_2$ to not vanish as 
$Q_1, Q_2 ~\to 0$, there will be a logarithmic divergence of the form
$$
\eqalign{ \int {d^2 Q  \over Q^2 }&~~\biggl( \int {d^2 Q \over Q^4} V(Q) 
T_{333}^{ {\cal F}_L,0}(Q,Q)\biggr)\biggl( \int {d^2 Q \over Q^4} 
V(Q) T_{333}^{ {\cal F}_R,0}(Q,Q)\biggr) \cr
& \equiv \int {d^2 Q  \over Q^2 }~~ K[T_{333}^{ {\cal F}_L,0}]
~~ K[T_{333}^{ {\cal F}_R,0}] }
\auto\label{rwv4}
$$
where the functional 
$$
K[T^ {\cal F}] ~=~ \int {d^2 Q \over Q^4} ~V(Q)~ T^ {\cal F}(Q)
\auto\label{rwv5}
$$
will occur again in the following. If $T^ {\cal F}(Q)$ satisfies
(\ref{sca2}), then $K[T^ {\cal F}]$ is logarithmically divergent in the 
infra-red region. 

Consider next a diagram with an additional $T^ {\cal F}$ vertex and
having the structure of Fig.~8.5. 
With the vertices appropriately chosen 
this diagram can be associated with the hexagraph of Fig.~5.28.
Again \raisebox{-0.7mm}{\epsfxsize=0.3in \epsffile{dss15.ps}} 
$~$represents the anomalous odderon reggeon state, 
\raisebox{-0.3mm}{\epsfxsize=0.3in \epsffile{dss16.ps}} $~$ is any normal 
reggeon state, and both can contain interactions. 
\begin{center}

\leavevmode
\epsfxsize=3in
\epsffile{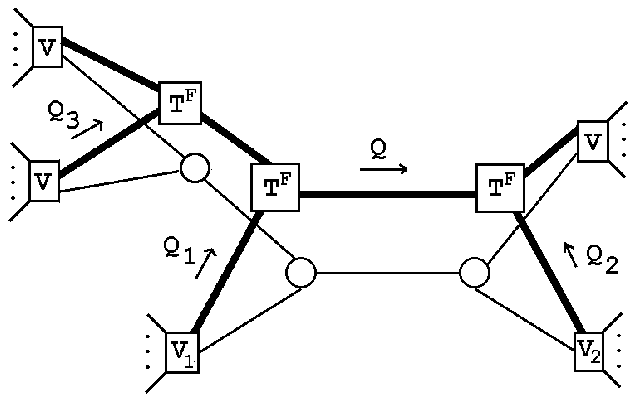}

Fig.~8.5 A Reggeon Diagram Containing Three $T^ {\cal F}$ Vertices. 
\end{center}
Now, as indicated, there
are four independent transverse momenta integrated over. 
If we again choose $V_1$ and $V_2$ to be finite when $Q_1, Q_2 ~\to 0$,
we obtain from Fig.~8.5 
$$
\int {d^2 Q  \over Q^2 }~~ K^3[V,T^{\cal F}]
\auto\label{rwv6}
$$
and, if we insert (\ref{sca2}), the overall 
logarithmic divergence persists.

Before proceeding further we consider how (\ref{sca1}) and (\ref{sca2})
are inter-related by the reggeon Ward identities as $m \to 0$. We discuss 
this in terms of a simple model that illustrates the general behavior
to be expected. 

\subhead{8.3 The Triangle Anomaly and Reggeon Ward Identities } 

We first make the separation 
$$ 
T_{333}^{\cal F}~=~T^{ {\cal F},m^2} ~+~
\tilde{T}^{ {\cal F}} ~,
\auto\label{sep}
$$
where $T^{ {\cal F},m^2}$ contains the contribution from
$\Gamma{\mu_1\mu_2\mu_3,m^2}$ and $\tilde{T}^{ {\cal F}}$ does not 
contain any singular behavior associated with 
the quark triangle diagram. If we write, in the region 
(\ref{sca}),
$$
T^{ {\cal F},m^2} (m^2,Q)~=~ T^{ {\cal F},0}~ F(Q/m) 
~=~C~Q ~F(Q/m)
\auto\label{sep1}
$$
an oversimplified  model for $F(x)$ which nevertheless gives 
the essential behavior of the triangle graph is
$$
F(x)~=~{ 1 \over (1 + x)^2 }
\auto\label{sep2}
$$
The Pauli-Villars quarks in 
$\tilde{T}^{ {\cal F}}$ will give the same singular behavior but 
with the opposite sign and with the light quark mass scale replaced by the
cut-off scale $m_{\lambda}$. Therefore we can take 
$$
\tilde{T}^{\cal F}(m,Q) ~=~ - C Q~ \left(
 {m_{\lambda}^2 \over (m_{\lambda} + Q)^2}\right) ~+ ~~\cdots  
\auto\label{sep3}
$$
and so for the full $T^{\cal F}$ we obtain
$$
\eqalign{ T^{\cal F}(m,Q) ~&=~ C Q ~\left({m^2 \over (Q + m )^2}
- {m_{\lambda}^2 \over (Q + m_{\lambda} )^2}\right) ~+ ~~\cdots  ~~\cr
& \centerunder{$\longrightarrow$}{\raisebox{-5mm}{$ Q \to 0$}}
~~C~Q^2 ~\left( { 2 \over m_{\lambda}} - {2 \over m} \right)   ~+ ~~\cdots }
\auto\label{sep4}
$$
Now consider 
$$
\eqalign {I(m)~&= ~K[V,T^{\cal F} G] \cr
& \equiv 
\int {d^2 Q \over Q^3} ~T^{\cal F}(m,Q) ~G(m,Q)}
\auto\label{sep44} 
$$
where $G(m,Q)$ is regular at $ m \sim Q 
\sim 0$ and represents the remainder of some reggeon diagram. 
Substituting our model for $T^{\cal F}(m,Q)$ we obtain
$$
\eqalign{ I(m) ~&=~ C~\int_0 {d^2 Q \over Q^2} \left( {m^2 \over (m +Q)^2 } ~ - 
 1 + {2Q \over m_{\lambda} } + \cdots \right) G(m,Q) \cr
&= ~- C ~ \int_0 {d^2 Q \over (m +Q)^2 }~G(m,Q) ~ - ~C~
\int_0 {d^2 Q \over Q^2} \left( {2Qm \over (Q +m)^2 } - {2Q \over m_{\lambda} 
} + \cdots \right) ~ G(m,Q) \cr
&  \equiv I_1(m) + I_2(m) }
\auto\label{sep5}
$$
$I_2(m)$ is finite as $m \to 0$, while $I_1(m)$ gives
$$
I_1(m) ~~ \to ~~ - C~ ln[m^2]~ G(0,0) 
\auto\label{sep6}
$$
Therefore we have a logarithmic divergence with the residue given by the 
remainder of the regggeon diagram evaluayed at $Q=0$.

In the above model we have 
$$ 
T^{\cal F}(0, Q) ~~\sim~~ - ~C  Q ~ + ~O(Q^2)
\auto\label{sep62}
$$
where the leading term can simply be identified with $- T^{{\cal F},0}$. 
The model illustrates simply the general situation. The use 
of a Pauli-Villars ultra-violet cut-off implies that in the infra-red 
region, where all transverse momenta are uniformly small, the reggeon Ward 
identities are satisfied by a simple cancellation between the light quark 
triangle graph and the corresponding regulator graph. However, the 
non-uniformity in the neighborhood of $Q \sim m \sim 0$ implies that the 
limits $Q \to 0$ and $m \to 0$ do not commute for the light quark graph.
Consequently, the satisfaction of the reggeon Ward identities when $m \neq
0$ implies that they are partially lost in the limit $m \to 0$. However, the
offending contribution, i.e. $T^{{\cal F},0}$, can be evaluated in terms of
a loop of on-shell massless quarks. As we discussed in Section 6, such a
contribution can violate the reggeon Ward identities while not violating the
underlying Ward identities that give the gauge invariance of the theory. 

It is apparent from (\ref{sep3}) - (\ref{sep62}) that we are seeing the 
infra-red presence\cite{cg} of the anomaly in the 
triangle graph reflected in reggeon interactions involving anomalous parity
reggeon states. 
This happens for the reasons discussed in sub-section 7.1. Gluon Ward
identities relate the longitudinal Regge limit interactions to 
transverse interactions that can be sensitive to the anomaly. In addition 
the anomalous color parity of, for example, the anomalous odderon
three-reggeon state determines that, effectively, 
it has an infra-red ``axial-vector coupling'' via on-shell 
quark states. (As we stated would be the case, in this Section we have 
considered only AAA couplings. We recall from the last Section that, as for 
the normal anomaly, we also have ANN couplings). 

We can also view our ultra-violet regularization 
procedure, using Pauli-Villars regulator fermions, as responsible for
introducing the 
anomaly in the infra-red region. From general arguments we expect 
the fermion anomaly to introduce an ambiguous interplay between infra-red 
and ultra-violet behavior in the massless quark theory. 
Our manipulations can be viewed as fixing 
this ambiguity by requiring a finite reggeon theory and 
reggeon Ward identities for the massive quark theory. In fact, as we discuss 
further in the next paper, this is very likely to be the only resolution of
this ambiguity that gives a unitary solution to the theory. 

\subhead{8.4 Infra-Red Divergence of Diagrams with Many Helicity-Flip Vertices} 

Consider now an arbitrary reggeon diagram containing many $T^{\cal F}$ 
vertices, for example the diagram shown in Fig.~8.6.
\begin{center}

\leavevmode
\epsfxsize=4in
\epsffile{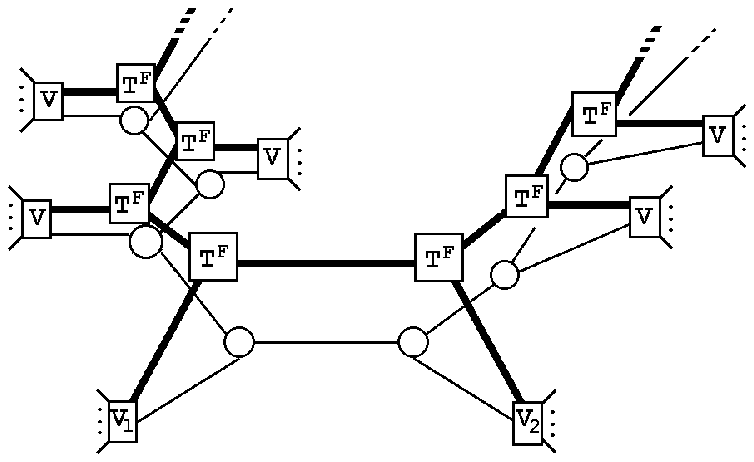}

Fig.~8.6 A Reggeon Diagram Involving Many Helicity-Flip Vertices

\end{center}
Once again \raisebox{-0.7mm}{\epsfxsize=0.3in \epsffile{dss15.ps}} 
$~$represents the anomalous odderon reggeon state and 
\raisebox{-0.3mm}{\epsfxsize=0.3in \epsffile{dss16.ps}} $~$ is any normal 
reggeon state.
As illustrated, for every new $T^{\cal F}$ vertex introduced there is 
inevitably an accompanying $V$ vertex which, from normal QCD interactions, 
will satisfy a reggeon 
Ward identity. Consequently, if we impose that $V_1$ and $V_2$ are non-zero 
when the anomalous reggeon state carries zero momentum and consider 
$T^{\cal F}$ to be the full massless quark vertex, the diagram will have an
overall infra-red logarithmic divergence of the form 
$$
\int {d^2 Q  \over Q^2 }~~\left(K[V,T^{\cal F}]\right)^{n_T}
\auto\label{rwv7}
$$
where $n_T$ is the number of $T^{\cal F}$ vertices in the diagram. 
From (\ref{sep5}), it is clear that the 
residue involves evaluating every $T^{\cal F}$, and therefore every 
anomalous reggeon state, at zero transverse momentum. 
As before, including interactions within the anomalous or the normal reggeon 
states does not change the discussion. Recall also that, as we 
emphasized in Section 5, because the divergence involves helicity-flip 
vertices, there is implicitly a zero longitudinal component also 
associated with the zero transverse momentum of the anomalous reggeon state.

Imposing that $V_1$ and $V_2$ are non-zero 
when the anomalous reggeon state carries zero momentum is equivalent to
choosing two initial reggeon scattering states that contain a zero 
momentum anomalous component. Fig.~8.6 shows that 
if these states are allowed to scatter (within QCD) into general reggeon
states, an overall logarithmic divergence selects final states having the 
same property. This is potentially a completeness property for this class of 
reggeon states. The crucial question is then whether the infra-red
divergence we have found in the class of diagrams we have studied can be
canceled by a similar divergence in some further class of diagrams. This is
the subject of the next two sub-sections. 

\subhead{8.5 Cancellation of Infra-Red Divergences} 

In this sub-section we will give an argument suggesting that if all reggeons 
are massless, i.e. SU(3) gauge symmetry is fully restored, then 
the infra-red divergence that we have discussed cancels when all diagrams
are summed over. We formulate the argument by discussing the reggeon diagram 
of Fig.~8.7. This is the lowest-order diagram that is most obviously of the
form we have discussed. 
All reggeon lines represent a single reggeized (massless)
gluon and, since this is a ``lowest-order diagram'', we specifically exclude 
interactions within either the anomalous odderon or the normal reggon 
states. The multi-reggeon states, for which reggeon propagators are
present, are indicated by the thin vertical line. To avoid
the exponentiation of infra-red divergences in higher-orders these states
must carry color zero. In lowest-order, 
the ``regular'' interaction R between the normal reggeon states will 
actually be disconnected. 
Fig.~8.7 clearly has the form illustrated in Fig.~8.4 once the
anomalous odderon three reggeon state is identified with 
\raisebox{-0.7mm}{\epsfxsize=0.3in \epsffile{dss15.ps}}$~$ and the
remaining two reggeon state is identified with
\raisebox{-0.3mm}{\epsfxsize=0.3in \epsffile{dss16.ps}}$~$. 
The logarithmic divergence is present (as $m \to 0$)
provided only that $V_1$ and $V_2$ are appropriately chosen. 
\begin{center} 

\leavevmode
\epsfxsize=3.2in
\epsffile{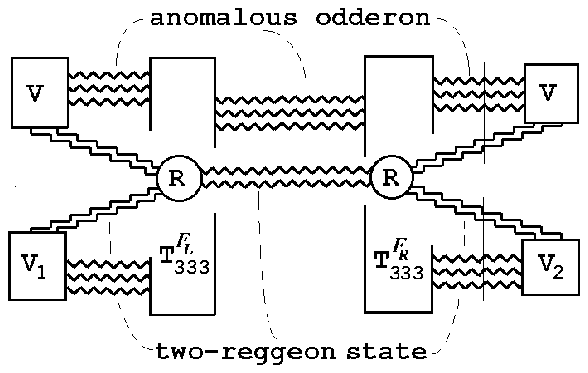}

Fig.~8.7 An Infra-red Divergent Diagram.

\end{center}

In the previous discussion of this Section we have assumed that the anomalous
odderon state separately carries zero color. In this case the two-reggeon state
must also carry zero color. Fig.~8.7 is then the lowest-order diagram
containing Fig.~8.3. However, as we discussed in the previous Section, in SU(3)
the anomalous odderon can also carry octet color. We also showed that 
helicity-flip $T^{{\cal F},0}$ triple odderon couplings exist when 
either all the odderons, or two of the three, carry octet color. In addition
there are anomalous reggeon states, with $T^{{\cal F},0}$ couplings, 
that contain only two reggeons and carry octet color. In fact, once we allow
anomalous reggeon states that are not color singlets, Fig.~8.7 is not the
lowest-order diagram containing the $m=0$ divergence. The lowest-order
diagrams involve combinations of 
normal one and two-reggeon states with anomalous two and three reggeon 
states. Because the lowest order R vertices contain only gluon internal
interactions the lowest-order diagrams involve only reggeized gluon reggeons. 
When internal quark interactions are included in the R vertices (or 
we consider the scattering of multi-quark reggeon states, as we will do in 
the next paper) the symmetric octet bound-state reggeon also appears. In
this case a particularly simple potential cancelation is between the reggeon
states illustrated in Fig.~8.8. 
\begin{center}

\leavevmode
\epsfxsize=3in
\epsffile{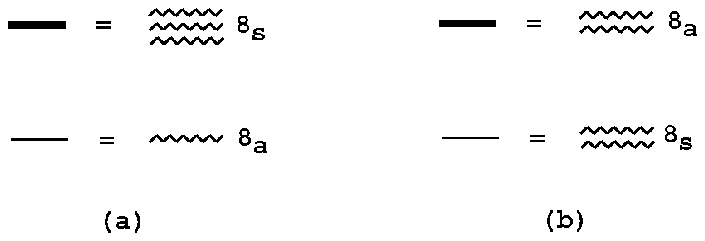}

Fig.~8.8 Potential Canceling Configurations. 

\end{center}
This cancelation will be particularly relevant for
our discussion of deep-inelastic scattering in the next paper.

Since, as we have already said, Fig.~8.7 is the lowest-order diagram that
fits specifically into our previous discussion, we will concentrate on 
finding diagrams that cancel the divergence of this particular diagram. 
Since the R vertices are lowest-order they can not 
involve internal quark interactions. Consequently the symmetric octet reggeon
can not appear in canceling diagrams. Note also that, since color parity is
conserved and we have chosen each of the reggeon channels in Fig.~8.7 to
carry anomalous color parity, we do not need to consider AVV vertices (in
addition to AAA vertices) when looking for cancelations. We 
proceed by considering possible alternative
couplings for the reggeons originating from $V_1$. 

If reggeons within the anomalous and normal reggeon states interact
additional reggeon propagators are introduced and a cancelation with
Fig.~8.7 is not possible. The most obvious possibility for a cancelation is
that a reggeon participating in the anomalous odderon interaction instead
participates in the regular reggeon interaction, as illustrated in Fig.~8.9.
To produce a zero quark mass divergence identical to that in Fig.~8.7 the
regular reggeon interaction must give an infra-red divergence involving the
indicated dashed lines. 
\begin{center}

\leavevmode
\epsfxsize=3in
\epsffile{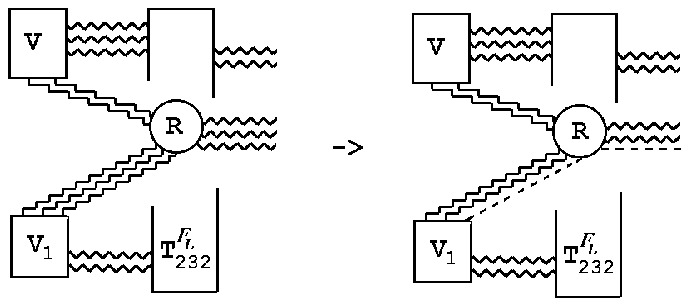}

Fig.~8.9 Divergences to be Produced by the Regular Reggeon Interaction.

\end{center}
Because the anomalous states with just two reggeons must carry octet color,
the regular reggeon interaction also carries net octet color. A normal 
reggeon interaction carrying non-zero color is necessarily divergent.
The simplest divergence will be produced by a massless $R_{22}$ interaction, 
as in Fig.~8.10(a). 
Since the anomalous odderon reggeons are participating in a helicity-flip 
interaction, it is also possible for an infra-red divergent interaction
to occur as in Fig.~8.10(b). 
As we discussed in sub-section 6.5, 
similar reggeon infra-red divergent interactions to those of Fig.~8.10 are
involved in producing the reggeon Ward identities for the $T^{F}$ vertices,
for example the third diagram of Fig.~6.6. 
\begin{center}

\leavevmode
\epsfxsize=4.5in
\epsffile{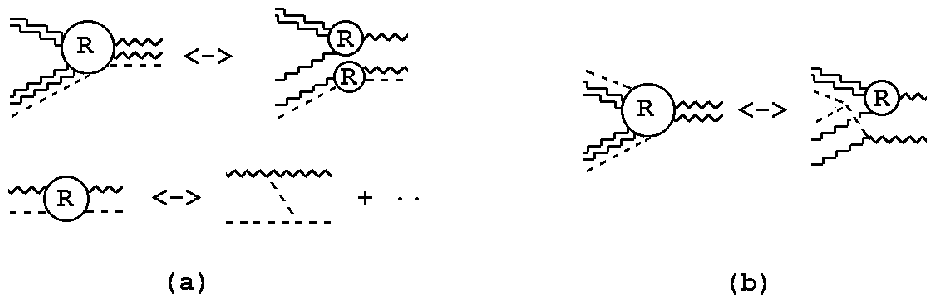}

Fig.~8.10 Regular Reggeon Interactions Producing Divergences.

\end{center}

For $m \neq 0$, the complete
cancelation of all divergences related to those of Fig.~8.10 will necessarily
involve all possible interactions 
between the color zero five reggeon states. This is achieved if we combine
all left and right side diagrams of the form of Fig.~8.9 with the
corresponding diagrams forming Fig.~8.7.  In this way we obtain a set of
diagrams containing triple anomalous reggeon vertices, which each have the 
$m = 0$ divergence and which, when $m\neq 0$, are related by the cancelation 
of divergences of the form of Fig.~8.9. In the infra-red region producing the 
$m= 0$ divergence the cancelation
of divergences related to Fig.~8.9 is between reggeon interactions having
the distinct forms shown in Fig.~8.11 (all dashed lines carry zero
transverse momentum). 
\begin{center}

\leavevmode
\epsfxsize=4in
\epsffile{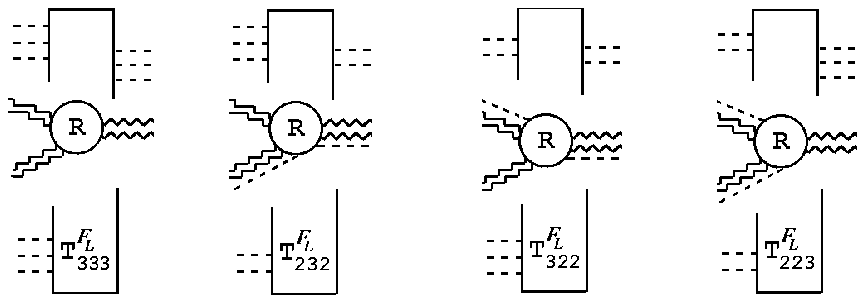}

Fig.~8.11 The Reggeon Interactions Producing the Infra-red Cancelation.

\end{center}

Each of the interactions in Fig.~8.11 contains the $m=0$ anomalous
interaction and scales appropriately to generate the logarithmic divergence
in individual diagrams. However, the additional infra-red cancelation
between the complete set of diagrams should survive the $m \to 0$ limit and
be sufficient to ensure that there is no $m = 0$ divergence. 
If we go to higher-order and incorporate reggeon interactions within and 
between the normal and anomalous states, we can expect more elaborate 
cancelations to hold. We can also expect the ANN vertices to play a 
role. We note that the crucial feature of the cancelation is the existence 
of infra-red divergent interactions between the reggeons in the anomalous 
odderon state and the reggeons in the normal state. This will be an 
important dynamical element of our further discussion.

\subhead{8.6 Symmetry Breaking and Confinement} 

Suppose now that the SU(3) gauge symmetry is only partially restored to 
SU(2). In this case five of the eight SU(3) gluons remain massive. There is 
one SU(2) singlet and two SU(2) doublets. We use the notation of Fig.~8.12.
\begin{center}

\leavevmode
\epsfxsize=3.5in
\epsffile{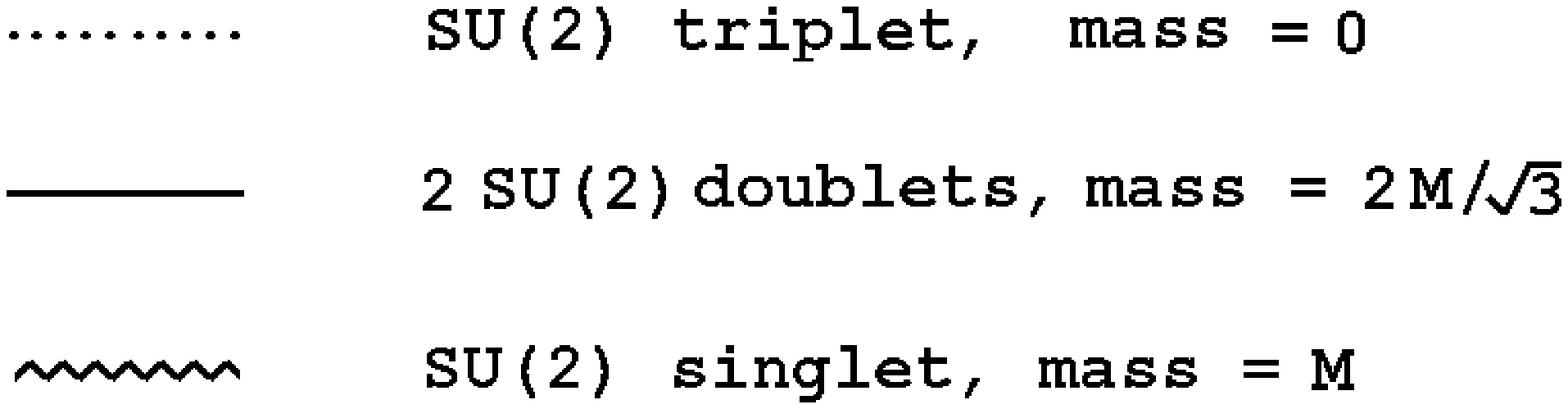}

Fig.~8.12 Notation for the Gluon Spectrum when the Gauge Symmetry is Broken. 

\end{center}
The $f$- and $d$- couplings of the different representations are
illustrated in Fig.~8.13.
\begin{center}

\leavevmode
\epsfxsize=4in
\epsffile{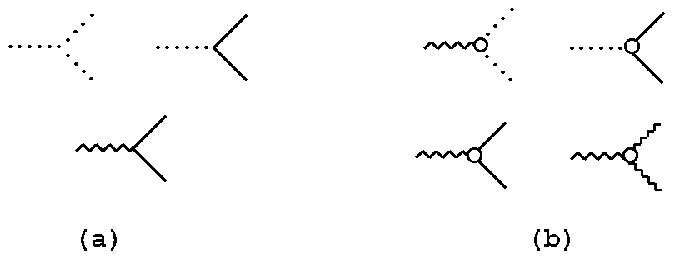}

Fig.~8.13 (a) $f$- Couplings and (b) $d$- Couplings after Symmetry Breaking. 

\end{center}
The resulting trajectory function transverse momentum diagrams are shown in 
Fig.~8.14.
\begin{center}

\leavevmode
\epsfxsize=3.5in
\epsffile{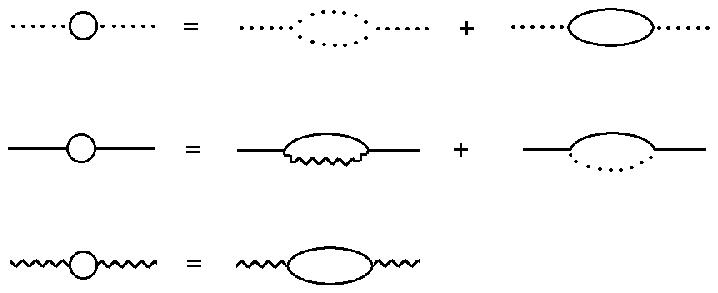}

Fig.~8.14 Trajectory Functions for the Different Representations. 

\end{center}

The SU(2) singlet trajectory function contains no massless reggeon 
contributions and so is manifestly infra-red finite. Therefore this gluon is 
a simple massive reggeon which, if color charge parity is carried over from
the unbroken theory, carries negative color parity. The two
SU(2) doublets form SU(2) singlets with both even 
and odd SU(3) color parity. The odd color parity combination gives 
the reggeization of the color singlet reggeon shown  in Fig.~8.14. The 
even color parity doublet forms a separate infra-red finite, even
signature, ``bound-state'' reggeon with a trajectory 
that is exchange degenerate with 
the singlet reggeized gluon trajectory.
The cancelation of Fig.~8.15
demonstrates simultaneously the infra-red finiteness and the reggeization of
this trajectory, provided we omit the contribution of the massless reggeons.
The reason for this omission will soon become apparent. 
\begin{center}

\leavevmode
\epsfxsize=4in
\epsffile{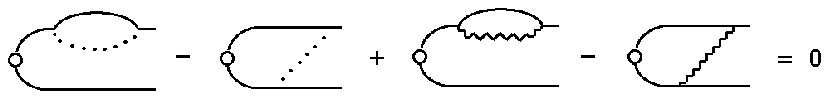}

Fig.~8.15 Reggeization of the Bound State Reggeon. 

\end{center}
In the massless limit, i.e. as the full SU(3) gauge symmetry is restored,
this bound-state trajectory becomes the even signature octet 
trajectory, that we referred to in sub-sections 5.3 and 6.2.

Initially we consider the complete set of reggeon diagrams containing 
both massless and massive reggeons. A-priori the $m=0$ logarithmic
divergence we have discussed will still be present 
in individual diagrams containing the relevant configurations of massless SU(2) 
reggeons. For example, if we consider Fig.~8.7 to be composed entirely of
SU(2) massless reggeons, then the divergence will be present. However, 
the reggeon infra-red cancelation of Figs.~8.9 - 8.11 also remains valid. In
fact the necessary infra-red divergent interactions will exist and so, 
presumably, an analagous cancellation will take place, 
provided only that one of the
normal reggeon states in the diagram contains massless reggeons. 

An obviously divergent class of diagrams is those of the form of Fig.~8.16.
This diagram is an SU(2) version of Fig.~8.7, except that the normal reggeon
states contain no massless reggeons. 
\begin{center}

\leavevmode
\epsfxsize=3.2in
\epsffile{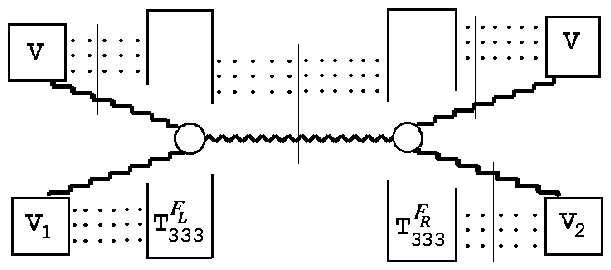}

Fig.~8.16 A Diagram Containing SU(2) Singlet Reggeons. 

\end{center}
The reggeons indicated by the dotted lines are massless and form an SU(2)
singlet anomalous odderon state. (From now on we use the 
dotted line notation to indicate reggeons that both belong to the massless 
SU(2) triplet and carry zero transverse momentum in the overall infra-red
divergence of the diagram.) All the multi-reggeon states cut by a thin
vertical line are SU(2) singlets if, in particular, the additional 
reggeon states indicated by a thick unbroken reggeon line are (some 
number of) the SU(2) singlet, massive, reggeized gluons. We recall from the
last Section that since we now discuss SU(2) color only, to contain
$T^{{\cal F}, 0}$ color zero interactions the $T^{\cal F}$'s must be AAA
couplings of anomalous odderons. Since there is no triple coupling
for the singlet reggeon, we can not take all the normal states to contain
only a single reggeon. (It would be sufficient for some normal states to be a
single bound state reggeon but these states require an internal quark regular
interaction in order to couple. As we shall see shortly, there is also an
additional subtlety involved.) 
 
Fig.~8.16 is obviously of the form of Fig.~8.4 and so contains the
logarithmic divergence. The anomalous odderon three 
reggeon state once again corresponds to \raisebox{-0.7mm}{\epsfxsize=0.3in
\epsffile{dss15.ps}} $~$ while 
\raisebox{-0.3mm}{\epsfxsize=0.3in \epsffile{dss16.ps}} $~$ is now
identified with reggeon states that, in lowest-order, consist of massive
SU(2) singlet gluons. From Fig.~8.13(a) 
it is clear that, at lowest-order, the singlet simply has no coupling to the 
massless sector. As a result there are no infra-red divergent
interactions analagous to Fig.~8.10 and no cancelation corresponding to
Fig.~8.11.  The analog of the interactions of Fig.~8.10 involve the exchange 
of a massive SU(2) doublet. That is the divergent interactions that were 
part of the cancelation with the SU(3) symmetry unbroken now contain massive 
propagators. This implies that the logarithmic divergence as $m \to 0$ is 
qualitatively of the form 
$$
\int_{m^2} d Q^2 ~ \left( {1 \over Q^2 }~ -~ {1 \over Q^2 + M^2}\right) 
~~~ \sim~~~ \ln \left( {M^2 \over m^2} \right)
\auto\label{mdv}
$$
and so is clearly a direct consequence of the symmetry breaking. 

All diagrams having the form of Figs.8.5, 8.6 etc. will similarly contain an
uncanceled overall logarithmic divergence (with $V_1$ and $V_2$
appropriately chosen) if the \raisebox{-0.7mm}{\epsfxsize=0.3in
\epsffile{dss15.ps}}$~$ state contains any number of (interacting) massless
reggeons forming a state with the quantum numbers of the anomalous odderon
and \raisebox{-0.3mm}{\epsfxsize=0.3in \epsffile{dss16.ps}} $~$ is any 
combination of (interacting) masssive SU(2) singlet reggeon states. 
Interactions between the massless and massive reggeons can take place but, 
since they are infra-red finite, they simply produce reggeon Ward identity
zeroes that eliminate the overall infra-red divergence. Therefore such 
interactions do not appear in the divergent diagrams. If the anomalous 
reggeon state carries color, interactions within this state will 
exponentiate the diagram to zero.

Clearly 
\raisebox{-0.3mm}{\epsfxsize=0.3in \epsffile{dss16.ps}} $~$ could also be a 
multi-quark reggeon state, but we will leave a discussion of quark reggeon 
states until the next paper. As preparation for our discussion of chiral 
symmetry-breaking it will be interesting to discuss here how the 
bound-state reggeon avoids an infra-red interaction of the form of Fig.~8.9.
At lowest-order the bound-state reggeon couples to infra-red divergences
via the two diagrams illustrated in Fig.~8.17. 
\begin{center}

\leavevmode
\epsfxsize=1.7in
\epsffile{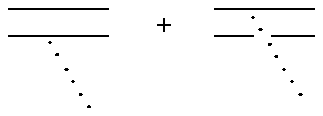}

Fig.~8.17 The Bound-state Coupling to Infra-red Divergences.

\end{center}
Because of the antisymmetry of the gauge coupling, if the two massive reggeons 
are in a completely symmetric state, the two diagrams of Fig.~8.17 cancel. 
Since it is even signature and symmetric with respect to color this requires 
that the
bound-state carry positive parity (which it does). As we shall enlarge on in
the second paper, this same argument implies that the quark component of
scalar bound-states must carry positive parity. The anomalous odderon
component then gives an overall negative parity and produces the massless
pseudoscalars associated with chiral symmetry breaking. 

We now take all the amplitudes containing the logarithmic divergence as our
physical amplitudes. We remove the divergence as a normalization factor and
also factorize off all the 
$V$ couplings. We are left with a set of multi-reggeon diagrams in which 
every reggeon state has the form shown in Fig.~8.18
\begin{center}
\leavevmode
\epsfxsize=1.5in 
\epsffile{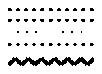}

Fig.~8.18 The Confinement Reggeon States

\end{center}
where now the wee parton component $~$
\raisebox{-0.7mm}{\epsfxsize=0.5in \epsffile{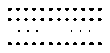}}$~~$ contains 
arbitrary numbers of massless reggeons with odd signature, color zero, and 
positive color parity. Each massless reggeon carries zero transverse
momentum. \raisebox{.1mm}{\epsfxsize=0.6in \epsffile{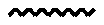}}$~$ is 
any combination of masssive SU(2) singlet reggeon states. Note that the 
odd-signature nature of the wee parton component switches the signature of 
the massive reggeon component of states. In particular, the odd-signature
elementary reggeon gives an even signature ``pomeron'' while the bound-state 
reggeon gives an odd-signature, exchange degenerate, partner to the pomeron.
Because of signature factors the pomeron will not generate a vector particle 
while the odd-signature bound-state Regge pole will give such a particle at
the mass of the SU(2) singlet. In effect, while the reggeized gluon becomes
the pomeron, the unconfined massive single gluon vector particle, that in
perturbation theory lies on the reggeized gluon trajectory, is replaced by a
composite bound-state of confined massive gluons. 

We have thus demonstrated the ``confinement phenomenon'' which we referred to
earlier. If we insist that two initial 
scattering reggeon states have the form of Fig.~8.18 then these states
scatter into arbitrary numbers of the same states only. Also, since the
wee parton component of the state acts like a background ``reggeon
condensate'' the dynamical properties of the reggeon states are identical to
that of the SU(2) singlet reggeon component of the state. Therefore we also
have confinement in the sense that we have only massive reggeon states
composed of elementary Regge pole constituents. 

As we have emphasized throughout this paper, the zero transverse momenta
involved in producing the infra-red divergences and reggeon condensate are 
implicitly accompanied by longitudinal zero momenta. 
The presence of this longitudinal component
implies that the condensate can potentially be understood as a light-cone
zero-mode contribution at finite momentum or, in the language of the
Introduction, as a ``wee parton'' component at infinite momentum. 

\subhead{8.7 The Supercritical Pomeron }

Finally we note that the divergent diagrams will also include those of the 
form illustrated in Fig.~8.19 in which the helicity-flip $T^{\cal F}$ 
vertices, in addition to coupling the zero tranverse momentum anomalous 
odderon massless reggeons, produce an additional pair of massive 
reggeons carrying zero net transverse momentum.
The $T^{\cal F}$ vertices involved will also contain the triangle anomaly 
we have discussed. 
\begin{center}

\leavevmode
\epsfxsize=4.5in
\epsffile{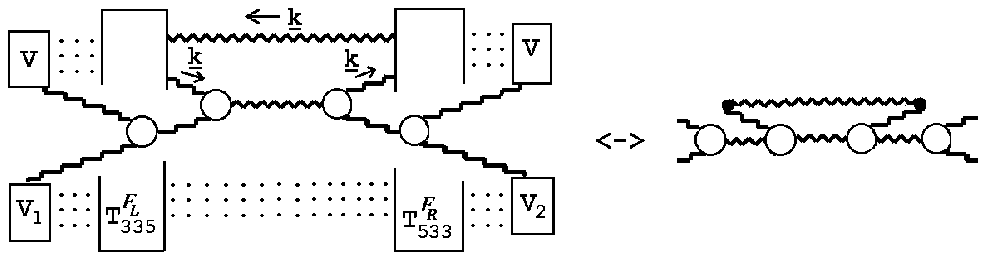}

Fig.~8.19 A Diagram With Vacuum Production of SU(2) Singlet Reggeons. 

\end{center}
The reggeon lines in the right-hand diagram of Fig.~8.19 are ``physical'', 
i.e. they correspond to either the pomeron or it's odd signature partner.
Diagrams such as Fig.~8.19, together with all the obvious 
generalizations are, effectively, responsible for ``vacuum production'' of 
massive reggeon states within the reggeon diagrams describing our confining
theory. 

We can, therefore, summarize our confining solution of partially-broken
QCD as containing exchange degenerate even and odd signature reggeons,
with vacuum production of
multi-reggeon states. These are the defining characteristics of
supercritical pomeron RFT\cite{arw1}. We have shown that the appearance of
this RFT phase is a consequence of the confinement produced by the infra-red
divergence associated with the massless quark anomaly. (Having derived the 
massless theory, it should be 
possible to add effective quark masses to the theory, for example by chiral 
perturbation theory, and still remain in the supercritical phase.) We have 
postponed discussion of the RFT formulation of the supercritical phase to 
the following paper because we want to emphasize the self-contained nature 
of the QCD infra-red analysis. 

We have explicitly associated the supercritical phase with the 
breaking of SU(3) gauge symmetry to SU(2). The restoration of SU(3) symmetry
should follow if we take the zero mass limit for the SU(2) singlet reggeon.
This is equivalent to setting the intercept of the pomeron to zero. The 
principle of complimentarity\cite{fs} implies that the symmetry can be 
smoothly restored provided only that an ultra-violet cut-off is introduced.
However, since the massless quark divergence has selected only a part of the
broken theory, restoration of full SU(3) symmetry is clearly non-trivial.
Nevertheless, provided we can completely identify our solution of
partially-broken QCD with the super-critical pomeron, setting the 
pomeron intercept to zero corresponds to taking the critical limit from 
within the super-critical phase. Note that two additional important features
of this limit must also be realized. That is, both the odd-signature
reggeon partner for the pomeron and the vacuum production of Fig.~8.18 must
simultaneously decouple as the pomeron intercept vanishes. The reinstatement
of the infra-red cancellation of Figs.~8.8 - 8.10 is presumably involved in
these effects in a subtle manner. 

An inescapable conclusion from our construction is that the pomeron 
carries odd color charge parity. The odd and even color parity of the
reggeized gluon and the wee parton component, respectively, combine 
to give overall odd color parity. This property will persist after the SU(3) 
gauge symmetry is restored. Note that to obtain an SU(3) color singlet, 
the anomalous odderon that appears in the pomeron has to be 
an SU(3) octet (rather than the singlet discussed initially in sub-section 
8.5). For an odd color-parity pomeron to describe total 
cross-sections, the scattering hadrons can not be eigenstates of color 
parity. We will show in our next paper that the pion is a mixture of states
with even and odd color parity (but odd physical parity). The
quark-antiquark and anomalous odderon components are, correspondingly, in
either a color singlet or a color octet state. The
pomeron scatters the odd(even) state into the even(odd) state. 

The RFT formalism also tells us 
that the transverse momentum cut-off is a relevant parameter for the 
critical limit. Therefore, if this (gauge-invariant) cut-off is
varied it is possible for the supercritical phase to appear even when 
the full gauge symmetry is restored. In this case the direction of 
the SU(2) wee parton component is effectively averaged over within 
SU(3). In the next paper, we will discuss how this can be understood
as an average over the
SU(2) direction of the anomaly (or instanton effects) in SU(3).
It is also possible to regard the large $Q^2$ of deep-inelastic scattering
as introducing a ``finite volume'' effect which removes the critical
phase-transition. As a result the theory remains in the ``single gluon
dominated'' supercritical phase as the SU(3) symmetry is restored. With the
wee-parton component included, this feature can be seen explicitly by
studying the reggeon/gluon diagrams involved\cite{arw3}. Deep-inelastic 
scattering is another subject that will be
covered in depth in the following paper. 

\vspace{0.5in}

\noindent {\Large \bf Acknowledgements}

I am particularly grateful to Mark W\"usthoff for extensive discussions of 
the contents of this paper. I have also benefited, over the years, from 
many discussions with Jochen Bartels and Lev Lipatov.

\end{document}